\documentclass[a4paper,11pt]{article}
\usepackage{jcappub} 
\usepackage{lineno}
\usepackage{lscape}
\usepackage{makecell}
\usepackage{geometry}

\usepackage{graphicx}
\usepackage{dcolumn}
\usepackage{bm}
\usepackage{xspace}
\usepackage{xcolor}
\usepackage{hyperref}
\hypersetup{colorlinks=true,citecolor=blue,filecolor=blue,urlcolor=blue,}
\usepackage[noabbrev]{cleveref}
\usepackage{orcidlink}

\newcommand{\lya}{Ly$\alpha$ }
\newcommand{\hMpc}{h^{-1}\,\mathrm{Mpc}}

\title{\boldmath Clustering of high-redshift quasars with DESI DR2}

\author[a,b]{{M.~Charles}\orcidlink{0009-0006-4036-4919},}
\author[b,c,a]{{P.~Martini}\orcidlink{0000-0002-4279-4182},}
\author[b,c,a]{{A.~J.~Ross}\orcidlink{0000-0002-7522-9083},}
\author[c,a]{{D.~H.~Weinberg}\orcidlink{0000-0001-7775-7261},}
\author[d]{{J.~Aguilar},}
\author[e]{{S.~Ahlen}\orcidlink{0000-0001-6098-7247},}
\author[f,g]{{D.~Bianchi}\orcidlink{0000-0001-9712-0006},}
\author[h]{{D.~Brooks},}
\author[i,j]{{F.~J.~Castander}\orcidlink{0000-0001-7316-4573},}
\author[d]{{T.~Claybaugh},}
\author[d]{{A.~Cuceu}\orcidlink{0000-0002-2169-0595},}
\author[k]{{A.~de la Macorra}\orcidlink{0000-0002-1769-1640},}
\author[h]{{P.~Doel},}
\author[d,l]{{S.~Ferraro}\orcidlink{0000-0003-4992-7854},}
\author[m,n]{{A.~Font-Ribera}\orcidlink{0000-0002-3033-7312},}
\author[o,p]{{J.~E.~Forero-Romero}\orcidlink{0000-0002-2890-3725},}
\author[q]{{S.~{Gontcho A Gontcho}}\orcidlink{0000-0003-3142-233X},}
\author[r]{{G.~Gutierrez},}
\author[d]{{J.~Guy}\orcidlink{0000-0001-9822-6793},}
\author[s]{{C.~Hahn}\orcidlink{0000-0003-1197-0902},}
\author[t,u]{{H.~K.~Herrera-Alcantar}\orcidlink{0000-0002-9136-9609},}
\author[b,v,a]{{K.~Honscheid}\orcidlink{0000-0002-6550-2023},}
\author[w]{{C.~Howlett}\orcidlink{0000-0002-1081-9410},}
\author[x]{{M.~Ishak}\orcidlink{0000-0002-6024-466X},}
\author[y]{{R.~Joyce}\orcidlink{0000-0003-0201-5241},}
\author[z]{{D.~Kirkby}\orcidlink{0000-0002-8828-5463},}
\author[d]{{A.~Kremin}\orcidlink{0000-0001-6356-7424},}
\author[d]{{M.~Landriau}\orcidlink{0000-0003-1838-8528},}
\author[aa]{{L.~Le~Guillou}\orcidlink{0000-0001-7178-8868},}
\author[d]{{M.~E.~Levi}\orcidlink{0000-0003-1887-1018},}
\author[ab,n]{{M.~Manera}\orcidlink{0000-0003-4962-8934},}
\author[y]{{A.~Meisner}\orcidlink{0000-0002-1125-7384},}
\author[m,n]{{R.~Miquel},}
\author[ac]{{J.~Moustakas}\orcidlink{0000-0002-2733-4559},}
\author[ad]{{A.~D.~Myers},}
\author[ae]{{S.~Nadathur}\orcidlink{0000-0001-9070-3102},}
\author[u,d]{{N.~Palanque-Delabrouille}\orcidlink{0000-0003-3188-784X},}
\author[af,ag,ah]{{W.~J.~Percival}\orcidlink{0000-0002-0644-5727},}
\author[ai]{{F.~Prada}\orcidlink{0000-0001-7145-8674},}
\author[aj]{{I.~P\'erez-R\`afols}\orcidlink{0000-0001-6979-0125},}
\author[ak]{{G.~Rossi},}
\author[al,am,an]{{L.~Samushia}\orcidlink{0000-0002-1609-5687},}
\author[ao]{{E.~Sanchez}\orcidlink{0000-0002-9646-8198},}
\author[d]{{D.~Schlegel},}
\author[ap,aq]{{M.~Schubnell},}
\author[y]{{D.~Sprayberry},}
\author[aq]{{G.~Tarl\'{e}}\orcidlink{0000-0003-1704-0781},}
\author[y]{{B.~A.~Weaver},}
\author[d]{{R.~Zhou}\orcidlink{0000-0001-5381-4372}}

\affiliation[a]{Department of Physics, The Ohio State University, 191 West Woodruff Avenue, Columbus, OH 43210, USA}
\affiliation[b]{Center for Cosmology and AstroParticle Physics, The Ohio State University, 191 West Woodruff Avenue, Columbus, OH 43210, USA}
\affiliation[c]{Department of Astronomy, The Ohio State University, 4055 McPherson Laboratory, 140 W 18th Avenue, Columbus, OH 43210, USA}
\affiliation[d]{Lawrence Berkeley National Laboratory, 1 Cyclotron Road, Berkeley, CA 94720, USA}
\affiliation[e]{Department of Physics, Boston University, 590 Commonwealth Avenue, Boston, MA 02215 USA}
\affiliation[f]{Dipartimento di Fisica ``Aldo Pontremoli'', Universit\`a degli Studi di Milano, Via Celoria 16, I-20133 Milano, Italy}
\affiliation[g]{INAF-Osservatorio Astronomico di Brera, Via Brera 28, 20122 Milano, Italy}
\affiliation[h]{Department of Physics \& Astronomy, University College London, Gower Street, London, WC1E 6BT, UK}
\affiliation[i]{Institut d'Estudis Espacials de Catalunya (IEEC), c/ Esteve Terradas 1, Edifici RDIT, Campus PMT-UPC, 08860 Castelldefels, Spain}
\affiliation[j]{Institute of Space Sciences, ICE-CSIC, Campus UAB, Carrer de Can Magrans s/n, 08913 Bellaterra, Barcelona, Spain}
\affiliation[k]{Instituto de F\'{\i}sica, Universidad Nacional Aut\'{o}noma de M\'{e}xico,  Circuito de la Investigaci\'{o}n Cient\'{\i}fica, Ciudad Universitaria, Cd. de M\'{e}xico  C.~P.~04510,  M\'{e}xico}
\affiliation[l]{University of California, Berkeley, 110 Sproul Hall \#5800 Berkeley, CA 94720, USA}
\affiliation[m]{Instituci\'{o} Catalana de Recerca i Estudis Avan\c{c}ats, Passeig de Llu\'{\i}s Companys, 23, 08010 Barcelona, Spain}
\affiliation[n]{Institut de F\'{i}sica d’Altes Energies (IFAE), The Barcelona Institute of Science and Technology, Edifici Cn, Campus UAB, 08193, Bellaterra (Barcelona), Spain}
\affiliation[o]{Departamento de F\'isica, Universidad de los Andes, Cra. 1 No. 18A-10, Edificio Ip, CP 111711, Bogot\'a, Colombia}
\affiliation[p]{Observatorio Astron\'omico, Universidad de los Andes, Cra. 1 No. 18A-10, Edificio H, CP 111711 Bogot\'a, Colombia}
\affiliation[q]{University of Virginia, Department of Astronomy, Charlottesville, VA 22904, USA}
\affiliation[r]{Fermi National Accelerator Laboratory, PO Box 500, Batavia, IL 60510, USA}
\affiliation[s]{Department of Astronomy, University of Texas at Austin, 2515 Speedway, TX 78712, USA}
\affiliation[t]{Institut d'Astrophysique de Paris. 98 bis boulevard Arago. 75014 Paris, France}
\affiliation[u]{IRFU, CEA, Universit\'{e} Paris-Saclay, F-91191 Gif-sur-Yvette, France}
\affiliation[v]{Department of Physics, The Ohio State University, 191 West Woodruff Avenue, Columbus, OH 43210, USA}
\affiliation[w]{School of Mathematics and Physics, University of Queensland, Brisbane, QLD 4072, Australia}
\affiliation[x]{Department of Physics, The University of Texas at Dallas, 800 W. Campbell Rd., Richardson, TX 75080, USA}
\affiliation[y]{NSF NOIRLab, 950 N. Cherry Ave., Tucson, AZ 85719, USA}
\affiliation[z]{Department of Physics and Astronomy, University of California, Irvine, 92697, USA}
\affiliation[aa]{Sorbonne Universit\'{e}, CNRS/IN2P3, Laboratoire de Physique Nucl\'{e}aire et de Hautes Energies (LPNHE), FR-75005 Paris, France}
\affiliation[ab]{Departament de F\'{i}sica, Serra H\'{u}nter, Universitat Aut\`{o}noma de Barcelona, 08193 Bellaterra (Barcelona), Spain}
\affiliation[ac]{Department of Physics and Astronomy, Siena University, 515 Loudon Road, Loudonville, NY 12211, USA}
\affiliation[ad]{Department of Physics \& Astronomy, University  of Wyoming, 1000 E. University, Dept.~3905, Laramie, WY 82071, USA}
\affiliation[ae]{Institute of Cosmology and Gravitation, University of Portsmouth, Dennis Sciama Building, Portsmouth, PO1 3FX, UK}
\affiliation[af]{Department of Physics and Astronomy, University of Waterloo, 200 University Ave W, Waterloo, ON N2L 3G1, Canada}
\affiliation[ag]{Perimeter Institute for Theoretical Physics, 31 Caroline St. North, Waterloo, ON N2L 2Y5, Canada}
\affiliation[ah]{Waterloo Centre for Astrophysics, University of Waterloo, 200 University Ave W, Waterloo, ON N2L 3G1, Canada}
\affiliation[ai]{Instituto de Astrof\'{i}sica de Andaluc\'{i}a (CSIC), Glorieta de la Astronom\'{i}a, s/n, E-18008 Granada, Spain}
\affiliation[aj]{Departament de F\'isica, EEBE, Universitat Polit\`ecnica de Catalunya, c/Eduard Maristany 10, 08930 Barcelona, Spain}
\affiliation[ak]{Department of Physics and Astronomy, Sejong University, 209 Neungdong-ro, Gwangjin-gu, Seoul 05006, Republic of Korea}
\affiliation[al]{Abastumani Astrophysical Observatory, Tbilisi, GE-0179, Georgia}
\affiliation[am]{Department of Physics, Kansas State University, 116 Cardwell Hall, Manhattan, KS 66506, USA}
\affiliation[an]{Faculty of Natural Sciences and Medicine, Ilia State University, 0194 Tbilisi, Georgia}
\affiliation[ao]{CIEMAT, Avenida Complutense 40, E-28040 Madrid, Spain}
\affiliation[ap]{Department of Physics, University of Michigan, 450 Church Street, Ann Arbor, MI 48109, USA}
\affiliation[aq]{University of Michigan, 500 S. State Street, Ann Arbor, MI 48109, USA}

\emailAdd{charles.185@osu.edu}

\abstract{We present clustering measurements for high-redshift quasars using data from the Dark Energy Spectroscopic Instrument Data Release 2. Our sample consists of quasars with $2.0 < z < 3.5$ in the luminosity range $M_{1450} \leq -19.94$\,mag. 
We measure the mean quasar bias $b_Q(\bar{z} = 2.48) = 3.61 \pm 0.01$ for the full sample of $\sim 715,000$ quasars and quantify the redshift evolution of quasar bias by dividing the sample into four equal redshift bins. There is strong evolution of the quasar bias with redshift that is well fit by the function $b_Q(z) = a [(1 + z)^2 - 6.565] + b$ with $a=0.230 \pm 0.007$ and $b=2.394 \pm 0.035$, and this fit is also a good match to lower redshift measurements in the literature. This bias evolution is consistent with a characteristic halo mass of $\bar{M}_{\mathrm{h}} \sim 10^{12}\,\mathrm{M_\odot}$ that does not vary significantly with redshift. The inferred duty cycles for quasars in our sample are $f_{\mathrm{duty}} \sim 10^{-2}$, staying mostly constant over redshifts. We investigate the luminosity dependence of quasar clustering by dividing each of our four redshift bins into three luminosity bins. The size of our quasar sample permits the first statistically significant measurement of the luminosity dependence of quasar bias at these redshifts. We measure weak dependence of quasar bias on luminosity at fixed redshift, inconsistent with no dependence, but weaker than predicted by a model in which quasar luminosity is tightly correlated with halo mass. These clustering measurements provide a stringent test for models of active black hole light curves and the black hole-halo connection at high redshift.}

\begin{document}
\maketitle
\flushbottom

\section{Introduction}

Quasars are the most luminous type of active galactic nuclei powered by the accretion of matter onto supermassive black holes (SMBHs) at the centers of galaxies.
This accretion releases enormous amounts of energy across the electromagnetic spectrum, enabling quasars to be observed out to high redshifts.
Quasar activity is believed to be short-lived relative to cosmological timescales (e.g.\,\cite{rees1984, richstone1998, yu2002}).
As a result, only a fraction of dark matter halos host an active quasar at any given time, and the relationship between quasar properties and their host halos is not uniquely determined.

The large-scale spatial distribution of quasars contains information about the dark matter halos in which they reside. 
Quasar clustering measurements are quantified by the quasar bias, a measure of how clustered quasars are compared to the underlying dark matter distribution. 
The quasar bias is related to the manner in which quasars occupy their host halos through theoretical models of the halo mass function (e.g.\,\cite{tinker2008}) and halo bias (e.g.\,\cite{tinker2010}). 
Clustering analyses across a wide range of redshifts consistently find that quasars occupy dark matter halos with similar masses, typically $M_{\mathrm{h}} \sim 10^{12}-10^{13}~ h^{-1} M_{\odot}$ (e.g.\,\cite{croom2005, shen2007, richardson2012, eftekh19, alam2020, yuan2024}).

Given these inferred host halo masses, we can compare the observed number density of quasars to the abundance of dark matter halos at the same mass scale to estimate the quasar duty cycle: the fraction of halos capable of hosting a quasar that contain an actively accreting SMBH at a given redshift and luminosity (e.g.\,\cite{cole1989, martini2001, haiman2001}). 
Because quasar activity is thought to be short-lived relative to halo lifetimes, the duty cycle provides a critical link between the observed quasar population and the underlying halo population traced by large-scale structure. 

Studies of quasar clustering span several decades, with the first attempt to measure quasar clustering by \cite{osmer1981}.
Early quasar clustering measurements were limited by small sample sizes (e.g.\,\cite{shanks1987, stephens1997}). 
The advent of large spectroscopic surveys, most notably the 2dF QSO Redshift Survey and the Sloan Digital Sky Survey (SDSS), dramatically increased the number of known quasars and enabled detailed investigations of the redshift evolution of quasar clustering. 
Using these datasets, studies at lower redshifts ($z \lesssim 2$) found moderate to strong evolution of the quasar bias with redshift (e.g.\,\cite{porciani2004, croom2005, porciani2006, daangela2008, ross2009, laurent2017}). 
Measurements at higher redshifts ($z \gtrsim 2$) have also been performed (e.g.\,\cite{shen2007, white2012, eftekh15}), although results in this regime have been inconsistent. 
Shen et al.\,\cite{shen2007} report evidence for strong redshift evolution in the quasar bias at $z > 2.9$. 
In contrast, White et al.\,\cite{white2012} found no convincing evidence for evolution over $2.2 < z < 2.8$, and similarly, Eftekharzadeh et al.\,\cite{eftekh15} reported no significant evolution over $2.2 < z < 3.4$.
At very high-redshift ($z \gtrsim 5$), Arita et al.\,\cite{arita+2023} found much higher clustering, by a factor of almost four compared to the bias measurements at $z \sim 3$, indicating that the quasar bias continues to grow at higher redshift.
In the standard picture of structure formation, an increase in quasar bias with redshift reflects the fact that halos of fixed mass are increasingly rare and therefore more strongly biased tracers of the matter density field at earlier cosmic times (e.g.\,\cite{kaiser1984, bardeen1986}). 

Early analytic models of quasars often adopted a simplified picture in which quasars were assumed to exist in one of two states--either `on,' during a phase of rapid accretion and high luminosity, or `off,' when inactive (e.g.\,\cite{kauffmann2000, wyithe2003}). 
These models, which were primarily motivated by their simplicity, typically assumed a tight correlation between quasar luminosity and host dark matter halo mass.
Under this assumption, more luminous quasars would reside in more massive, more strongly biased halos, leading to an expectation of luminosity-dependent quasar clustering. 
More recent models incorporating physically motivated quasar light curves and extended accretion histories suggest instead that quasars spend the majority of their lifetimes at luminosities well below their peak values, with only a small fraction of their time spent near peak luminosity (e.g.\,\cite{lidz2006}). 
In such models, instantaneous quasar luminosity is only weakly correlated with host halo mass, naturally suppressing any luminosity dependence in quasar clustering.
Additionally, a weak or absent luminosity dependence could arise from scatter in stellar mass at fixed halo mass, or from scatter in the Eddington ratio at fixed stellar mass, both of which would weaken the relation between quasar luminosity and host halo mass.
Consistent with this picture, most observational studies have found little to no evidence for luminosity-dependent quasar clustering (e.g.\,\cite{croom2005, daangela2008, white2012, eftekh15}). 
However, at moderate to high redshifts, samples have usually not been large enough to measure the bias as a function of luminosity at fixed redshift, leaving this question unresolved.

The Dark Energy Spectroscopic Instrument (DESI) \cite{desisurvey, desiinstrumentdesign, desiinstrument} has collected an unprecedented volume of spectroscopic data during its first three years of operation, with the primary goal of constraining cosmology \cite{desi2025VII} and investigating the nature of dark energy through baryon acoustic oscillation (BAO) measurements.
At lower redshifts, galaxies are the primary tracers of the BAO feature \cite{desi2024III, desi2024VI}. 
At higher redshifts $(z>2.1)$, observing galaxies becomes increasingly difficult because of survey flux limits.
Instead, \lya quasars are used as the primary tracers for BAO measurements. 
Quasars act as background light sources, illuminating the distribution of neutral Hydrogen through absorption features in their spectra. 
These features, known as the \lya forest, appear at wavelengths shorter than the \lya emission line (rest frame 1216 \AA) and are visible in quasars with $z>2.1$, where the forest is redshifted into the observable wavelength range.
Correlation analyses between the \lya forest and quasars \cite{desidr2lya, desi2024IV} have been essential for pushing BAO measurements to early cosmic times. 
Mock catalogs play a critical role in these analyses by enabling validation of analysis pipelines, estimation of covariance matrices, and characterization of systematic uncertainties.
Accurate modeling of the quasar distribution in these mocks is therefore increasingly important.
Since quasar clustering evolves with redshift, implementing more realistic quasar clustering in mock catalogs will contribute to improving the precision and robustness of future BAO measurements. 

We focus on the auto-correlation function of quasars observed by DESI.
Using a sample of 713,706 quasars from DESI Data Release 2 (DR2) spanning the redshift range $2.0 \leq z \leq 3.5$, this represents the largest sample to date used for such measurements. 
The DESI quasar sample is highly complete (93.9\% targeting completeness) and extends to $r \leq 23$\,mag, probing a fainter sample than prior spectroscopic surveys (e.g., compare to SDSS BOSS at $r < 21.85$\,mag). 
These survey improvements enhance the statistical power to constrain the redshift evolution of the quasar bias and allow more robust tests of the luminosity dependence of quasar clustering. 

We describe the data and sample selection in Section~\ref{sec:data}. 
In Section~\ref{sec:clust} we discuss the methodology for measuring the two-point quasar auto-correlation function.
We present our measurement of the quasar auto-correlation function over the redshift range $2.0 \leq z \leq 3.5$ and investigate the redshift evolution and luminosity dependence in Section~\ref{sec:cluster}. 
Section~\ref{sec:halomass+dutycycle} provides our inferred minimum and characteristic host halo masses and quasar duty cycle, and includes comparisons of our results to simple model predictions that incorporate scatter in the $L-M_{\mathrm{h}}$ relation. 
We summarize our results in the final section.
Throughout this paper we adopt a Planck 2018 flat $\Lambda$CDM cosmological model with $\Omega_m=0.31$, $\Omega_\Lambda=0.69$, $\sigma_8=0.81$, and $h=0.6766$ \cite{planck2018}.

\section{Data}
\label{sec:data}

We use data from the first three years (May 2021 - April 2024) of the DESI survey, which will become publicly available as part of the DESI collaboration's second data release (DR2).
DR2 includes observations from the first data release (DR1) \cite{desiDR1}, new observations from the second and third years of survey operations, re-observations of objects from DR1 to improve the signal-to-noise ratio (SNR) of the spectra, and new processing of the entire dataset. 
The first subsection below presents an overview of the DESI instrumentation, operations, survey, and data. 
Section \ref{sec:desiqso} describes the properties of the quasar catalogs, and the data splits used in the analysis are outlined in Section \ref{sec:datasplits}.

\subsection{DESI data} \label{sec:desidata}

The Dark Energy Spectroscopic Instrument (DESI) is installed on the 4-meter Mayall Telescope at Kitt Peak National Observatory in Arizona.
The eight-year DESI survey \cite{desioperations} will measure redshifts for more than 63 million galaxies and quasars \cite{guy2023} and cover over 17,000 $\deg^2$.
Light entering the telescope is focused by a wide-field ($3^\circ$ diameter field of view) prime-focus corrector \cite{desicorrector} onto the focal plane system with 5,020 robotic fiber positioners \cite{desifibers}. 
The light collected in the fiber system is then dispersed by ten bench-mounted spectrographs, which split the light into three wavelength bands that cover a total wavelength range of $360-980$\,nm \cite{desiinstrument}. 
Data is processed through a spectroscopic pipeline which generates wavelength- and flux-calibrated spectra as well as spectroscopic classifications and redshifts \cite{guy2023}. 
DESI aims to achieve a highly complete sample by covering each area on the sky with multiple tiles. 
A tile represents the region on the sky that is in the field of view of the focal plane.
On average, a given region on the sky has about five tiles of coverage in dark time in the DR2 dataset. 

DESI quasar targets are selected from photometry in Data Release 9 of the Legacy Imaging Surveys \cite{legacysurvey}, which covers an area of more than 19,700\,deg$^2$ in three optical bands ($g$, $r$, $z$) \cite{desitargetval}, with a magnitude limit of $r \approx 23$\,mag and an average density of $310$ deg$^{-2}$ \cite{desitargetval}. 
The imaging survey includes over 9,000\,deg$^2$ of imaging in the $g$, $r$, and $z$ bands provided by the Dark Energy Camera Legacy Survey (DECaLS) and $\sim$5,000\,deg$^2$ of the Northern Galactic Cap from the Beijing-Arizona Sky Survey (BASS) in $g$ and $r$, and from the Mayall z-band Legacy Survey (MzLS) in $z$. 
In addition to the DECaLS, BASS, and MzLS imaging, full-sky imaging from the Wide-Field Infrared Survey Explorer (WISE) is used in the target selection process. 

Quasar selection in DESI incorporates near-infrared photometry from the all-sky WISE data in the $W1$ $(3.4 \mu m)$ and $W2$ $(4.6 \mu m)$ bands. 
These infrared measurements improve the distinction between quasars and stars, as quasars are typically about two magnitudes brighter than stars of similar optical color in these bands \cite{desisurvey}.
Quasar targets in DESI are selected using a combination of optical and infrared photometry, color cuts, and machine-learning classifiers designed to maximize completeness while controlling stellar contamination.
The selection function is inherently anisotropic and redshift-dependent, and is influenced by several factors such as observing conditions, hardware performance, and the intrinsic properties of the observed targets. 
More specifically, it quantifies the probability that a target is both observed and yields a successful redshift measurement. 
Details on the selection function applied to DESI quasars are provided in \cite{yeche2020, desitargetval}. 

DESI executed a survey validation period prior to the start of the main survey that included visual inspection of a random subset of quasar spectra \cite{desisurveyval}. 
In addition to assessing whether the quality of the spectra allow for confident spectroscopic redshift measurements, the visual inspection was used to confirm classifications of the targets as either quasars, galaxies, or stars.
Within DESI, \texttt{Redrock} \cite{redrock} is the standard template-fitting code utilized to identify redshifts. 
\texttt{Redrock} fits templates for quasars, galaxies, and stars and determines both the best-fitting (lowest $\chi^2$ value) redshift and template solutions for each spectrum. 
Spectra for quasar targets are obtained as described in Section 6 of \cite{desioperations}.
Other redshift warning flags are determined in the spectroscopic pipeline, flagging low-quality observations for a single object or an entire petal and setting all objects to be re-observed. 
Two additional algorithms beyond \texttt{Redrock} are relevant for DESI quasar identification. 
QuasarNet \cite{busca2018, farr2020, desitargetval} is used to help identify \lya quasars for additional observations. 
In addition, when final quasar catalogs are constructed, QuasarNet and an MgII algorithm \cite{desisurveyval, desitargetval} are used as `afterburners' to search for quasars that \texttt{Redrock} may have missed from the wider DESI spectroscopic sample, or that \texttt{Redrock} may have assigned an incorrect redshift.

The DESI survey strategy is designed to optimize observing efficiency. 
DESI uses an algorithm to automatically select which fields to observe, in what order, and adjusts the selection if necessary based on the observing conditions. 
DESI observes several target classes (e.g., QSOs, ELGs, and LRGs), each with different priority levels for observation. 
Quasars are observed during dark time and provide the highest-redshift coverage in the DESI survey \cite{desisurvey}.
Unobserved quasars are assigned the highest priority, followed by quasars at $z \geq 2.1$, which are prioritized due to their rarity and the need for repeated observations to increase the signal-to-noise ratio of the \lya forest absorption features. 
This higher priority for unobserved quasars contributes to the high completeness of our sample, especially for quasar targets with small angular separations. 
Further details about the DESI survey operations can be found in \cite{desioperations}.

\subsection{DESI quasar catalog}
\label{sec:desiqso} 

We use the DESI DR2 v2.1 PIP quasar large-scale structure (LSS) catalog for all of the clustering analyses in this paper (PIP is pairwise-inverse probability). 
This catalog contains 713,706 quasars in the redshift range $2.0 \leq z \leq 3.5$ with absolute magnitudes $M_{1450} \leq -19.94$. 
This sample is substantially larger than those used in previous studies at similar redshifts, most notably SDSS \cite{eftekh15}, which included 73,884 quasars.

In spectroscopic surveys with fiber-fed instruments there are physical constraints on the spacing between fibers that limits the ability to observe objects that lie very close together (often $\lesssim 1^{\prime}$) on the sky, which results in separation-dependent incompleteness.
Since close pairs are preferentially missed, not accounting for this effect results in underestimated clustering measurements, particularly on small scales.
To correct for this incompleteness, we apply PIP weights \cite{bianchi2017, bianchi2025}, which weight each observed pair by the inverse of the fraction of fiber-assignment realizations in which both quasars were observed.
Pairs that are rarely assigned fibers receive larger weights, thereby recovering an unbiased estimate of the clustering signal. 
The assignment probabilities are estimated from alternate realizations of the DESI fiber assignment process using the alternate `merged target list' (altMTL), which records the full assignment history of each target \cite{lasker+2025}.
The PIP version of the LSS catalog uses this altMTL information to compute the PIP weights applied in our clustering measurements. 
PIP weights are closely related to individual-inverse-probability (IIP) weights; \cite{bianchi2017} demonstrates that the two weighting schemes yield equivalent results once the angular separation scale exceeds the fiber collision scale.

Clustering measurements require a well-defined survey footprint and selection function to accurately model observational systematics. 
The DESI footprint is defined using the DESI \texttt{FIBERASSIGN} software \cite{lasker+2025}, which simulates the fiber assignment process. 
\texttt{FIBERASSIGN} assigns targets to available fibers on each tile, subject to survey priorities and instrument constraints. 
In addition to all the information relevant for clustering measurements, the LSS catalogs include space densities of the sample with corrections for assignment incompleteness.
Ross et al.\,\cite{ross2025} provide a detailed description of the construction of the LSS catalogs for DESI.

Random catalogs used in the clustering analysis are constructed by sampling the redshifts, weights, and TARGETIDs (unique object identifier) from the observed quasar population such that the weighted redshift distribution, $dN/dz$, of the random catalog matches the data. 
This matching is preserved even with restricted subsamples, such as individual redshift or luminosity bins. 
These catalogs are initially constructed to match the sky coverage of the Legacy Imaging Surveys \cite{legacysurvey, myers2023}, with coordinates uniformly sampled over that footprint, and then downsampled to match the footprint of the corresponding data release. 
We constructed separate random catalogs for each of the data splits described in Section \ref{sec:datasplits}, ensuring that both the redshift distribution and survey footprint is matched independently for each subsample.
For each subsampled catalog, the imaging systematic weights are recomputed from the objects within that selection, so that the corrections reflect the imaging property dependencies specific to that subsample's redshift and luminosity range.
Further details on the construction of DESI random catalogs are also in \cite{ross2025}. 

In the next section, we quantify how quasar clustering depends on their luminosity. We characterize luminosity in terms of the absolute magnitude at $1450$\,\AA\,($M_{1450}$), derived from the rest-frame integrated luminosity $L_{1450}$ estimated by \texttt{FastSpecFit} \cite{fastspecfit} from the best-fit spectrophotometric model. 
\texttt{FastSpecFit} is a spectrophotometric fitting code that models rest-frame UV/optical spectra and broadband photometry using physically motivated emission-line and continuum templates.

\subsection{Data splits}
\label{sec:datasplits}

\begin{figure}[tbp]
    \centering
    \includegraphics[width=1.0\linewidth]{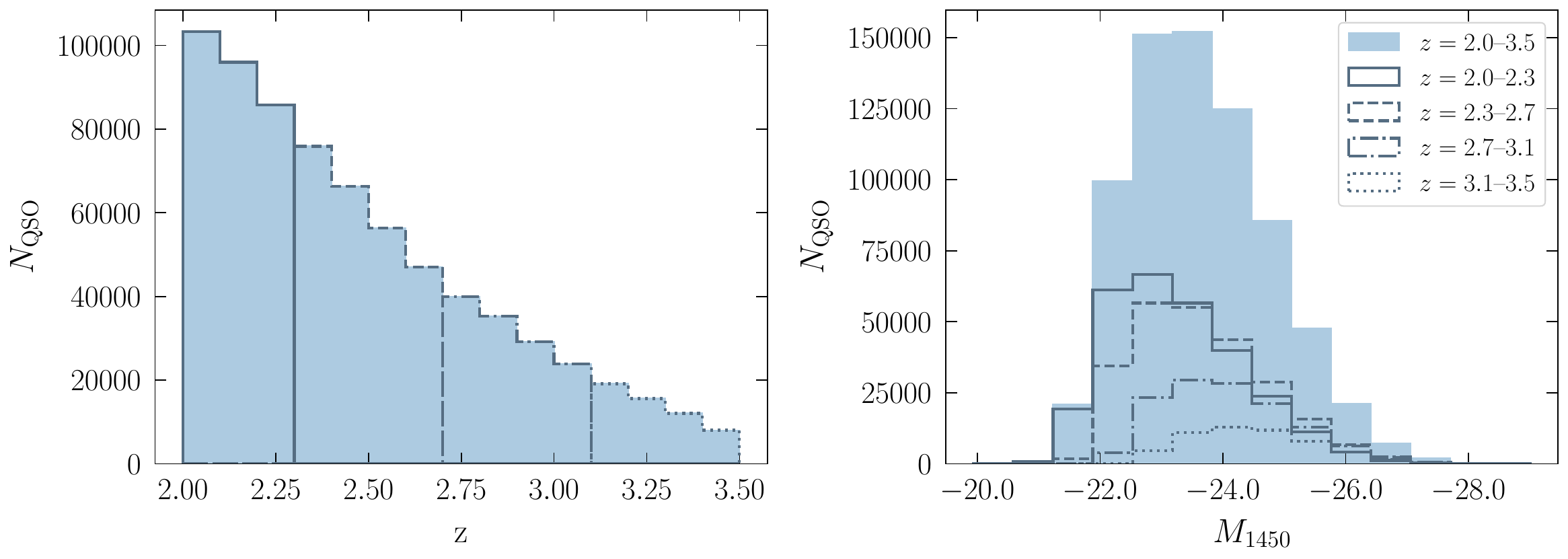}
    \caption{Redshift distribution of the DESI quasar sample split into four redshift bins (left) and the luminosity distribution for the corresponding redshift bins (right). The shaded blue histograms represent the distributions for the entire sample.}
    \label{fig:zdist}
\end{figure}

Measurements of quasar clustering as a function of redshift and luminosity enable constraints on the evolution of clustering with time and the relationship between quasar luminosity and the mass of the host dark matter halos. 
To study the redshift evolution, we divide the sample into four redshift bins. 
The redshift distribution of the sample is shown in the left panel of Figure \ref{fig:zdist},  with vertical lines differentiating the four redshift bins. 
The right panel shows the luminosity distribution for the entire sample (shaded blue region) as well as the distribution for each of the redshift bins.

\begin{figure}[tbp]
    \centering
    \includegraphics[width=1.0\linewidth]{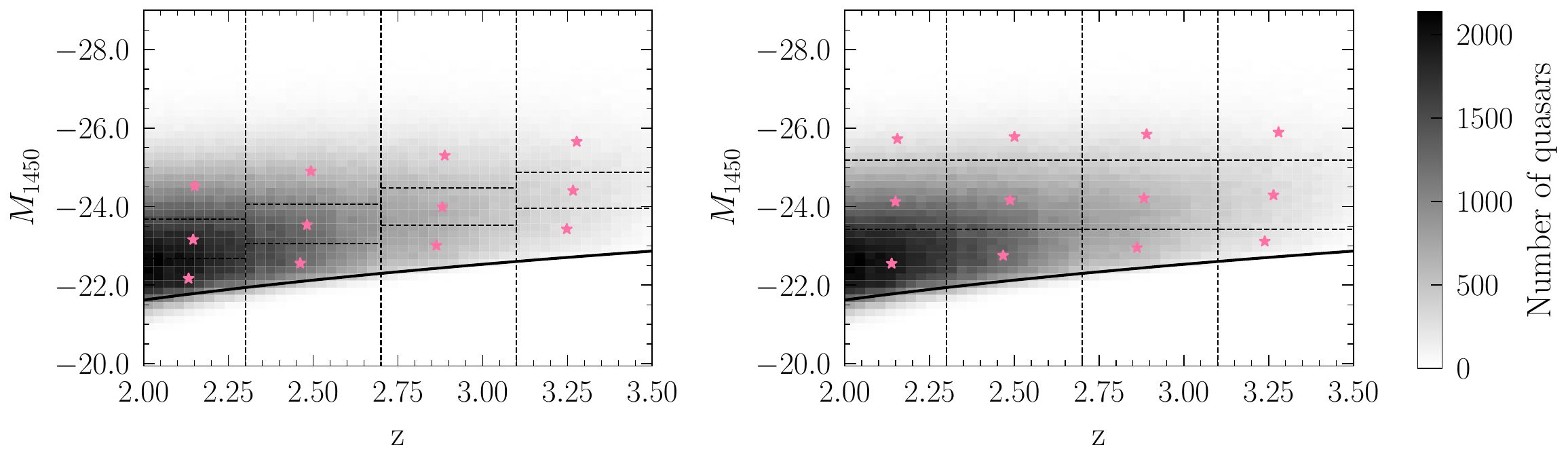}
    \caption{The binned redshift-luminosity distribution of the quasar catalog. Darker shades of gray correspond to higher densities. Dashed lines denote the redshift and luminosity bin edges. The left panel shows the `equal numbers' binning scheme and the right panel shows the `same boundaries' binning scheme. The pink stars mark the mean redshift and mean luminosity of each redshift-luminosity bin. The solid black line shows the magnitude limit of the DESI survey at each redshift (Eq. \ref{eq:M1450}).}
    \label{fig:zLdist}
\end{figure}

Luminosity-dependent quasar clustering is more challenging to model because flux-limited target selection introduces a degeneracy between redshift and luminosity. 
To lessen this effect, we split each redshift bin into three luminosity bins. 
This binning scheme allows us to isolate luminosity-dependent trends at fixed redshift while minimizing the effects of redshift evolution within each bin. 
We perform the luminosity splits in two ways, using the same redshift bins in each case: (1) equal-number luminosity bins defined independently within each redshift bin (`equal numbers') and (2) equal-width luminosity bins across all redshift bins (`same boundaries'). 
Figure \ref{fig:zLdist} shows the bin edges plotted as dashed lines over the redshift-luminosity distribution of our sample.
Dividing the sample according to the `same boundaries' binning scheme allows for a wider luminosity range to be covered, but some regions contain a small number of quasars compared to the other bins, which can lead to higher uncertainties on the measurements. 
The `equal numbers' binning scheme has a similar number of quasars in each bin at a given redshift bin, which provides more statistically robust measurements. 
The solid black line in Figure \ref{fig:zLdist} shows the magnitude limit of the survey, computed from 
\begin{equation}
    M_{1450} = m_r - \mu(z) - K(z),
    \label{eq:M1450}
\end{equation}
where $m_r = 23.0$ mag, $\mu(z) = 5 \log_{10} (d_L(z) / 10\, \text{pc})$ is the distance modulus with luminosity distance $d_L(z)$, and $K(z)$ is the $K$-correction.
Here we use the $K$-correction from \cite{richards2006},
\begin{equation}
    K(z) = -2.5 (1 + \alpha_\nu) \log_{10} (1 + z) - 2.5 \alpha_\nu \log_{10} \left( \frac{1450 \text{\AA}}{6260 \text{\AA}} \right),
\end{equation}
where we use a continuum slope of $\alpha_\nu=-0.5$. 
The first term is the standard $K$-correction to $z=0$ for a power-law continuum and the second term converts the $r$-band magnitude to $1450$\,\AA.
Table~\ref{tab:results} details these different data splits along with properties of each of the subsamples. 

\section{Clustering measurements}
\label{sec:clust}

Quasar clustering measurements quantify the extent to which quasars trace the large-scale structure of the universe by comparing their spatial distribution to that of the underlying matter field. 
In this section, we outline the methodology for computing these measurements.
Section~\ref{sec:2pcf} introduces the two-point correlation function, and Section~\ref{sec:zcf} discusses the redshift-space correlation function and the procedure used to estimate the quasar bias.

\subsection{Two-point correlation function}
\label{sec:2pcf}

The two-point quasar auto-correlation function quantifies the excess probability of finding a quasar pair at a given separation $r$ compared to a random distribution. 
This measurement probes the large-scale structure of the universe by characterizing the spatial clustering of quasars.

Mathematically, the spatial two-point correlation function (2PCF), $\xi(r)$, is defined by \cite{peebles1980} through the probability $\delta P = n[1 + \xi(r)]\delta V$, where $\delta P$ is the probability of finding a quasar pair at some separation $r$, $n$ is the mean number density of quasars, and $\delta V$ is the differential volume element.
However, this formal definition does not specify how to account for survey geometry, selection effects, or the shot noise present in real data.

To address these limitations, we use the Landy \& Szalay \cite{landyszalay} estimator
\begin{equation}
    \xi(s,\mu) = \frac{DD(s,\mu) - 2 DR(s,\mu) + RR(s,\mu)}{ RR(s,\mu)},
    \label{eq:landyszalay}
\end{equation}
where $DD(s,\mu)$, $DR(s,\mu)$, and $RR(s,\mu)$ are the number of data-data, data-random, and random-random pairs, respectively, at a separation $s$ and cosine of the angle to the line of sight $\mu$. 
This estimator corrects for edge effects and selection biases by incorporating random catalogs that match the survey geometry and selection function. 

The random catalogs are typically created with a much higher density than the data sample to reduce Poisson noise in the estimation of $RR(s,\mu)$ and $DR(s,\mu)$. 
Random catalogs provide a statistical baseline for clustering measurements by representing a uniform distribution of objects that follow the same angular selection function, survey geometry, and completeness as the data. 
This ensures that the measured 2PCF describes the intrinsic structure of the sources rather than observational or instrumental systematics.
To mitigate the statistical fluctuations caused by Poisson noise, especially at small scales, we generate three independent random catalogs for each data split, each with $40$ times the number of objects as the full sample.
Eftekharzadeh et al. \cite{eftekh15} found that random catalogs beyond $\sim 20$ times the size of the data did not significantly improve the statistics in the clustering signal, suggesting our choice is more than sufficient; any residual variance from the number of randoms used is propagated through the jackknife covariance. 
Section \ref{sec:desiqso} contains a brief description of how these random catalogs are created, see \cite{ross2025} for further details.

\subsection{Redshift-space correlation function and quasar bias}
\label{sec:zcf}

Peculiar velocities distort the observed positions of quasars along the line of sight in redshift space, introducing anisotropy into the clustering signal.
To account for these redshift-space distortions (RSD), the correlation function must depend on both the separation $s$ between quasar pairs and the orientation of that separation relative to the line of sight, commonly expressed as $\mu = \cos \theta$, where $\theta$ is the angle between the separation vector and the line of sight. 

To compare with the measured correlation function, we focus on the monopole $(\ell = 0)$, which is the angle-averaged component of the Legendre polynomial expansion:  
\begin{equation}
	\xi_0(s) = \frac{1}{2} \int_{-1}^1 \xi(s, \mu) \mathcal{L}_0 (\mu) \ d\mu,
\end{equation} 
where $\mathcal{L}_0 (\mu) = 1$.
We substitute $\xi(s, \mu)$ into Eq.~\ref{eq:landyszalay} to determine the monopole: 
\begin{equation}
    \xi(s) = b_Q^2 \left(1 + \frac{2}{3} \beta + \frac{1}{5} \beta^2 \right) \xi_{\mathrm{mat}}(r),
    \label{eq:zcf}
\end{equation}
where $\xi_{mat}(r)$ is the real-space matter correlation function, $b_Q$ is the quasar bias, $\beta = f / b_Q$ is the RSD parameter, and the linear growth rate of structure, $f$, is approximated as $f \approx \Omega_m(z)^{0.55}$ \cite{linder2005}.
This approximation is valid on large scales, typically for separations $s \gtrsim 10\, \hMpc$, where linear theory holds.  

To estimate the quasar bias, $b_Q$, we first compute the monopole of the redshift-space correlation function using the Landy \& Szalay estimator (Eq. \ref{eq:landyszalay}), as implemented in \texttt{pycorr} \cite{corrfunc2019, corrfunc2020}.
The \texttt{pycorr} package is a Python wrapper for the high-performance \texttt{CORRFUNC} two-point statistics package, which provides efficient pair counting and other tools for the calculation of correlation functions.
We estimate the covariance matrix for the auto-correlation measurement using jackknife resampling with $128$ spatial sub-samples, following the method in \cite{mohammad2022}.
In this approach, we divide the survey area into $128$ equal-area regions, and recompute the correlation function $128$ times, omitting one region each time. 
We then derive the covariance matrix from the variations among these jackknife realizations, providing an empirical estimate of the statistical uncertainties and correlations between bins in the correlation function.
We compute the two-point correlation function $\xi(s, \mu)$ using 48 logarithmically spaced bins in $s \in [0.01, 100]\, \hMpc$, and 200 linearly spaced bins in $\mu \in [-1,1]$.

We compare the measured clustering to the predicted matter clustering at the same mean redshift with the \texttt{CAMB} \cite{camb} software package. Specifically, we generate the linear matter power spectrum for a flat $\Lambda$CDM cosmology that is consistent with Planck 2018 \cite{planck2018} and then take the Fourier transform to obtain $\xi_{\mathrm{mat}}(r)$ at the mean redshift of the sample. We then fit Eq. \ref{eq:zcf} to the monopole with a one-parameter $\chi^2$ minimization over the range $10 \leq s \leq 80~\mathrm{h}^{-1}$Mpc to determine $b_Q$. 
The lower bound of our fit range avoids nonlinear effects that violate the assumptions of the Kaiser model and the upper bound avoids the BAO peak and larger scales that are more susceptible to systematics and observational uncertainties. 
Appendix \ref{app:fitrange} shows the robustness of our bias measurements in each redshift bin to the fit region chosen.
The fit incorporates the full covariance matrix to account for correlations between spatial bins, which tend to be highly correlated.
As a robustness check, we repeat the fit using only the diagonal elements of the covariance matrix and find consistent results, indicating the off-diagonal terms do not significantly impact our bias measurements.

\section{Clustering results}
\label{sec:cluster}

\begin{figure}[tbp]
\centering
    \includegraphics[width=0.75\textwidth]{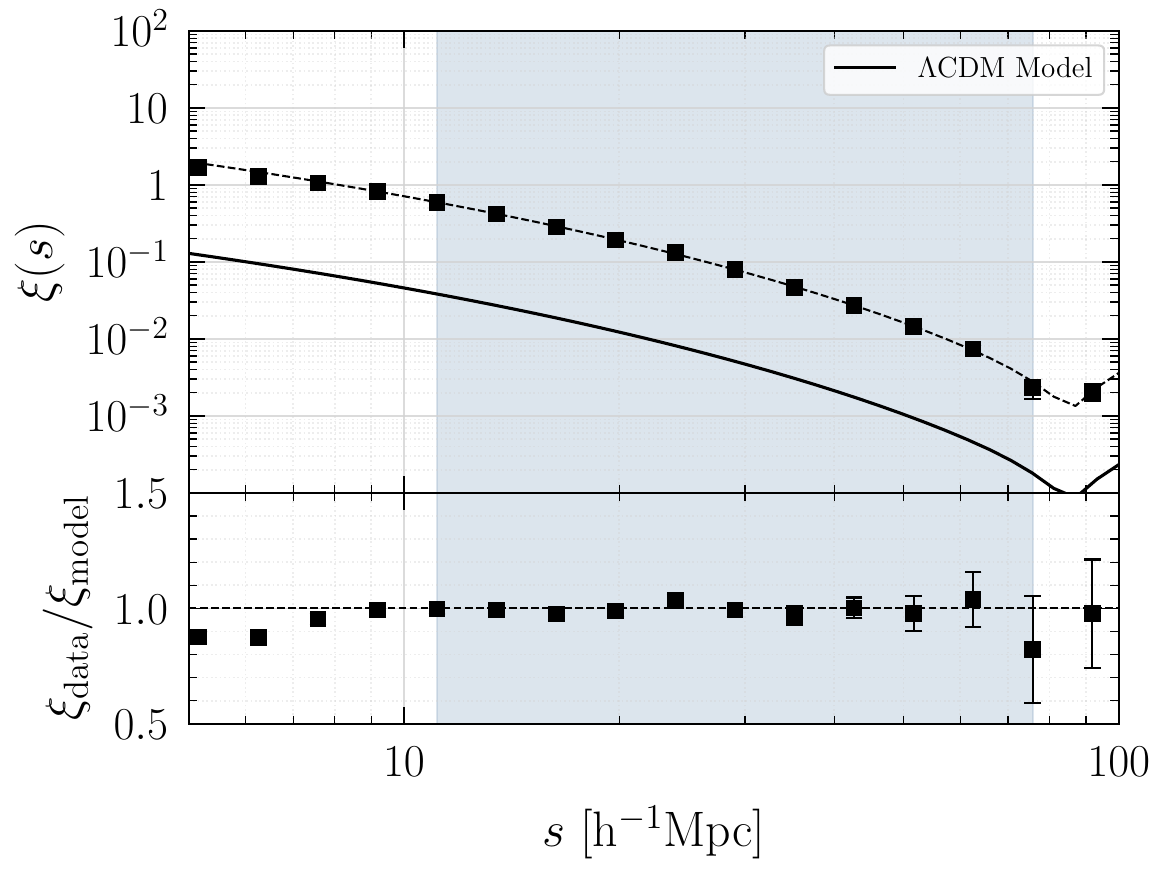}
    \hfill
    \caption{Redshift-space quasar auto-correlation function with the full sample of DESI DR2 quasars at $z>2.1$. The measured quasar auto-correlation function is shown with black squares and the matter correlation function calculated assuming a Planck 2018 flat $\Lambda$CDM cosmology at the mean redshift of our sample ($z=2.477$) is shown in the solid black line. The black dashed line is the fit of the matter correlation function to the measured auto-correlation function. The fit region ($10 \leq s \leq 80$ h$^{-1}$Mpc) is shown by the shaded region. This quasar bias is $b_Q= 3.61 \pm 0.01$ and the fit has $\chi_{\mathrm{red}}^2 = 0.80$ (for 10 degrees of freedom). The bottom panel shows the ratio of the measured correlation function to the model fit.}
    \label{fig:acf}
\end{figure}

We measure the redshift-space quasar auto-correlation function using $713,706$ quasars in the redshift range $2.0 \leq z \leq 3.5$ with absolute magnitudes $M_{1450} \leq -19.94$\,mag. 
Figure \ref{fig:acf} shows the quasar auto-correlation function and the best-fit theoretical model for the full sample; details of the fit are listed in Table \ref{tab:results}. 
Our one-parameter fit yields a quasar bias of $b_Q=3.61 \pm 0.01$ at $\bar{z}=2.477$ with $\chi^2_{\mathrm{red}} = 0.80$ (for 10 degrees of freedom), assuming a flat $\Lambda$CDM cosmology consistent with Planck 2018 \cite{planck2018}. 
Additionally, we estimate a systematic error of $\sim 1.4-5.9 \times$ the reported statistical errors for the bias measurements in each of our subsamples; the details of the systematic error estimation can be found in Appendix \ref{app:syserr}.
This is the most precise measurement of the quasar bias at these redshifts to date. 

We fit the correlation function over the range $10-80\,\hMpc$. 
At smaller scales $(s \lesssim \,5-10\,\hMpc$), density perturbations grow and linear theory is no longer able to model the correlation function. 
These scales are where Halo Occupation Distribution models (e.g.\,\cite{jing1998, peacock2000, berlind2002, zheng2005}) typically define the `one-halo regime,' which includes sources within the bounds of a typical dark matter halo, compared to the `two-halo regime' at larger scales which includes sources in separate dark matter halos. 
Several quasar clustering studies (e.g.\,\cite{hennawi2006, eftekh19}) have measured down to scales within the one-halo regime and have found excess clustering on small scales, clearly implying that a different model is necessary to model the clustering at these scales. 
Measurements that extend to smaller scales can provide interesting insights into quasar triggering and evolution, however, modeling these scales is beyond the scope of this paper. 
We chose the fit range $10-80\,\hMpc$ to remain at scales large enough that linear theory holds, but small enough that we avoid the large scale regions where statistical uncertainties are larger. 
Previous studies have used different ranges to fit the correlation function; for example, Shen et al.\,\cite{shen2007} fit between $4-150\,\hMpc$, Eftekharzadeh et al.\,\cite{eftekh15} used the region $4-25\,\hMpc$, and Laurent et al.\,\cite{laurent2017} fit over $10-85\,\hMpc$.
In Appendix \ref{app:fitrange}, we demonstrate the robustness of our measurement to the fit region. 

In the first subsection below we present our measurements of the redshift evolution of quasar clustering. Then in Section \ref{sec:zlumevol} we investigate the potential luminosity dependence of the clustering.

\subsection{Redshift evolution of quasar clustering}
\label{subsec:zevol}

\begin{figure}[tbp]
\centering
    \includegraphics[width=1\textwidth]{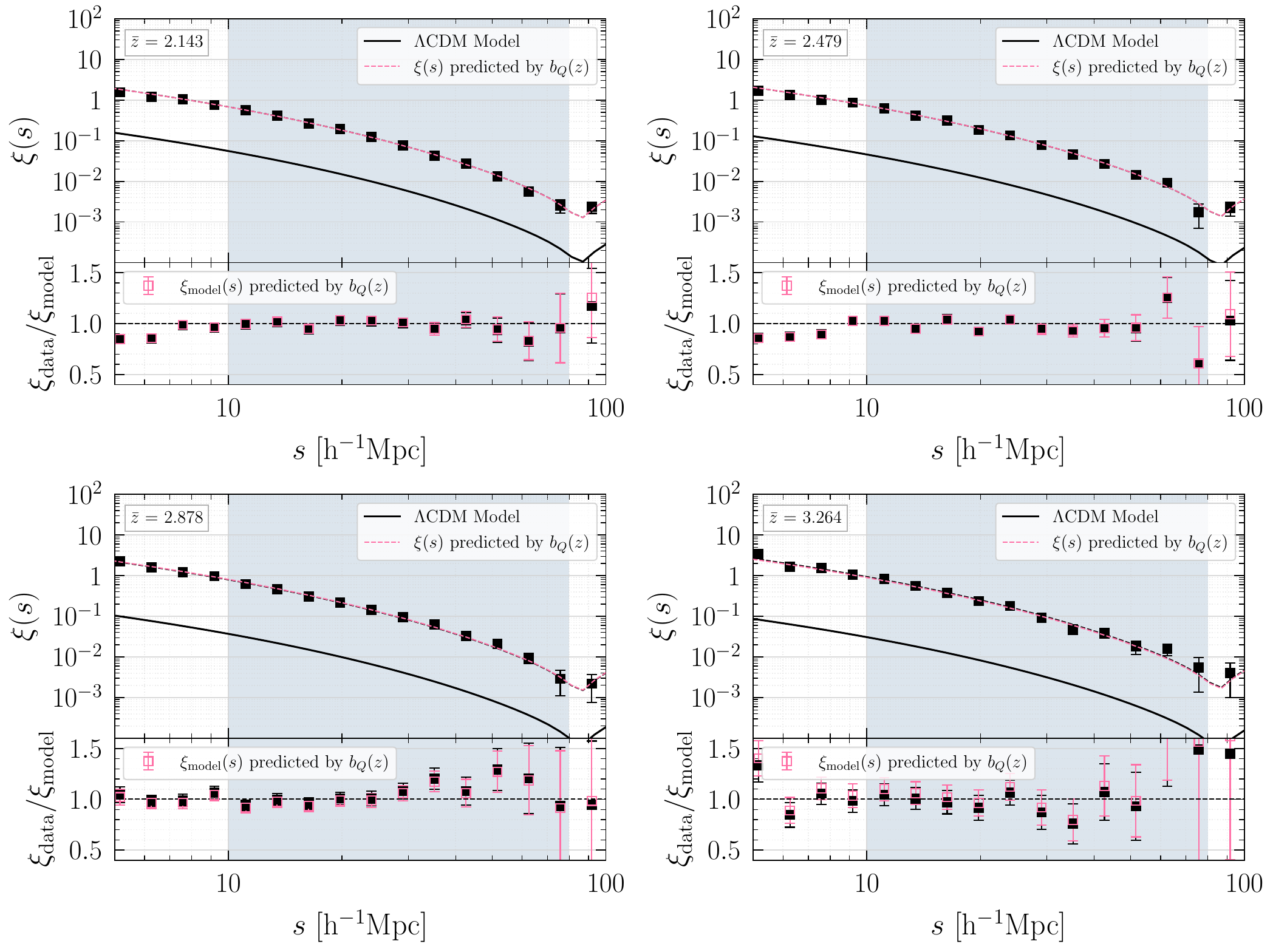}
    \hfill
    \caption{Quasar auto-correlation function (black squares) measured in four redshift bins. The solid black line is the matter correlation function evaluated at the mean redshift of each bin using the Planck 2018 flat $\Lambda$CDM cosmology. The black dashed line shows the best fit of the matter correlation function to the data, while the pink dashed line indicates the correlation function predicted by the best-fit $b_Q(z)$ model (Eq. \ref{eq:bqz}) at each mean redshift. The shaded region ($10 \leq s \leq 80$ h$^{-1}$Mpc) denotes the fitting range. The quasar bias values are listed in Table \ref{tab:results}. The lower part of each panel shows the ratio of the measured correlation function to the best-fit model (black squares), as well as the ratio between the data and the $\xi(s)$ inferred from our best-fit $b_Q(z)$ model evaluated at the mean redshift of each bin (open pink squares).}
    \label{fig:acfz}
\end{figure}

We divide the DESI sample into four equal-width redshift bins to investigate the redshift evolution of quasar clustering. 
The measured redshift-space auto-correlation functions for each bin are shown in Figure \ref{fig:acfz}, and the corresponding numerical values are listed in Table \ref{tab:results}. 
The solid black line in each panel represents the matter correlation function computed at the mean redshift of the bin using a Planck 2018 flat $\Lambda$CDM cosmology and the dashed line corresponds to the best-fit model. 
As in Figure \ref{fig:acf}, the shaded region shows the $10 \leq s \leq 80~h^{-1}$Mpc fit range. 

\begin{figure}[tbp]
\centering
	\includegraphics[width=0.75\textwidth]{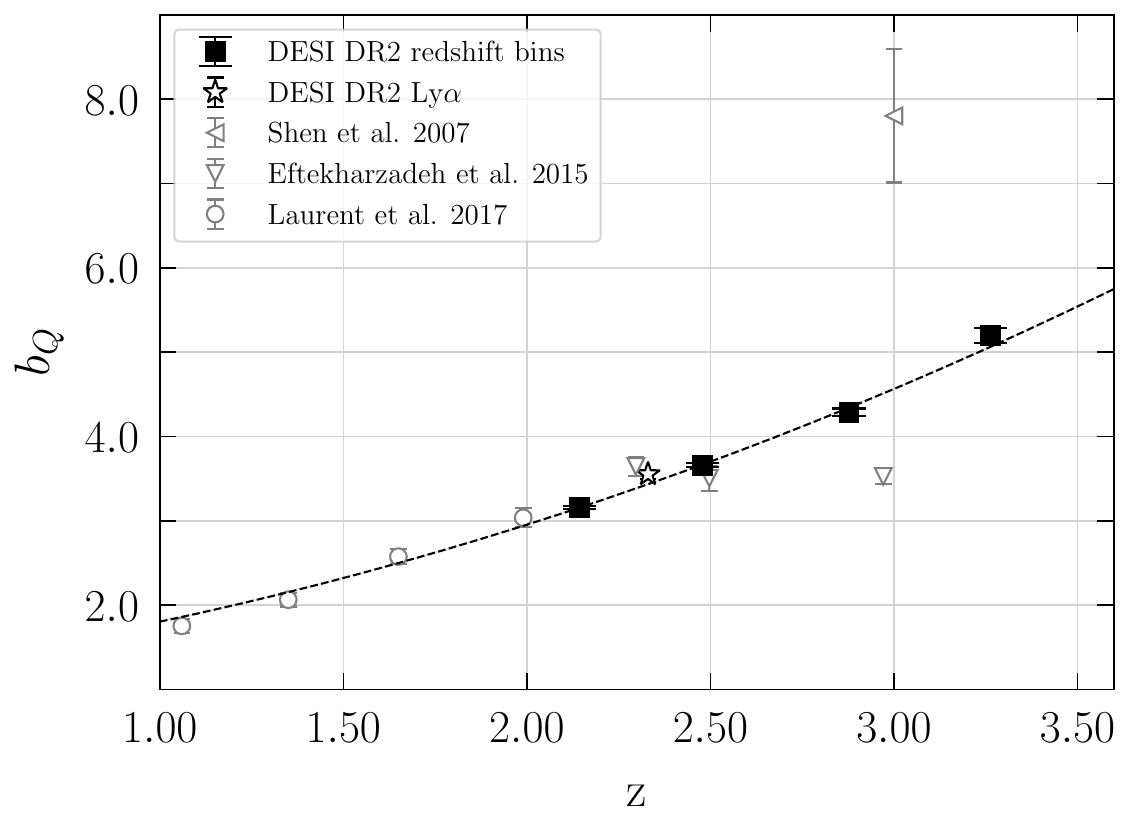}
    \hfill
    \caption{Quasar bias as a function of redshift for the four redshift bins described in Table~\ref{tab:results} (filled squares), the quasar bias derived as part of the DESI DR2 \lya BAO analysis (open star), the Shen et al.\,\cite{shen2007} results from SDSS DR5 (open left triangle), the Eftekharzadeh et al.\,\cite{eftekh15} results from the SDSS-III BOSS sample (open down triangles), and the Laurent et al.\,\cite{laurent2017} results from the SDSS-IV eBOSS quasars at lower redshifts (open circles). The dashed line shows the fit of the function defined in \ref{eq:bqz} to the four DESI DR2 redshift bins points. We chose to include only one point from Shen et al.\,\cite{shen2007} bias measurements for visual clarity, as they are significantly higher than our measurements.}
    \label{fig:bqz}
\end{figure}

The quasar bias measurements versus redshift are shown in Figure \ref{fig:bqz}. 
We find that quasar clustering evolves strongly with redshift.
We model the redshift dependence using the empirical relation introduced by \cite{laurent2017}:
\begin{equation}
        b_Q(z) = a [(1 + z)^2 - 6.565] + b,
        \label{eq:bqz}
\end{equation}
and fit this function to our measurements, obtaining best-fit parameters $a=0.230 \pm 0.007$ and $b=2.394 \pm 0.035$ ($\chi^2 = 4.14$; 2 degrees of freedom).
The dashed line in Figure \ref{fig:bqz} shows this fit, which provides an excellent description of the data. This line also provides a good match to the lower-redshift measurements from \cite{laurent2017}, although those points were not part of our fit. 
This bias evolution refers to quasars above an approximate constant apparent magnitude threshold, so the luminosity threshold increases with redshift. 
When comparing these results, it should be noted that DESI's deeper magnitude limit ($r<23$\,mag) yields a sample with more faint objects than the SDSS sample, which was limited to $r<21.85$\,mag.
Chaussidon et al.\,\cite{chaussidon2024} applied the same relation to DESI DR1 over $0.8 \leq z \leq 3.5$ and found $a=0.237 \pm 0.010$ and $ b = 2.328 \pm 0.026$.

Figure \ref{fig:bqz} demonstrates the high precision of our measurements compared to previous quasar bias determinations in this redshift range. 
The error bars on the DESI DR2 measurements are substantially smaller than those from earlier surveys, reflecting the much larger sample size and improved observational coverage.
Our results are broadly consistent with prior studies, including the DESI DR2 \lya forest measurements \cite{desidr2lya}, SDSS-III BOSS\,\cite{eftekh15}, and SDSS-IV eBOSS\,\cite{laurent2017}, after rescaling their bias values to our fiducial cosmology using $\sigma_8$ corrections. 
This correction is implemented through a correction factor $\sigma_{8, \mathrm{lit}} / \sigma_{8, \mathrm{fid}}$, where $\sigma_{8, \mathrm{lit}}$ is the value assumed in the original study ($\sigma_8$ is evaluated at $z=0$). 

Our measurements are systematically lower than those reported by \cite{shen2007} at similar redshifts. 
For example, at redshifts $z > 2.9$ Shen et al.\ found $b_Q > 7$, compared to $b_Q=5.20 \pm 0.08$ in our analysis at $\bar{z}=3.264$. 
This discrepancy may reflect a combination of factors, including the smaller sample size in SDSS DR5, differences in selection, and the higher fraction of luminous quasars in the earlier sample. We explore the evidence for luminosity-dependent clustering in Section \ref{sec:zlumevol}.

\subsection{Luminosity dependence of quasar clustering}
\label{sec:zlumevol}

We bin the sample in both redshift and luminosity to isolate luminosity-dependent clustering from the strong redshift evolution. 
Relative to previous studies of high redshift quasar clustering, the most important advantage of our analysis is that the large sample enables precise measurements at different luminosity in a fixed, narrow redshift bin, so that we do not conflate redshift evolution and luminosity dependence.
Within each redshift bin, we divide the sample into three luminosity bins in two ways:  
\begin{enumerate}
    \item `equal numbers': equal numbers of objects per redshift bin
    \item `same boundaries': the same boundaries for all redshift bins
\end{enumerate}
These binning schemes are detailed in Table \ref{tab:results}. 
The redshift bins are narrow enough that any redshift evolution of the quasar bias within a single bin should be small compared to the redshift evolution observed across the full sample, though not entirely negligible. 

We extend the redshift-only quasar bias model (Eq. \ref{eq:bqz}) by including a luminosity term to distinguish the contributions from redshift and luminosity, 
\begin{equation}
    b_Q(z,M_{1450}) = a [ (1 + z)^2 - 6.565 ] + b - m(M_{1450} + 24),
    \label{eq:bQzM1450}
\end{equation}
where $z$ and $M_{1450}$ are the mean redshift and luminosity values in each bin, $a$ controls how steep the bias increases with redshift, $b$ sets the overall normalization of the bias, and $m$ controls how steeply bias changes with luminosity at a fixed redshift.
This assumes a linear dependence on luminosity, or equivalently that the shape of the redshift evolution is the same for bright and faint quasars. 
We fit this function to our quasar bias measurements in the 12 redshift-luminosity bins, for each binning scheme.
We find statistically significant evidence for weak luminosity dependence; Table \ref{tab:bQzL} contains the best-fit parameters.
The calculation in this section is using just the statistical errors, but the results do not change significantly if we include the systematic error described in Appendix \ref{app:syserr}.

Figure \ref{fig:z-L_acf_zbins-eqnum} shows the measured correlation function in each redshift bin for `equal numbers' luminosity split. 
\begin{figure}[tbp]
	\includegraphics[width=1.0\textwidth]{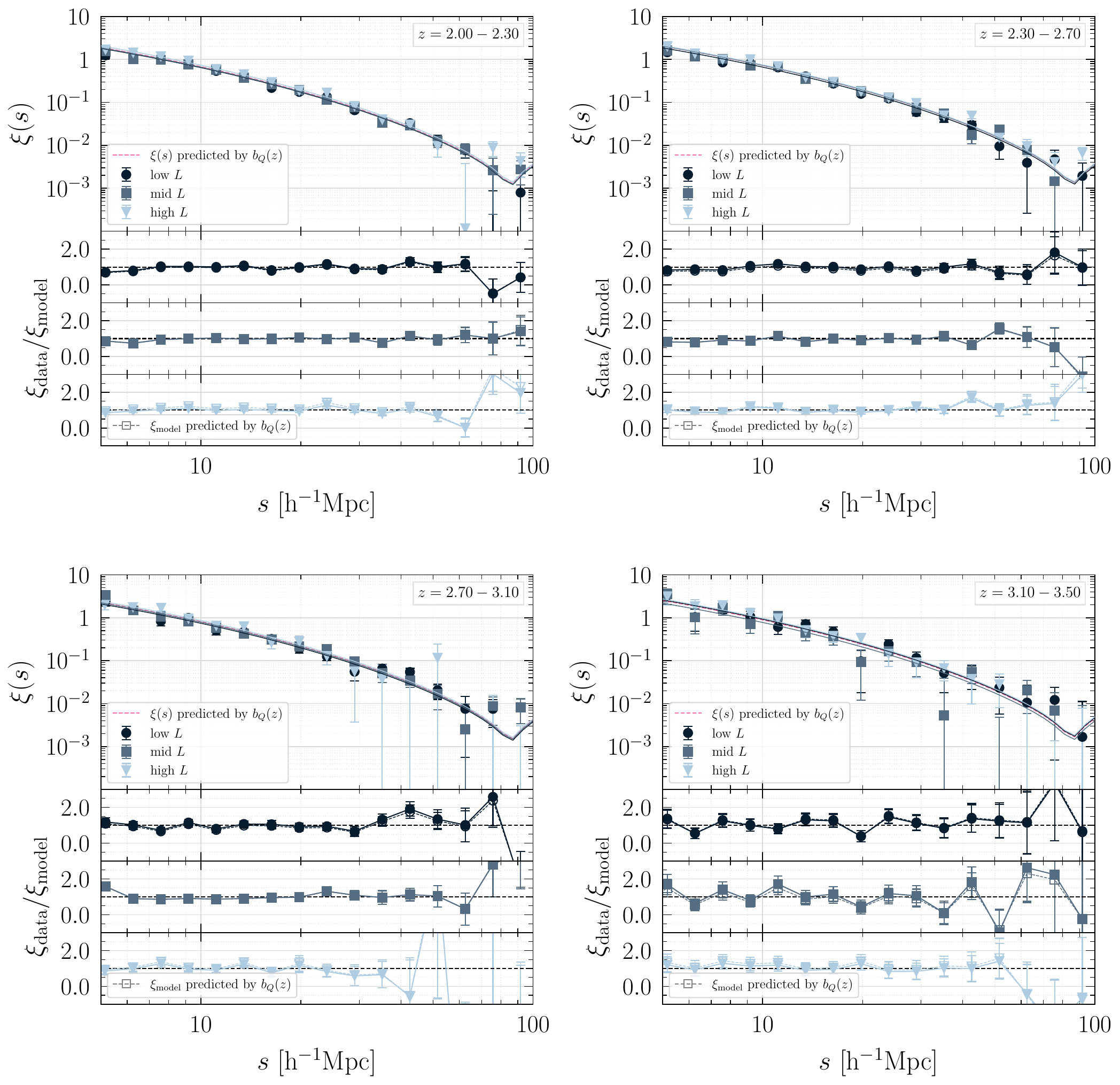}
    \caption{Quasar auto-correlation function measurements with each redshift bin separated into three `equal number' luminosity bins. The redshift range is in the upper right of each panel. The bias values are shown in Figure~\ref{fig:bqevolzLbins-eqnum} and listed in Table \ref{tab:results}. The pink dashed line in the top panels shows the $\xi(s)$ predicted by our best-fit $b_Q(z)$ (Eq. \ref{eq:bqz}). The lower panels show the ratio of the measured correlation function in each luminosity bin to the corresponding model fit in solid points. The open faced points (connected by dashed lines) show the data compared to the model predicted by our best-fit $b_Q(z)$.}
    \label{fig:z-L_acf_zbins-eqnum}
\end{figure}
Each panel shows the correlation functions of the three luminosity bins,`low $L$,' `mid $L$,' and `high $L$,' within that redshift range. 
The left panel of Figure \ref{fig:bqevolzLbins-eqnum} shows the redshift evolution of the quasar bias for the `equal numbers' luminosity bins. 
\begin{figure}[tbp]
	\includegraphics[width=1.0\textwidth]{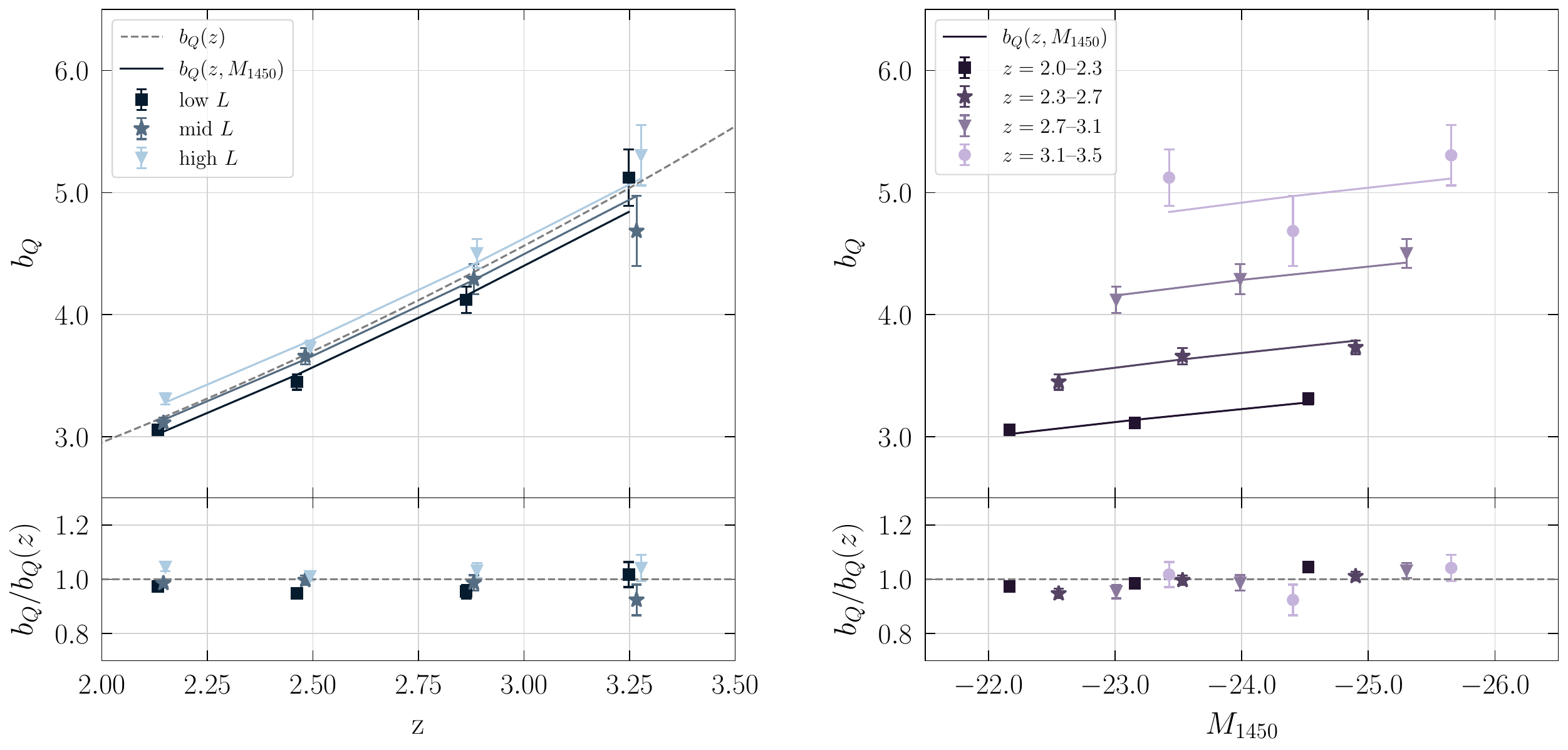}
    \caption{Quasar bias measurements in the 12 redshift and luminosity bins according to the `equal numbers' binning scheme outlined in Table \ref{tab:results}. (left) Quasar bias as a function of redshift in different luminosity bins. The $b_Q(z)$ model (black dashed line) fit to the redshift-only bins is shown for comparison. (right) Quasar bias as a function of luminosity in the different redshift bins. The solid lines in each panel show the $b_Q(z,M_{1450})$ model fit to all 12 data points. The lower panel shows the ratio of the measured quasar bias to the quasar bias predicted by our best-fit $b_Q(z)$ model (Eq. \ref{eq:bqz}).}
    \label{fig:bqevolzLbins-eqnum}
\end{figure}
We find that the quasar bias evolves strongly with redshift in all luminosity bins, as indicated by the positive slope of the quasar bias as a function of redshift.
The right panel of Figure \ref{fig:bqevolzLbins-eqnum} shows the quasar bias as a function of luminosity in each of the redshift bins. 
There is only a very slight dependence of quasar bias on luminosity; much weaker than the redshift evolution. 
We find similar results with the `same boundaries' luminosity split. The correlation functions and bias evolution are shown in Figures \ref{fig:z-L_acf_zbins-grid} and \ref{fig:bqevolzLbins-grid}, respectively. 
These results show more evidence for a dependence on luminosity, specifically in the highest luminosity bin at $z=2.7-3.1$ and $z=3.1-3.5$. 

\begin{table}[]
    \centering
    \begin{tabular}{cccccc}
        \hline
         & $a$ & $b$ & $m$ & $\chi_{\text{red}}^2$ & $p$-value \\
        \hline
        \makecell{Equal \\ numbers} & $0.2056 \pm 0.0112$ & $2.5381 \pm 0.0594$ & $0.0998 \pm 0.0207$ & $0.76$ & $0.65$ \\
        \makecell{Same \\ boundaries} & $0.2017 \pm 0.0106$ & $2.5867 \pm 0.0618$ & $0.1471 \pm 0.0237$ & $4.16$ & $0.00$ \\ 
        \hline
    \end{tabular}
    \caption{Best-fit parameters for the fit of Eq. \ref{eq:bQzM1450} to the 12 redshift-luminosity binned data (9 degrees of freedom) for each of the binning schemes. The same boundaries fit is very poor because the model does not capture the substantially higher bias of the most luminous subsample (see Figure \ref{fig:bqevolzLbins-grid}).}
    \label{tab:bQzL}
\end{table}

\begin{figure}[tbp]
    \includegraphics[width=1.0\textwidth]{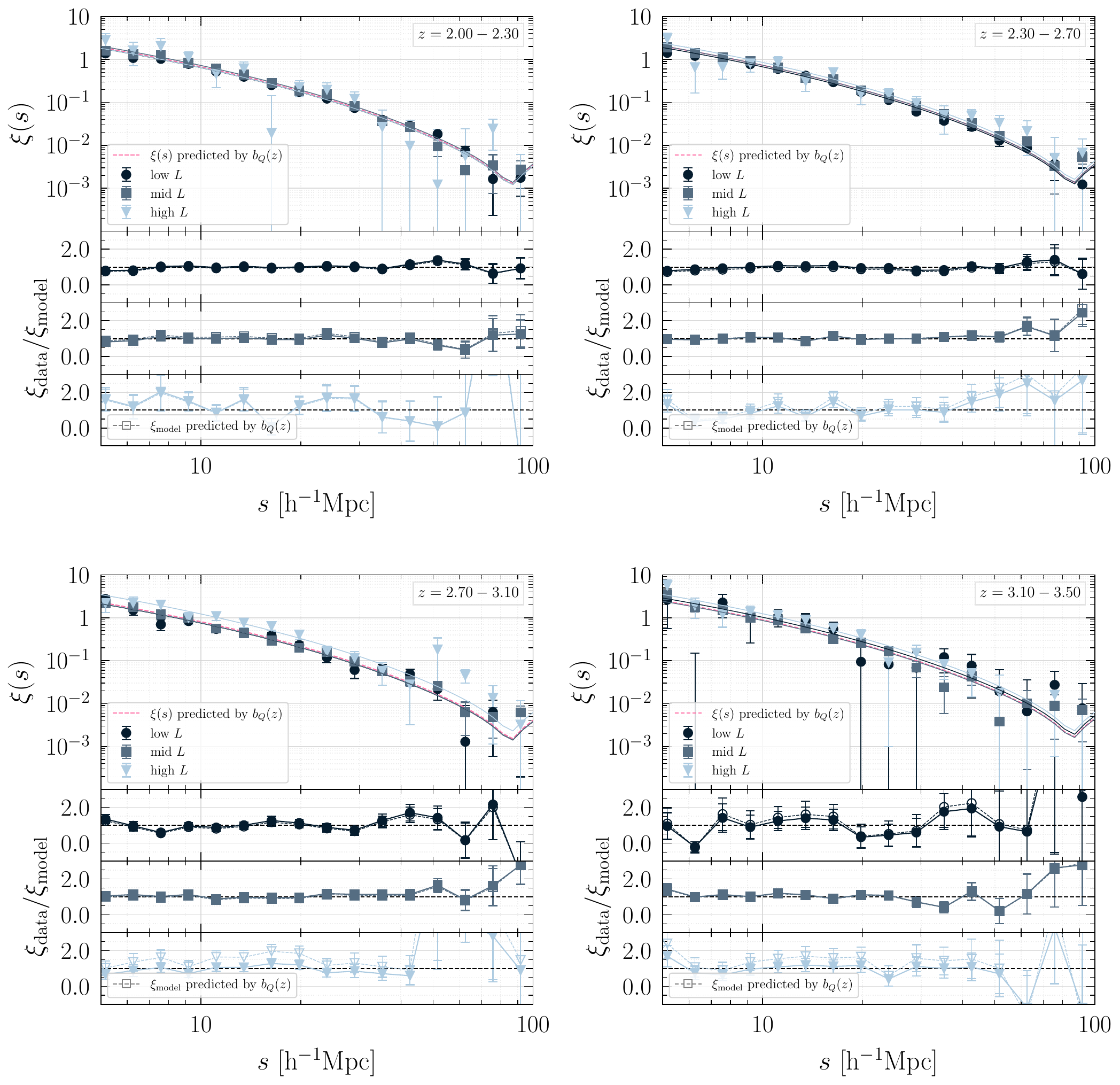}
    \caption{As in Figure~\ref{fig:z-L_acf_zbins-eqnum} for quasars split into the same luminosity bin boundaries at each redshift. The bias values are shown in Figure~\ref{fig:bqevolzLbins-grid} and listed in Table \ref{tab:results}.}
    \label{fig:z-L_acf_zbins-grid}
\end{figure}
\begin{figure}[tbp]
    \includegraphics[width=1.0\textwidth]{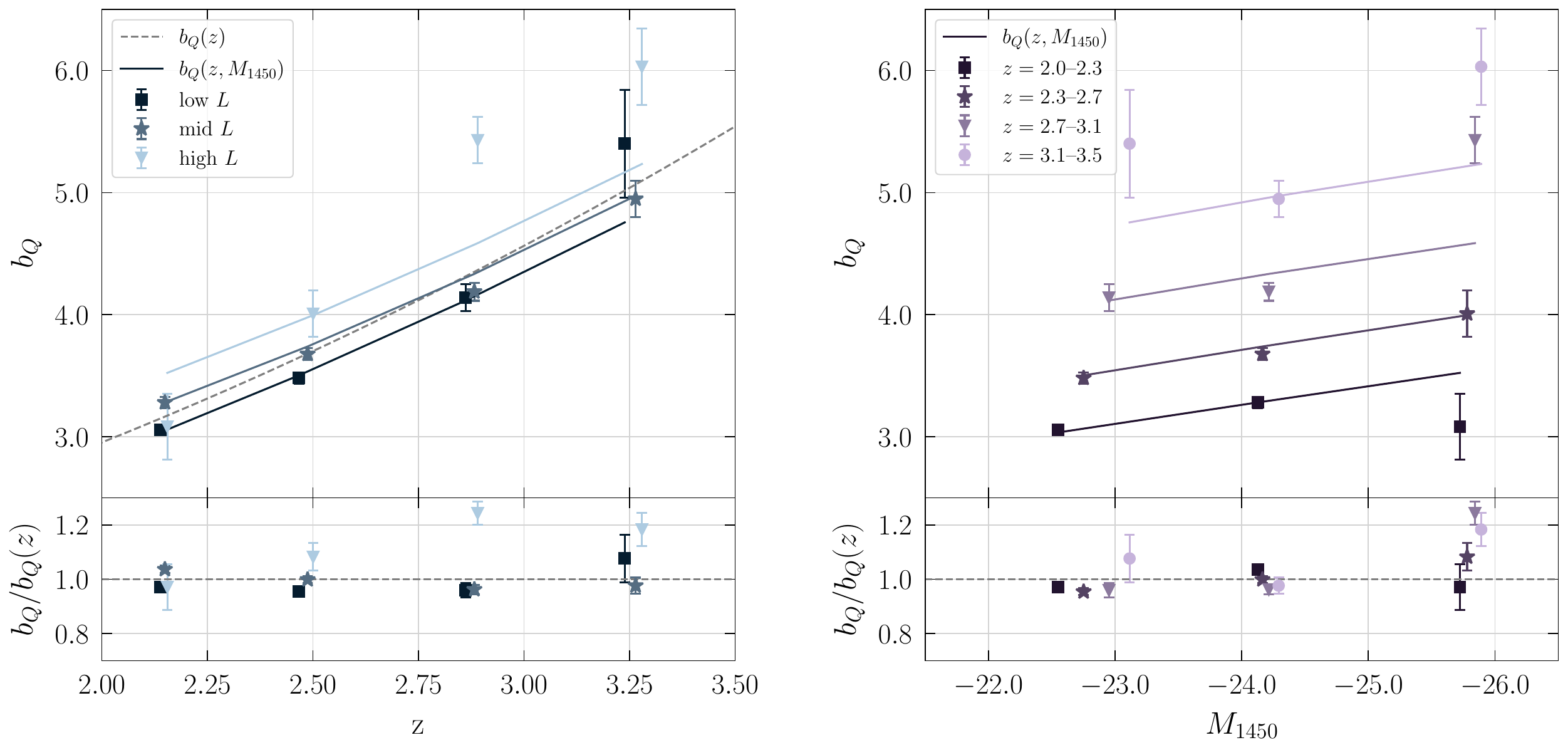}
    \caption{As in Figure~\ref{fig:bqevolzLbins-eqnum} for quasars binned according to the `same boundaries' binning at each redshift. The bias values are listed in Table \ref{tab:results}.}
    \label{fig:bqevolzLbins-grid}
\end{figure}

\begin{figure}[tbp]
    \includegraphics[width=1.0\textwidth]{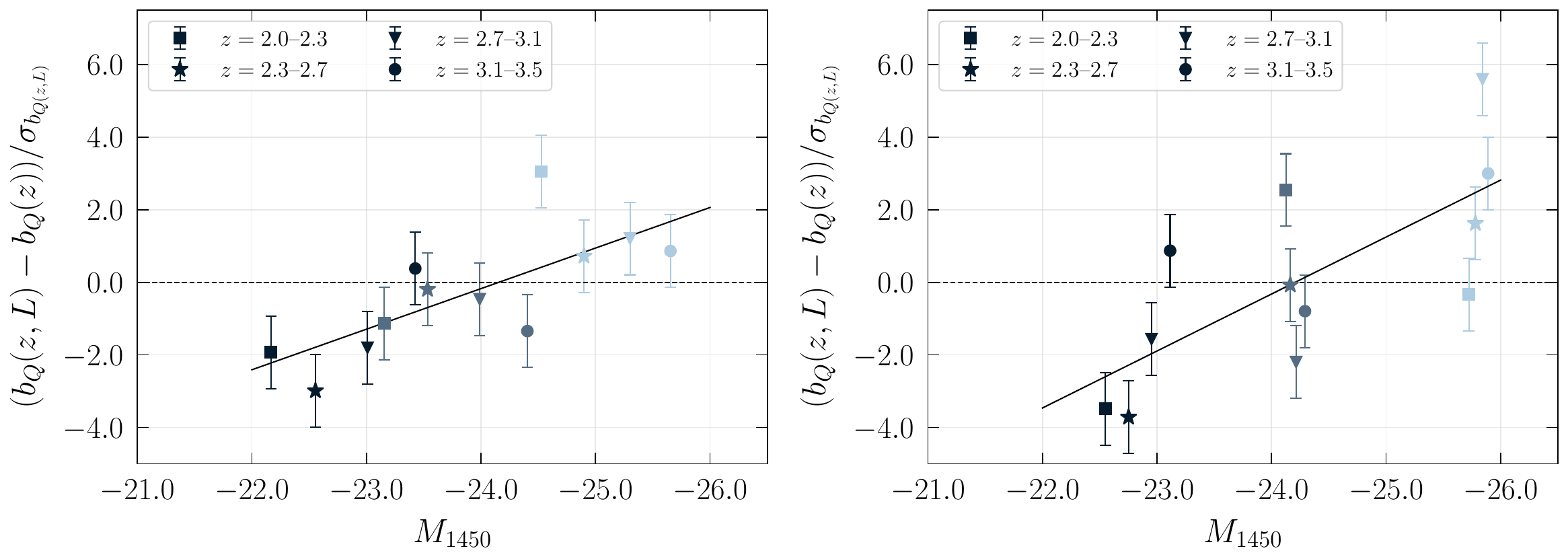}
    \caption{Pulls of the luminosity-binned bias measurements relative to the redshift-only bias model for the `equal numbers' (left) and `same boundaries' (right) binning. The pull is defined as the deviation of $b_Q(z,L)$ from $b_Q(z)$ in units of the statistical uncertainty. The points denote different redshift bins, and the three colors correspond to the `low', `mid', and `high' luminosity bins. The solid black line shows the best-fit linear relation between pull and luminosity. In both cases, the slope clearly deviates from zero which shows our results are inconsistent with a redshift-only model.}
    \label{fig:pull_dist}
\end{figure}

We compare the measured quasar bias values in each redshift-luminosity bin to the redshift-only bias model obtained from fitting Equation \ref{eq:bqz} to the redshift-binned data to assess whether there is evidence for luminosity dependence.
Figure \ref{fig:pull_dist} shows the results of this comparison.
The plotted values represent the deviation of each luminosity-binned measurement from the redshift-only model in units of the statistical uncertainty (i.e., pull values). 
In each panel, the darker shades correspond to lower-luminosity bins, with luminosity increasing toward lighter shades, while different marker styles denote different redshift ranges. 
If the bias depended only on redshift, we would expect the pull values to be consistent with zero, indicated by the dashed horizontal line. 
Instead, we observe a mild trend with luminosity such that higher-luminosity objects appear more clustered than if there were not luminosity dependence, while lower-luminosity objects appear less clustered. 
A $\chi^2$ test of the pull variance yields $\chi^2 = 30.31$ with 11 degrees of freedom for the `equal numbers' binning and $\chi^2 = 83.55$ with 11 degrees of freedom for the `same boundaries' binning.
In both cases, the $p$-values are $\ll 0.05$, suggesting that the pull distributions are inconsistent with the unit-variance expectation of the simple model where clustering only depends on redshift. 
To further explore the luminosity trend, we perform a linear fit to the data in each binning scheme and find the the best-fit linear relations are $m=-1.12 \pm 0.28$ ($\chi^2 = 13.82$, $p=0.18$ for 10 degrees of freedom) for the `equal numbers' binning and $m=-1.57 \pm 0.24$ ($\chi^2 = 39.80$, $p=1.84 \times 10^{-5}$ for 10 degrees of freedom) for the `same boundaries' binning. 
This is a fairly significant deviation from the $m=0$ slope we would expect for the redshift-only model. 
To compare the significance of the linear fit compared to the $m=0$ case with no luminosity dependence, we divide the slope by the slope error and find that the `same boundaries' binning slope is a $6.6 \sigma$ deviation from flat and the `equal numbers' binning slope is a $4.1 \sigma$ deviation.
These results provide strong evidence that quasar clustering is not a function of redshift alone.

The luminosity dependence of the quasar bias has previously been investigated, but these studies have generally lacked the statistical power to robustly constrain this dependence, especially at higher redshifts.
For example, \cite{porciani2006} measured the luminosity and redshift dependence of quasars in the redshift range $0.8 < z < 2.1$ and found only marginal evidence for luminosity dependence, concluding that a larger sample was needed to draw definitive conclusions. 
Similarly, \cite{myers2007} and \cite{daangela2008} reported little or no luminosity dependence in their clustering measurements at $0.75 < \bar{z} < 2.28$ and $z \sim 1.5$, respectively.
At higher redshifts, \cite{white2012} ($2.2 \leq z \leq 2.8$) and \cite{eftekh15} ($2.2 \leq z \leq 3.4$) also found no evidence for luminosity dependence to the quasar bias.

Recently, \cite{knox2025} used the empirical \texttt{TRINITY} model to predict clustering as a function of quasar luminosity and found only a $\gtrsim 0.3$ dex change in quasar bias between a sample with $L_{\mathrm{bol}} = 10^{42}$ erg s$^{-1}$ compared to $L_{\mathrm{bol}} = 10^{46}$ erg s$^{-1}$ across redshifts $z \sim 0-5$.
They found the luminosity dependence to be strongest at higher redshifts ($z \gtrsim 3$) and in the highest luminosity bins ($L_{\mathrm{bol}} > 10^{45}$ erg s$^{-1}$). 
Our sample contains quasars on the higher luminosity end of this range ($L_{\mathrm{bol}} \sim 10^{44} - 10^{48}$ erg s$^{-1}$ using the bolometric correction $BC_{1350} = 3.81$ from \cite{richards2006b}), in the luminosity regime where they showed the luminosity dependence is more significant.
They concluded that the lack of luminosity dependence is due to most quasars living in halos with a small mass range ($10^{12} - 10^{13}\ M_{\odot}$), and that current surveys have uncertainties that are too large to measure the (likely) slight luminosity dependence of quasar clustering.

Quasar samples tend to be magnitude limited, which produce a correlation between the typical luminosity and redshift. 
In addition, the quasar population evolves significantly with redshift such that more luminous quasars are much more common at higher redshifts, at least up to $z \sim 2-3$.  
These effects complicate the measurement and interpretation of any evidence for luminosity-dependent clustering and may explain why previous work did not find evidence for luminosity-dependent clustering. 
Fortunately, the substantial increase in sample size with DESI provides us with sufficient numbers of quasars to increase the precision of luminosity dependence compared to previous studies.
We therefore conclude that quasar bias evolves strongly with redshift, consistent with expectations from large-scale structure growth, but shows only a weak dependence on luminosity over the ranges probed, with the strongest variation found in the highest luminosity bin at $z>2.7$.

\section{Halo masses and duty cycles}
\label{sec:halomass+dutycycle}

The spatial clustering of astrophysical objects provides a long-established way to connect a population to the masses of their host dark matter halos, based on the fact that more massive halos are more strongly clustered than the underlying matter field (e.g.\,\cite{kaiser1984, mo1996}). 
While not every halo hosts a quasar at a given time, quasar clustering measurements can nevertheless be used to infer the typical masses of their host halos and their duty cycles \cite{cole1989, martini2001}. 
In simple terms, the clustering measurement yields the characteristic halo mass, and the ratio of the quasar density to the space density of such halos gives the duty cycle.
A number of studies have used quasar clustering measurements to derive halo masses and duty cycles (e.g.\,\cite{shen2007, eftekh15, laurent2017}).

Some models predict a correlation between quasar luminosity and halo mass (e.g.\,\cite{martini2001, haiman2001}), which could contribute to the higher clustering observed in the smaller, more luminous SDSS sample by \cite{shen2007}. 
White et al.\,\cite{white2008} investigated the implications of that measurement and conservatively set an upper limit on the dispersion in the $L-M_{\mathrm{h}}$ relation; for $f_{\mathrm{duty}} \sim 1$ they found a dispersion in $\ln L$ at fixed $M_{\mathrm{h}}$ of $\sigma_M < 0.5$ (or $< 0.22$ dex).
This already implies a quite tight constraint on the allowed scatter in this $L-M_{\mathrm{h}}$ relation, with smaller duty cycle values requiring even less scatter. 
Shankar et al.\,\cite{shankar2010} found that they could reproduce the observed clustering with either a small duty cycle and small, constant scatter (e.g., $f_{\mathrm{duty}} \sim 6 \times 10^{-4}$ with scatter $\sigma_M=0.1$ dex) or with larger duty cycle and scatter values (e.g., $f_{\mathrm{duty}} \sim 2 \times 10^{-3}$ with scatter $\sigma_M = 0.5$ dex).
Our clustering measurements can help distinguish between these scenarios.

In this section, we assume that each host halo contains at most one quasar.
We describe our methods for estimating the halo masses in Section~\ref{sec:halomass}.
In Section \ref{sec:qsodutycycle} we derive quasar duty cycles, and in Section \ref{subsec:L-Mh_model_predictions} we compare our results to model predictions that incorporate different amounts of scatter in the $L-M_{\mathrm{h}}$ relation.

\subsection{Halo masses}
\label{sec:halomass}

We estimate the masses of dark matter halos hosting quasars using the minimum and characteristic halo masses to quantify the quasar host halo population.
The concepts of characteristic and minimum halo masses, as used in previous quasar clustering studies (e.g.\,\cite{lidz2006, eftekh15, laurent2017}), allow one to infer the typical mass scale of halos hosting quasars and to relate quasar activity to halo properties. 

The minimum halo mass, $M_{\mathrm{h},\min}$, is the lower limit of the halo mass distribution that has the observed bias. 
The definition of this quantity is based on the assumption that the probability of hosting a quasar is independent of halo mass and there is no scatter that would allow quasars below this minimum to host quasars above the luminosity limit. 
To compute the minimum halo mass, we use the bias-weighted mean halo bias from halos above $M_{\mathrm{h},\min}$:
\begin{equation}
\label{eq:mmin}
    b_Q = b_{\mathrm{h}}(M > M_{\mathrm{h}, \min}) = \frac{\int_{M_{\mathrm{h}, \min}}^\infty \frac{dn}{dM} b_{\mathrm{h}}(M) dM}{\int_{M_{\mathrm{h}, \min}}^\infty \frac{dn}{dM} dM}.
\end{equation}
The observed quasar bias is interpreted here as the mean bias of halos in a given redshift bin with a mass larger than $M_{\mathrm{h}, \min}$, weighted by the halo mass function. 

We compute the halo mass function using the model from \cite{tinker2008}. 
Throughout this work, halo masses are defined with respect to a spherical overdensity of $\Delta = 200$ relative to the mean matter density. 
The overdensity is defined as 
\begin{equation}
    \Delta = \frac{M_{\Delta}}{(4/3) \pi R_{\Delta}^3 \bar{\rho}_m (z)}
\end{equation}
where $M_{\Delta}$ is the mass enclosed within the radius $R_{\Delta}$, and $\bar{\rho}_m (z) = \Omega_m (z) \rho_{\mathrm{crit}} (z)$ is the mean matter density at redshift $z$. 

The differential halo mass function can be written as 
\begin{equation}
    \frac{dn}{d\ln M} = f(\sigma)\, \frac{\bar{\rho}_m (0)}{M}\, \frac{d\ln \sigma^{-1}}{d\ln M},
    \label{eq:hmf}
\end{equation}
where $\bar{\rho}_m (0)$ is the present-day ($z=0$) mean matter density and $\sigma(M,z)$ is the rms fluctuation of the linear density field on the mass scale $M$.
The quantity $dn/d \ln M$ gives the comoving number density of halos per logarithmic mass interval, with units of $(\mathrm{Mpc}/h)^{-3}$. 

The function $f(\sigma)$ is given by 
\begin{equation}
	f(\sigma) = A \left[ \left( \frac{\sigma}{b} \right)^{-a} + 1 \right] e^{-c/\sigma^2} ,
\end{equation}
where the parameters $A$, $a$, $b$, and $c$ depend on the overdensity and redshift as specified in \cite{tinker2008}. 
The variance is computed as 
\begin{equation}
    \sigma^2(M,z) = \int P(k)\,\hat{W}(kR)\,k^2\,dk,
\end{equation}
where $P(k)$ is the linear matter power spectrum and $\hat{W}(kR)$ is the Fourier transform of a real-space top-hat window function of radius $R$, with $M = (4/3) \pi R^3 \bar{\rho}_m (0)$.

We compute the halo bias using the function from \cite{tinker2010}, who define the linear halo bias as a function of peak height $\nu$:
\begin{equation}
	b_{\mathrm{h}}(\nu) = 1 - A \frac{\nu^a}{\nu^a + \delta_c^a} + B \nu^b + C \nu^c,
\end{equation}
where $\nu = \delta_c / \sigma(M,z)$ is the peak height, $\delta_c \approx 1.686$ is the critical overdensity for collapse, and $\sigma(M,z)$ is the rms fluctuations of the linear matter density field as defined above.
The bias parameters $A$, $a$, $B$, $b$, $C$, $c$ are taken from \cite{tinker2010} for the same overdensity definition. 
We evaluate both the halo mass function and halo bias function using the \texttt{Colossus}\,\cite{colossus2018} package. 

The characteristic halo mass, $\bar{M}_{\mathrm{h}}$, is the halo mass whose clustering matches the observed bias of the population, that is 
\begin{equation}
    b_Q = b_{\mathrm{h}}(\bar{M}_h). 
\end{equation}
This quantity gives an ``average'' halo mass for a given quasar population, under the assumption that quasars live in halos with similar clustering properties. 

\begin{figure}[tbp]
    \centering
	\includegraphics[width=0.49\textwidth]{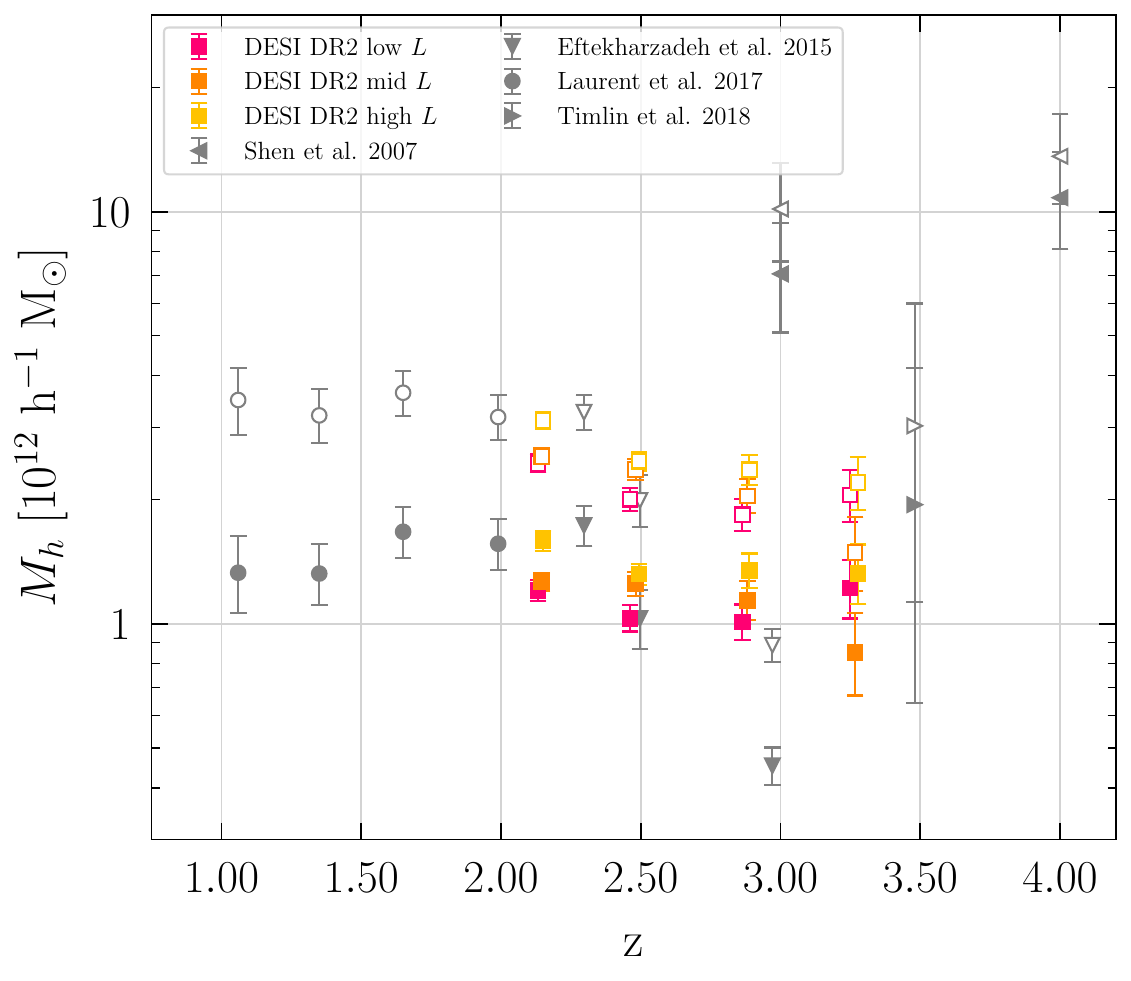}
    \includegraphics[width=0.49\textwidth]{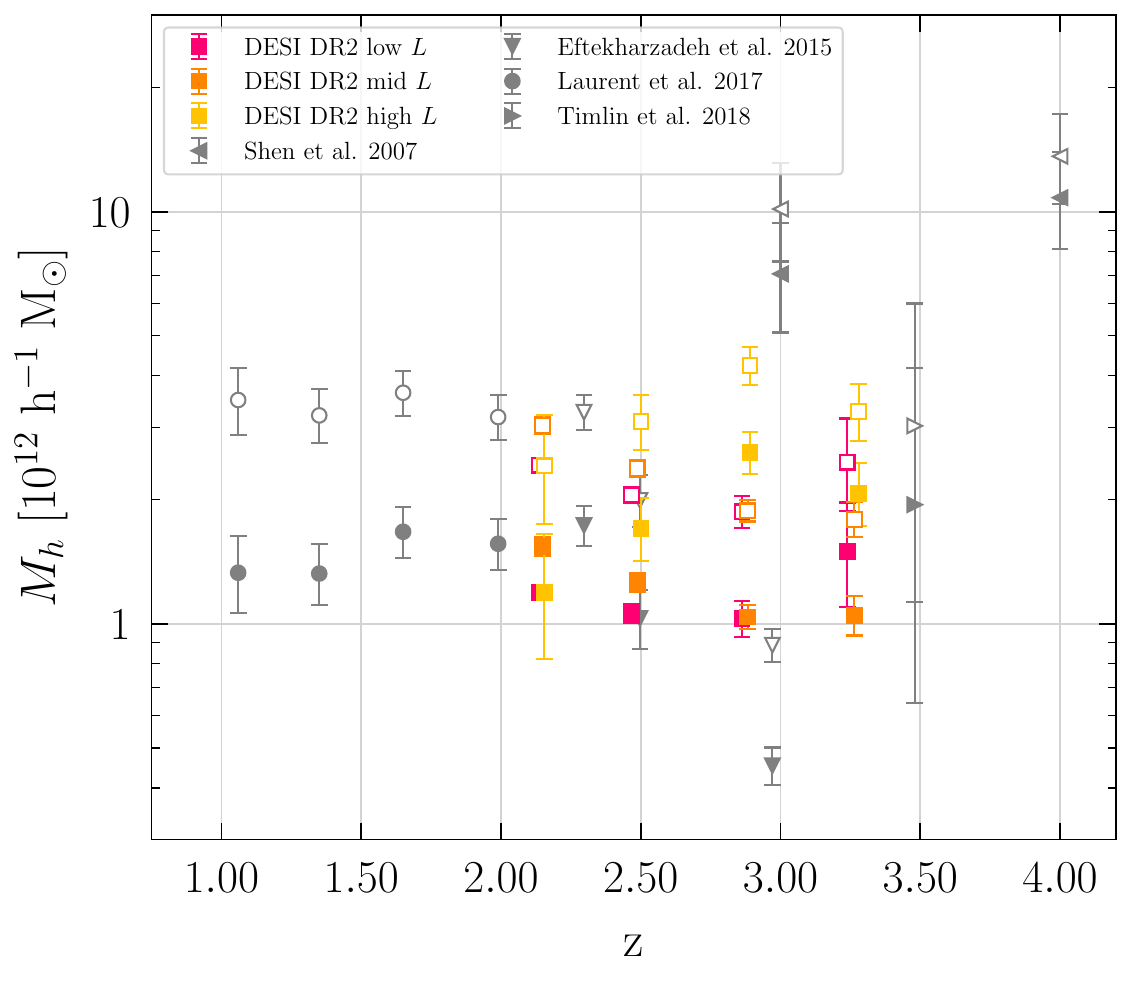}
    \caption{Characteristic (open points) and minimum (filled points) halo masses as a function of redshift for our `equal numbers' luminosity binning scheme (left) and `same boundaries' binning scheme (right). The pink, orange, and yellow squares correspond to the `low $L$', `mid $L$', and `high $L$' bins, respectively. We find only modest evolution in the halo mass and weak evidence for a correlation between halo mass and luminosity at lower redshift and luminosities. There is stronger evidence for luminosity dependence in the highest luminosity bin, especially at higher redshifts. We have also included previous measurements in the literature for comparison (gray points). See Section \ref{sec:halomass} for further details.}
    \label{fig:Mh}
\end{figure}

Figure \ref{fig:Mh} shows our estimates for the host halo masses for each of the redshift-luminosity splits described in Section \ref{sec:datasplits}. 
The open points represent characteristic halo masses and the solid points the minimum halo masses. 
We find that quasars are hosted by dark matter halos within a narrow $M_{\mathrm{h}, \min}$  (or $\bar{M}_{\mathrm{h}}$) range of just a few $10^{12} h^{-1} M_{\odot}$ over the entire redshift range we probed.
The number density of halos changes rapidly over this redshift range, and the inferred halo mass for each subsample should be interpreted as the host halo mass for quasars at the mean redshift of the corresponding subsample.
Our `equal numbers' binning shows a slight decrease in halo mass as redshift increases and weak dependence on luminosity for the host halo mass. 
The halo masses inferred in the `same boundaries' binning scheme shows evidence for stronger luminosity dependence, especially in the highest luminosity bin at higher redshifts.
We compare to previous studies by estimating the halo masses from their reported bias values.
We find our results are consistent with previous studies in this redshift range, such as \cite{eftekh15}, though their results show a more significant redshift evolution for both the minimum and characteristic halo masses, with smaller minimum halo masses required at higher redshifts compared to lower redshifts. 
There does appear to be a slight dependence on luminosity for the host halo mass; lower luminosity quasars seem to slightly favor lower mass halos more than higher luminosity quasars, which is consistent with what we found in our analysis in Section \ref{sec:zlumevol}. 
We explore the physical implications of this trend in Section \ref{subsec:L-Mh_model_predictions}. 

\subsection{Quasar duty cycle}
\label{sec:qsodutycycle}

We constrain the fraction of halos that host an active quasar using the quasar duty cycle. 
The quasar duty cycle compares the number density of quasars over a redshift and luminosity range to the number density of halos that have similar clustering strength.
This formulation, as introduced by \cite{martini2001}, assumes a monotonic relationship between luminosity and halo mass, such that there is only one quasar per host halo, and that quasars are at least as clustered as the dark matter halos that host them. 
Mathematically, the quasar duty cycle is defined as
\begin{equation}
	f_\mathrm{duty} = \frac{n_{\mathrm{QSO}}}{n_{\mathrm{h}}} = \frac{\int_{L_{\min}}^{L_{\max}} \Phi (L) \ dL}{\int_{M_{\mathrm{h}, \min}}^\infty \frac{dn_{\mathrm{h}}}{dM_{\mathrm{h}}} \ dM_{\mathrm{h}}},
\end{equation} \label{eqn:duty}
where $\Phi(L)$ represents the quasar luminosity function integrated over each luminosity bin, and $\frac{dn_{\mathrm{h}}}{dM_{\mathrm{h}}}$ is the halo mass function integrated above the minimum halo mass $M_{\mathrm{h},\min}$ for the bin. 
We use the space densities computed in the LSS catalogs for our duty cycle calculations \cite{ross2025}. 

\begin{figure}[tbp]
    \centering
	\includegraphics[width=1.0\textwidth]{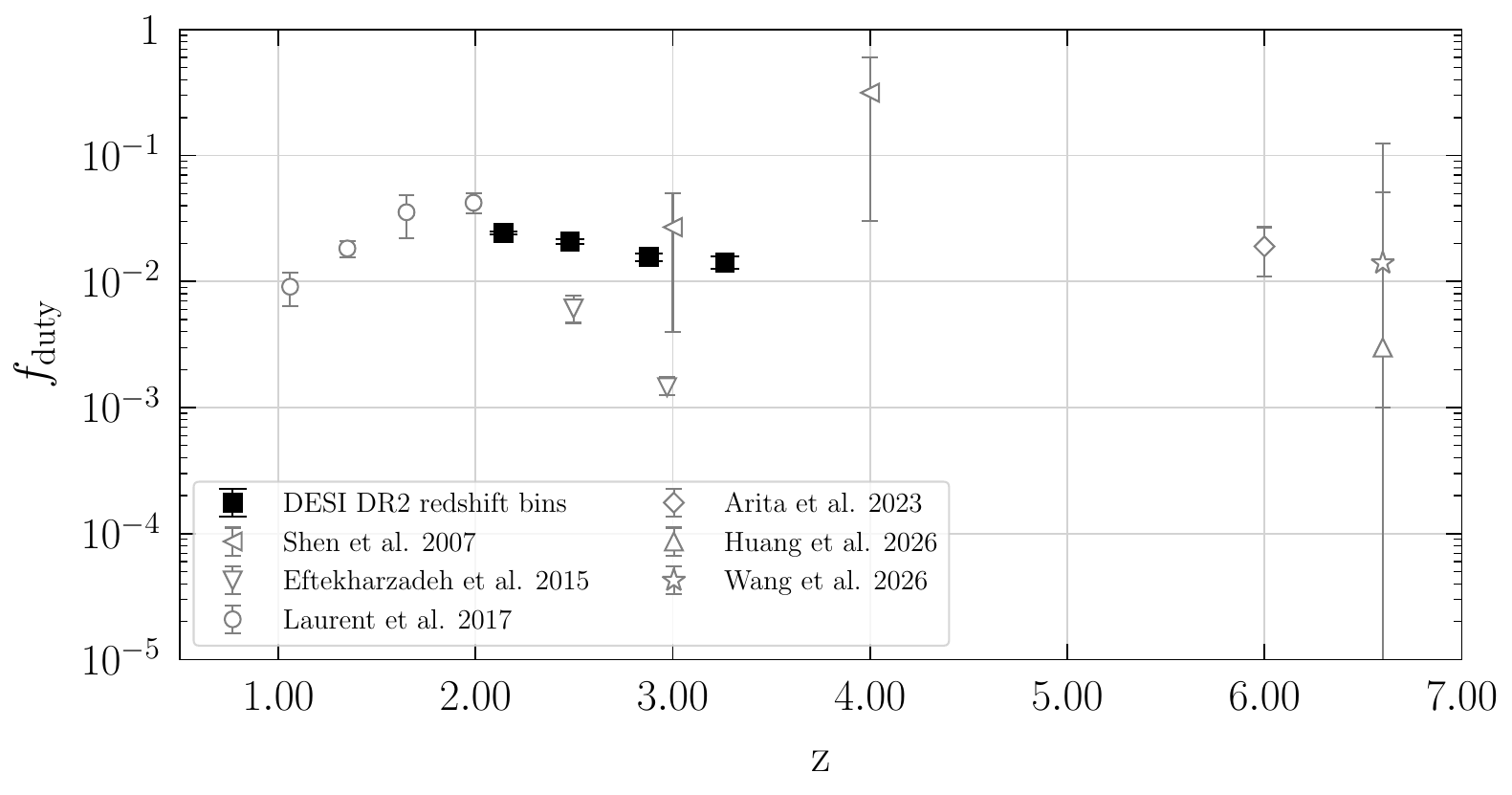}
    \caption{Quasar duty cycle as a function of redshift with the same symbols as in Figure~\ref{fig:Mh}. The duty cycle is the ratio of the quasar space density to the halo space density above the minimum halo mass. The values are nearly $10^{-2}$ and therefore only a small fraction of halos of similar mass host quasars. See Section~\ref{sec:qsodutycycle} for further details.}
    \label{fig:fduty}
\end{figure}

We find that the duty cycle for our sample is small and changes by only a factor of two between $z=3.26$ and $z=2.14$.
Figure \ref{fig:fduty} shows our duty cycle measurements for the individual redshift bins (solid squares), compared to other results in the literature. The corresponding values are listed in Table \ref{tab:nqsofduty}. 
Small duty cycle values imply that quasars appear short-lived and episodic, compared to the lifetime of dark matter halos. 
A flat or constant duty cycle implies that quasar triggering is not dependent on redshift.
The duty cycles we measure suggest that quasars are a relatively brief phase in the lifetime of massive halos. 
Since these duty cycle values are for the entire sample at each redshift, they should be interpreted as the probability that a halo of mass $M_{\min} \gtrsim 10^{12} h^{-1}M_{\odot}$ hosts a quasar above the redshift-dependent $M_{1450}$ threshold visible in Figure 2, which corresponds to $L_{\mathrm{bol}} \gtrsim 10^{45}$ erg s$^{-1}$ at $z=2.7$.

\begin{table}[]
    \centering
    \renewcommand{\arraystretch}{1.25} 
    \begin{tabular}{cccc}
        \hline
        \noalign{\vspace{3pt}}
        $\Delta z$ & $\bar{z}$ & \shortstack{$n_{\mathrm{QSO}}$ \\ $(10^{-6} \,h^{-1} \,\mathrm{Mpc})^{-3}$} & $f_{\mathrm{duty}}$ \\
        \hline
        \multicolumn{4}{l}{\underline{Full Sample}:} \\
        $2.0 \leq z \leq 3.5$ & $2.477$ & $11.92 \pm 0.014$ & $0.0150 \pm 0.0004$ \\
        \multicolumn{4}{l}{\underline{Redshift subsamples}:} \\
        $2.0 \leq z \leq 2.3$ & $2.143$ & $23.64 \pm 0.043$ & $0.0243 \pm 0.0008$ \\
        $2.3 \leq z \leq 2.7$ & $2.479$ & $15.13 \pm 0.029$ & $0.0207 \pm 0.0008$ \\
        $2.7 \leq z \leq 3.1$ & $2.878$ & $7.996 \pm 0.022$ & $0.0157 \pm 0.0011$ \\
        $3.1 \leq z \leq 3.5$ & $3.264$ & $3.494 \pm 0.014$ & $0.0142 \pm 0.0016$ \\
        \hline
    \end{tabular}
    \caption{Quasar number densities and duty cycle values for the full sample and each of the redshift bins in our sample.}
    \label{tab:nqsofduty}
\end{table}

\subsection{Simple models}
\label{subsec:L-Mh_model_predictions}

Our results indicate that quasar clustering shows little dependence on luminosity. 
One possible explanation is that the bias is changing too slowly with halo mass over the luminosity range of our sample. 
To test whether our data could detect luminosity-dependent clustering, we first assumed a monotonic $L-M_{\mathrm{h}}$ relation with no scatter.
In this scenario, the minimum halo mass of a quasar sample corresponds to the lowest-luminosity objects. 
Using this model, we estimate the halo mass function at a given redshift and divide it into three parts following the two binning schemes used to probe luminosity dependence, allowing us to predict the expected bias in each luminosity bin.
This procedure implicitly assumes that the duty cycle is independent of halo mass.

For both binning schemes we begin by estimating the halo mass function at the mean redshift of each redshift bin. 
We then determine the halo mass $M_{\mathrm{h},\mathrm{frac}}$ corresponding to a given fraction of the cumulative halo mass function above the minimum halo mass of the sample, $M_{\mathrm{h},\min}$, using
\begin{equation}
    \mathrm{frac} = \frac{\int_{M_{\mathrm{h,frac}}}^{\infty} \frac{dn_{\mathrm{h}}}{dM_{\mathrm{h}}}\,dM_{\mathrm{h}}}{\int_{M_{\mathrm{h},\min}}^{\infty} \frac{dn_{\mathrm{h}}}{dM_{\mathrm{h}}}\,dM_{\mathrm{h}}},
    \label{eq:modelMh}
\end{equation}
where $dn/dM$ is the halo mass function, and frac is the fraction of the total sample contained in each bin.
For the `equal numbers' luminosity split, we divide the area under the halo mass function above $M_{\mathrm{h},\min}$ into three equal parts (terciles) within each redshift bin, corresponding to divisions at $\mathrm{frac}=1/3$ and $\mathrm{frac}=2/3$ for the most luminous third and most luminous two-thirds of the sample, respectively.  
To reproduce the `same boundaries' luminosity split, we compute the fraction of quasars in each luminosity bin within a given redshift bin and use these fractions as the input values for $\mathrm{frac}$ in Equation \ref{eq:modelMh}. 

\begin{figure}
    \centering
    \includegraphics[width=1.0\linewidth]{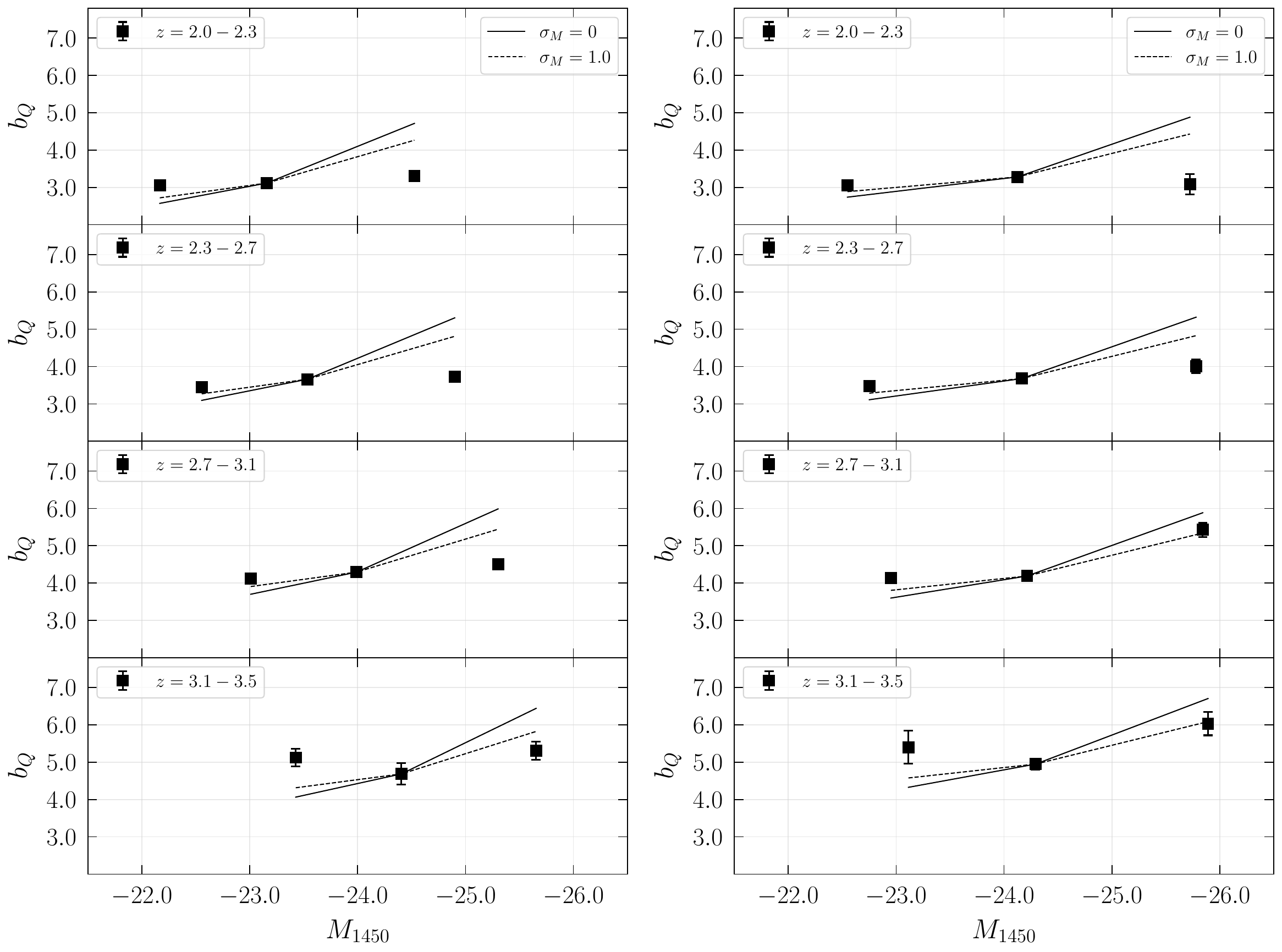}
    \caption{Predicted quasar bias values for models with no scatter (solid line) and with scatter $\sigma_M=1.0$ (dashed line) compared to measured quasar bias values (black squares). The left panel shows the results for the `equal numbers' luminosity bins and the right panel for the `same boundaries' luminosity bins. The models have been shifted so the model matches the $b_Q$ value for the `mid $L$' point for visual clarity.}
    \label{fig:L-Mh_predictions}
\end{figure}

We apply Eq. \ref{eq:modelMh} at each fraction boundary to obtain the halo mass limits $M_{h,\mathrm{frac}}$ separating the bins, and use these as the integration limits to compute the expected quasar bias in each bin:
\begin{equation}
    b_Q(M_{\mathrm{h},\max} < M_\mathrm{h} < M_{\mathrm{h},\min}) = \frac{\int_{M_{\mathrm{h},\min}}^{M_{\mathrm{h},\max}} \frac{dn_{\mathrm{h}}}{dM_{\mathrm{h}}} b(M_{\mathrm{h}}) ~dM_{\mathrm{h}}}{\int_{M_{\mathrm{h},\min}}^{M_{\mathrm{h},\max}} \frac{dn_{\mathrm{h}}}{dM_{\mathrm{h}}} ~dM_{\mathrm{h}}}.
    \label{eq:bQ_model}
\end{equation}
Here, $M_{\mathrm{h},\min}$ and $M_{\mathrm{h},\max}$ are replaced by the three mass ranges $[M_{\mathrm{h},\min},M_{\mathrm{h},2/3}]$, $[M_{\mathrm{h},2/3}, M_{\mathrm{h},1/3}]$, and $[M_{\mathrm{h},1/3}, \infty]$ for the `equal numbers' binning and $[M_{\mathrm{h},\mathrm{low}}, M_{\mathrm{h},\mathrm{mid}}]$, $[M_{\mathrm{h},\mathrm{mid}}, M_{\mathrm{h},\mathrm{high}}]$, $[M_{\mathrm{h},\mathrm{high}}, \infty]$ for the `same boundaries' binning. 
Table \ref{tab:Mh_model} lists these boundaries for each binning scheme.
Figure \ref{fig:L-Mh_predictions} shows the predicted quasar bias for a perfectly monotonic $L-M_{\mathrm{h}}$ relationship (solid lines) alongside the measured bias values (black squares). These points are labeled $\sigma_M = 0$ to indicate that they are calculated from a model that has a perfectly monotonic (zero scatter) relation between halo mass and luminosity.
The left panel corresponds to the `equal numbers' luminosity bins and the right panel to the `same boundaries' bins. 
Errors on the predicted bias are derived from the uncertainties on the computed $M_{\mathrm{h},\min}$ values. The variation in predicted bias with luminosity is clearly much greater than we measure. 

\begin{table}[]
    \centering
    \renewcommand{\arraystretch}{1.25} 
    \begin{tabular}{ccccccc}
        \hline
         & \multicolumn{3}{c}{$\log_{10} M_h$} & \multicolumn{3}{c}{frac} \\
        \hline
        $\Delta z$ & $\min$ & mid & $\max$ & low $L$ & mid $L$ & high $L$ \\
        \hline
        \multicolumn{7}{l}{\underline{`equal numbers'}:} \\
        $2.0-2.3$ & $12.13$ & $12.25$ & $12.44$ & $0.33$ & $0.33$ & $0.33$ \\
        $2.3-2.7$ & $12.10$ & $12.21$ & $12.38$ & $0.33$ & $0.33$ & $0.33$ \\
        $2.7-3.1$ & $12.06$ & $12.16$ & $12.32$ & $0.33$ & $0.33$ & $0.33$ \\
        $3.1-3.5$ & $12.10$ & $12.19$ & $12.33$ & $0.33$ & $0.33$ & $0.33$ \\
        \hline
        \multicolumn{7}{l}{\underline{`same boundaries'}:} \\
        $2.0-2.3$ & $12.13$ & $12.39$ & $12.87$ & $0.60$ & $0.34$ & $0.05$ \\
        $2.3-2.7$ & $12.10$ & $12.27$ & $12.67$ & $0.47$ & $0.44$ & $0.09$ \\
        $2.7-3.1$ & $12.06$ & $12.15$ & $12.48$ & $0.30$ & $0.54$ & $0.16$ \\
        $3.1-3.5$ & $12.10$ & $12.14$ & $12.40$ & $0.15$ & $0.60$ & $0.24$ \\
        \hline
    \end{tabular}
    \caption{Boundaries corresponding to the integration limits in Eq. \ref{eq:bQ_model} to match each of the luminosity binning schemes. The $M_{h,\min}$ value is the minimum halo mass inferred from the measured quasar bias for each redshift bin. For the `equal numbers' binning, $M_{h,\mathrm{mid}}$ and $M_{h,\max}$ are determined by dividing the area under the halo mass function (above $M_{h,\min}$) into equal area thirds. For the `same boundaries' binning, $M_{h,\mathrm{mid}}$ and $M_{h,\max}$ are determined by dividing the area under the curve into fractions to match the fraction of objects in each bin in the data. These fractions are listed in the columns `low $L$ frac' which refers to the range $[M_{h,\min}$ to $M_{h,\mathrm{mid}}]$, `mid $L$ frac' for the range $[M_{h,\mathrm{mid}}, M_{h,\max}]$, and `high $L$ frac' for the range $[M_{h,\max}, \infty]$, corresponding to the three luminosity bins in our sample.}
    \label{tab:Mh_model}
\end{table}

A monotonic $L-M_{\mathrm{h}}$ relation is the simplest case to model; however, it is also possible that scatter exists in this relation. 
Scatter may arise from variations in Eddington ratio (e.g.\,\cite{kollmeier2006}), the black hole mass--galaxy mass relation (e.g.\,\cite{haring2004, hopkins2007}), and the stellar mass-halo mass relation (e.g.\,\cite{behroozi2010, tinker2017}).
We model this scatter following the prescription outlined in \cite{white2008} where the probability that a quasar has luminosity $L$ is
\begin{equation}
    P(L|M_{\mathrm{h}}) d\ln L = \frac{1}{\sqrt{2\pi} \sigma_L} \exp \left[ - \frac{\ln^2 (L/L_0)}{2 \sigma_L^2} \right] d \ln L,
\end{equation}
$L_0(M_h)$ is the central luminosity and $\sigma_L$ is the scatter of the lognormal distribution. 
Then the probability that a halo hosts a quasar above some limiting luminosity, $L_{\min}$, is 
\begin{equation}
    P(L>L_{\min}|M_{\mathrm{h}}) = \int_{L_{\min}}^\infty P(L|M_{\mathrm{h}}) ~d \ln L.
\end{equation}
We assume that at most a fraction $f_{\mathrm{on}}$ of halos host an active quasar, and that $L_0 \propto M_{\mathrm{h}}^{\alpha}$. 
To add scatter to this $L-M_{\mathrm{h}}$ relation, we follow \cite{white2008}
\begin{equation}
    N(M_{\mathrm{h}}) = \frac{f_\mathrm{on}}{2} \mathrm{erfc} \left[ \frac{\ln M_t / M_{\mathrm{h}}}{\sqrt{2} \sigma_M} \right],
\end{equation}
where $N(M_{\mathrm{h}})$ is the mean number of quasars above $L_{\min}$ in halos of mass $M_{\mathrm{h}}$. $M_t$ here represents the halo mass at which $L_0(M_t) \equiv L_{\min}$ and $\sigma_M = \sigma_L/\alpha$ with $\alpha = 1$.
For simplicity, and based on our results for the duty cycle measurements, we further assume that the duty cycle is a constant rather than a function of halo mass. 
We then compute the effective quasar bias at a given halo mass using 
\begin{equation}
    \langle b \rangle = \frac{\int dM_{\mathrm{h}} \frac{dn_{\mathrm{h}}}{dM_{\mathrm{h}}} b_{\mathrm{h}}(M_{\mathrm{h}}) N(M_{\mathrm{h}})}{\int dM_{\mathrm{h}} \frac{dn_{\mathrm{h}}}{dM_{\mathrm{h}}} N(M_{\mathrm{h}})},
\end{equation}
where $dn_{\mathrm{h}}/dM_{\mathrm{h}}$ is the (comoving) number density of halos per mass interval and $b_{\mathrm{h}}(M_{\mathrm{h}})$ is the bias associated with halos of that mass. 

The dashed lines in Figure \ref{fig:L-Mh_predictions} show the predicted quasar bias values after adding a scatter of $\sigma_M=1.0$ ($\sim 0.4$ dex). 
Compared to the no-scatter case, we find that adding scatter reduces both the range and the overall value of the predicted bias.
This is consistent with physical expectations: if there is a tight relation between $L-M_{\mathrm{h}}$, then only the rarest halos would host the highest luminosity quasars, producing higher bias values. 
Scatter allows lower-mass halos to contribute, reducing the mean bias, especially at high redshift where the halo mass function declines steeply; even a small contribution from lower-mass halos has a large impact on the average bias. 
While the model with $L-M_{\mathrm{h}}$ scatter produces a better match to the data, it nevertheless has a greater trend with luminosity than the data. 

We perform a $\chi^2$ test to test whether either model provides a sufficient description of our data. 
We compute $\chi^2$ by comparing the measured bias $b_Q(z_i, L_{i,j})$ to the model prediction $b_{Q,\mathrm{model}}(z_i, \sigma_{M\ i,j})$ in each $(z,L)$ bin, weighting by the squared measurement uncertainty,  
$$
\chi^2 = \sum_{i,j} \frac{(b_Q(z_i,L_{i,j}) - b_{Q,\mathrm{model}}(z_i,\sigma_{M\,i,j}))^2}{\sigma_{b_Q(z_i,L_{i,j})}^2}.
$$
The $\sigma_M = 1.0$ case is a slightly better match than $\sigma_M = 0$, although for both $\sigma_M$ values and both binning schemes the $\chi^2$ indicate extremely poor fits. 
The two cases we considered represent fairly extreme assumptions about the amount of scatter, yet neither captures the trends seen in our measurements. 
This suggests that the relationship between quasar luminosity and host halo mass is likely more complicated than the simple power-law form with slope $\alpha=1$ assumed in our model.
If quasar luminosity is more fundamentally connected to stellar or black hole mass, then the nonlinear stellar mass--halo mass relation would naturally produce a non-power-law (or steeper power-law) mapping between quasar luminosity and halo mass \cite{conroy2013}. 
Consistent with this, recent empirical modeling of high-redshift quasar clustering finds that the data prefer a steeper relation, with a power law slope of $\gamma \gtrsim 2$ with very little scatter $(\sigma \lesssim 0.3 \text{ dex})$ for quasars at $z \sim 4$ \cite{pizzati2024}.
This suggests that a more sophisticated treatment of the $L-M_{\mathrm{h}}$ relation (e.g., adding redshift dependence to the model, or correlations with additional halo or quasar properties) is necessary to fully explain the observed bias behavior. 
Both models considered here assume that the quasar duty cycle is independent of halo mass over the range probed by our sample, with the luminosity of an active system connected to halo mass either with $\sigma_M=0.0$ or $\sigma_M=1.0$. 
If the duty cycle instead grows rapidly with increasing halo mass, then the most luminous quasars could reside in just slightly more massive halos, reducing the predicted luminosity dependence and improving agreement with the DESI measurements. 
An additional consideration when interpreting the weak luminosity dependence we are detecting, is that both redshift and luminosity are challenging to determine precisely. 
Measurement uncertainties can introduce scatter into the redshift--luminosity relation, which would tend to wash out any intrinsic luminosity dependence.
At higher redshift, evolving or imperfect k-corrections may introduce additional scatter, further suppressing the observed luminosity dependence. 

\newpage
\newgeometry{top=0.5in, bottom=0.65in, left=0.5in, right=0.5in}

\begin{table}[htbp]
\small
\centering
\renewcommand{\arraystretch}{1.25} 
\begin{tabular}{ccccccccc}
\hline
\noalign{\vspace{3pt}}
$\Delta z$ & $\bar{z}$ & $\Delta M_{1450}$ & $\bar{M}_{1450}$ & $b_Q$ & $\chi^2_\mathrm{red}$ & $N_{QSO}$ & \shortstack{$M_{h,\min}$ \\ ($10^{12}\,h^{-1}\,\mathrm{M}_\odot$)} & \shortstack{$\bar{M}_h$ \\ ($10^{12}\,h^{-1}\,\mathrm{M}_\odot$)} \\
\hline
\multicolumn{9}{l}{\underline{Full sample}:} \\
$2.00-3.50$ & 2.477 & $[-19.94, -29.00]$ & -23.64 & $3.61 \pm 0.01$ & 0.80 & 713706 & $1.192^{+0.018}_{-0.018}$ & $2.273^{+0.030}_{-0.030}$ \\
\multicolumn{9}{l}{\underline{Redshift subsamples}:} \\
$2.00-2.30$ & 2.143 & $[-19.94, -28.19]$ & -23.26 & $3.16 \pm 0.02$ & 0.86 & 284985 & $1.349^{+0.030}_{-0.030}$ & $2.700^{+0.053}_{-0.053}$ \\
$2.30-2.70$ & 2.479 & $[-19.95, -28.65]$ & -23.66 & $3.66 \pm 0.02$ & 1.42 & 245546 & $1.257^{+0.031}_{-0.030}$ & $2.381^{+0.051}_{-0.051}$ \\
$2.70-3.10$ & 2.878 & $[-19.97, -28.92]$ & -24.10 & $4.29 \pm 0.05$ & 0.92 & 128371 & $1.144^{+0.047}_{-0.045}$ & $2.049^{+0.073}_{-0.072}$ \\
$3.10-3.50$ & 3.264 & $[-20.06, -29.00]$ & -24.50 & $5.20 \pm 0.08$ & 0.61 & 54804 & $1.257^{+0.075}_{-0.072}$ & $2.104^{+0.111}_{-0.108}$ \\
\multicolumn{9}{l}{\underline{Redshift-Luminosity subsamples (same boundaries)}:} \\
$2.00-2.30$ & 2.138 & $[-19.94, -23.43]$ & -22.55 & $3.06 \pm 0.03$ & 1.37 & 171694 & $1.193^{+0.041}_{-0.040}$ & $2.427^{+0.074}_{-0.072}$ \\
 & 2.149 & $[-23.43, -25.18]$ & -24.13 & $3.28 \pm 0.05$ & 1.08 & 97951 & $1.539^{+0.084}_{-0.081}$ & $3.028^{+0.144}_{-0.141}$ \\
 & 2.155 & $[-25.18, -28.19]$ & -25.72 & $3.08 \pm 0.27$ & 1.27 & 15340 & $1.192^{+0.461}_{-0.371}$ & $2.418^{+0.802}_{-0.674}$ \\
$2.30-2.70$ & 2.467 & $[-19.95, -23.43]$ & -22.75 & $3.48 \pm 0.04$ & 0.92 & 114894 & $1.060^{+0.054}_{-0.053}$ & $2.054^{+0.092}_{-0.090}$ \\
 & 2.487 & $[-23.43, -25.18]$ & -24.16 & $3.68 \pm 0.05$ & 0.80 & 107353 & $1.262^{+0.067}_{-0.065}$ & $2.386^{+0.111}_{-0.108}$ \\
 & 2.500 & $[-25.18, -28.65]$ & -25.78 & $4.01 \pm 0.19$ & 0.87 & 23299 & $1.704^{+0.311}_{-0.280}$ & $3.099^{+0.493}_{-0.453}$ \\
$2.70-3.10$ & 2.862 & $[-19.97, -23.43]$ & -22.95 & $4.14 \pm 0.11$ & 0.97 & 38283 & $1.030^{+0.109}_{-0.102}$ & $1.871^{+0.174}_{-0.165}$ \\
 & 2.882 & $[-23.43, -25.18]$ & -24.21 & $4.19 \pm 0.07$ & 1.11 & 69272 & $1.040^{+0.071}_{-0.068}$ & $1.881^{+0.113}_{-0.109}$ \\
 & 2.890 & $[-25.18, -28.92]$ & -25.84 & $5.43 \pm 0.19$ & 1.29 & 20816 & $2.605^{+0.322}_{-0.300}$ & $4.234^{+0.460}_{-0.435}$ \\
$3.10-3.50$ & 3.239 & $[-20.06, -23.43]$ & -23.11 & $5.40 \pm 0.44$ & 0.56 & 8480 & $1.498^{+0.473}_{-0.397}$ & $2.467^{+0.682}_{-0.590}$ \\
 & 3.264 & $[-23.43, -25.18]$ & -24.29 & $4.95 \pm 0.15$ & 0.95 & 32922 & $1.049^{+0.120}_{-0.112}$ & $1.793^{+0.181}_{-0.171}$ \\
 & 3.279 & $[-25.18, -29.00]$ & -25.89 & $6.03 \pm 0.31$ & 0.61 & 13402 & $2.069^{+0.384}_{-0.346}$ & $3.273^{+0.537}_{-0.492}$ \\
\multicolumn{9}{l}{\underline{Redshift-Luminosity subsamples (equal numbers)}:} \\
$2.00-2.30$ & 2.133 & $[-19.94, -22.68]$ & -22.17 & $3.06 \pm 0.04$ & 1.68 & 96910 & $1.206^{+0.069}_{-0.067}$ & $2.453^{+0.123}_{-0.120}$ \\
 & 2.146 & $[-22.68, -23.68]$ & -23.16 & $3.12 \pm 0.04$ & 0.60 & 96099 & $1.265^{+0.065}_{-0.063}$ & $2.551^{+0.115}_{-0.112}$ \\
 & 2.151 & $[-23.68, -28.19]$ & -24.53 & $3.31 \pm 0.05$ & 1.64 & 91976 & $1.591^{+0.089}_{-0.086}$ & $3.117^{+0.152}_{-0.148}$ \\
$2.30-2.70$ & 2.462 & $[-19.95, -23.05]$ & -22.55 & $3.45 \pm 0.06$ & 0.89 & 81764 & $1.032^{+0.077}_{-0.073}$ & $2.007^{+0.131}_{-0.126}$ \\
 & 2.482 & $[-23.05, -24.05]$ & -23.53 & $3.66 \pm 0.07$ & 1.11 & 83092 & $1.251^{+0.088}_{-0.084}$ & $2.370^{+0.146}_{-0.141}$ \\
 & 2.493 & $[-24.05, -28.65]$ & -24.90 & $3.73 \pm 0.06$ & 1.55 & 80690 & $1.321^{+0.079}_{-0.076}$ & $2.481^{+0.130}_{-0.126}$ \\
$2.70-3.10$ & 2.863 & $[-19.97, -23.53]$ & -23.01 & $4.12 \pm 0.11$ & 1.31 & 42997 & $1.010^{+0.104}_{-0.098}$ & $1.840^{+0.167}_{-0.158}$ \\
 & 2.881 & $[-23.53, -24.48]$ & -23.99 & $4.29 \pm 0.12$ & 0.68 & 42238 & $1.142^{+0.127}_{-0.118}$ & $2.044^{+0.199}_{-0.189}$ \\
 & 2.888 & $[-24.48, -28.92]$ & -25.30 & $4.50 \pm 0.12$ & 1.40 & 43136 & $1.349^{+0.133}_{-0.125}$ & $2.366^{+0.205}_{-0.195}$ \\
$3.10-3.50$ & 3.248 & $[-20.06, -23.95]$ & -23.43 & $5.12 \pm 0.23$ & 0.82 & 18175 & $1.221^{+0.211}_{-0.191}$ & $2.056^{+0.313}_{-0.288}$ \\
 & 3.267 & $[-23.95, -24.88]$ & -24.41 & $4.69 \pm 0.29$ & 1.36 & 18190 & $0.853^{+0.211}_{-0.183}$ & $1.491^{+0.324}_{-0.288}$ \\
 & 3.277 & $[-24.88, -29.00]$ & -25.65 & $5.31 \pm 0.25$ & 0.34 & 18439 & $1.326^{+0.233}_{-0.210}$ & $2.202^{+0.341}_{-0.314}$ \\
\hline
\end{tabular}
\caption{Quasar bias results for each of the different subsamples described in Section \ref{sec:datasplits}. We have listed the redshift range $\Delta z$, mean redshift $\bar{z}$, luminosity range $\Delta M_{1450}$, quasar bias $b_Q$, reduced chi-squared $\chi_{\mathrm{red}}^2$ (for 10 degrees of freedom), the number of quasars $N_{\mathrm{QSO}}$ for each subsample, and the minimum ($M_{h,\min}$) and characteristic ($\bar{M}_h$) host halo masses.}
\label{tab:results}
\end{table}

\newpage
\restoregeometry 

\section{Conclusion}

We measured the redshift-space quasar auto-correlation function using 713,706 quasars from DESI DR2 over the redshift range $2.0 \leq z \leq 3.5$. 
At $\bar{z} = 2.48$ we measured a quasar bias of $b_Q = 3.61 \pm 0.01$. 
This value is consistent with previous results at this redshift, although is about an order of magnitude more precise than pre-DESI measurements due to the substantial increase in sample size. 

Our measurements demonstrate that the strong evolution of the quasar bias with redshift is very well fit by a functional form $b_Q(z) = a \left[ (1 + z)^2 - 6.565 \right] + b$ with $a=0.230 \pm 0.007$ and $b=2.394 \pm 0.035$. 
We inferred the masses of the host dark matter halos from the bias evolution and found that quasars are hosted by halos in a narrow mass range around $M_{\mathrm{h}} \sim 10^{12}\,h^{-1}M_{\odot}$, with not much variation with redshift.
We combined the minimum halo mass and the measured quasar space density to estimate the quasar duty cycle, which is defined to be the fraction of halos that host active quasars in a given redshift and luminosity range. 
The result is that these quasars have a very low duty cycle, with $f_{\mathrm{duty}} \sim 10^{-2}$, and the duty cycle remains nearly constant with redshift.  
This is consistent with models in which quasars are a relatively short-lived phase of their host halos. 

We employed the much larger sample of DESI quasars, which also has a larger dynamic range in luminosity, to re-examine the evidence for luminosity-dependent quasar clustering. 
We divided the sample into luminosity bins at each redshift with two binning schemes and found small but statistically significant evidence for luminosity-dependent clustering. 
Specifically, we fit $b_Q(z,M_{1450}) = a [(1 + z)^2 - 6.565] + b - m (M_{1450} + 24)$ and found $m \sim 0.1 - 0.2$ at $> 4.8 \sigma$ significance for both binning schemes. 
Additionally, we compared the bias vs. luminosity in each redshift bin relative to the mean bias for that redshift with a pull distribution. 
The $\chi^2$ test of the pull variance yielded $p \ll 0.05$, which rules out the hypothesis of a redshift-only model.
We also performed a linear fit to the pull values and found slopes deviating from zero at $>4 \sigma$ for both binning schemes.

Some models of quasar clustering (e.g.\,\cite{martini2001, haiman2001}) explored the implications of a monotonic $L-M_{\mathrm{h}}$ relation. 
While we see mild evidence for luminosity dependence, the observations rule out the significant luminosity-dependent clustering predicted by these simple models. 
It is more physical to assume that there is some scatter in the $L-M_{\mathrm{h}}$ relation, as there is scatter between quasar luminosity and black hole mass, between black hole mass and galaxy mass, and between galaxy mass and halo mass. 
We therefore added log-normal scatter with $\sigma_M = 1$ following the prescription in \cite{white2008}. 
This provides a better, although still not adequate fit to the data. 
The data may be a better match to models motivated by hydrodynamical simulations in which quasars undergo substantial luminosity evolution during their lifetime such that they only spend a small fraction of their luminous phase near their peak luminosity (e.g.\,\cite{dimatteo2005, hopkins2005}). 
In this scenario, which was developed in detail by \cite{lidz2006}, the most luminous quasars are the most massive halos with the most massive black holes radiating near Eddington, yet they spend the majority of their luminous phase at lower luminosities, and the vast majority of bright and faint quasars are in similar mass halos. 
We defer a detailed comparison to such models to future work. 
The precise measurements of luminosity-dependent clustering here offer a clear target for such models. 

In addition to models with more realistic luminosity evolution scenarios, studies of quasar clustering at smaller scales using halo occupation distribution (HOD) modeling could further illuminate how quasars are connected to the underlying halo population and provide a clearer picture of how quasars are triggered in halos and evolve over time. 
The mechanisms behind quasar triggering remain uncertain, and potential measurements of quantities such as central vs.\ satellite duty cycles may lead to new insights from small-scale clustering measurements (e.g.\,\cite{shen2010}). 
HOD modeling will also aid the creation of more realistic mock catalogs for future cosmological studies that use quasars as large-scale structure tracers. 

The third DESI data release, which will include the first five years of data collection, will have approximately a million quasars at $2.0 < z < 3.5$, and the final eight-year dataset should be on order 20\% larger. 
The proposed DESI-II survey\,\cite{DESI-II} will further expand the sample, in particular by increasing the dynamic range in luminosity by targeting fainter quasars and better small-scale clustering information through more passes over the survey area and a higher surface density of targets. 
DESI-II also plans to observe a large sample of Lyman Break Galaxies and Lyman-Alpha Emitters at similar redshifts. 
That data could be used to explore an even larger dynamic range in luminosity through cross-correlation studies with galaxies (e.g.\,\cite{adelberger2005}).

\section{Data Availability}
The data used in this work will be made public as part of DESI Data Release 2 (details at \url{https://data.desi.lbl.gov/doc/releases/}). The data points corresponding to the figures
are available at \url{https://doi.org/10.5281/zenodo.20722125}. 

\appendix 

\section{Robustness of measurements to fit range}
\label{app:fitrange}

In this Appendix we demonstrate the robustness of our quasar bias measurements to the choice of fitting range used for the quasar autocorrelation function. 
In addition to our baseline fit range of $10-80\,\hMpc$, we compute the quasar bias using three alternative ranges: $10-50\,\hMpc$, $50-80\,\hMpc$, and $20-80\,\hMpc$.

We consider the $10-50\,\hMpc$ range because it lies between the scales used in \cite{eftekh15}, who fit over $4-25\,\hMpc$, and \cite{laurent2017}, who fit over $10-85\,\hMpc$. 
Bault et al.\,\cite{bault2025} found that redshift uncertainties contribute an effective velocity dispersion of $\sigma_v \sim 5-8\,\hMpc$ across our sample's redshift range, potentially affecting clustering measurements near our lower fitting bound of $10\,\hMpc$. 
The $50-80\,\hMpc$ and $20-80\,\hMpc$ ranges were therefore chosen to test for possible systematic effects arising from nonlinear clustering and redshift uncertainties that primarily impact smaller scales. 

\begin{figure}[tbp]
    \centering
    \includegraphics[width=1.0\linewidth]{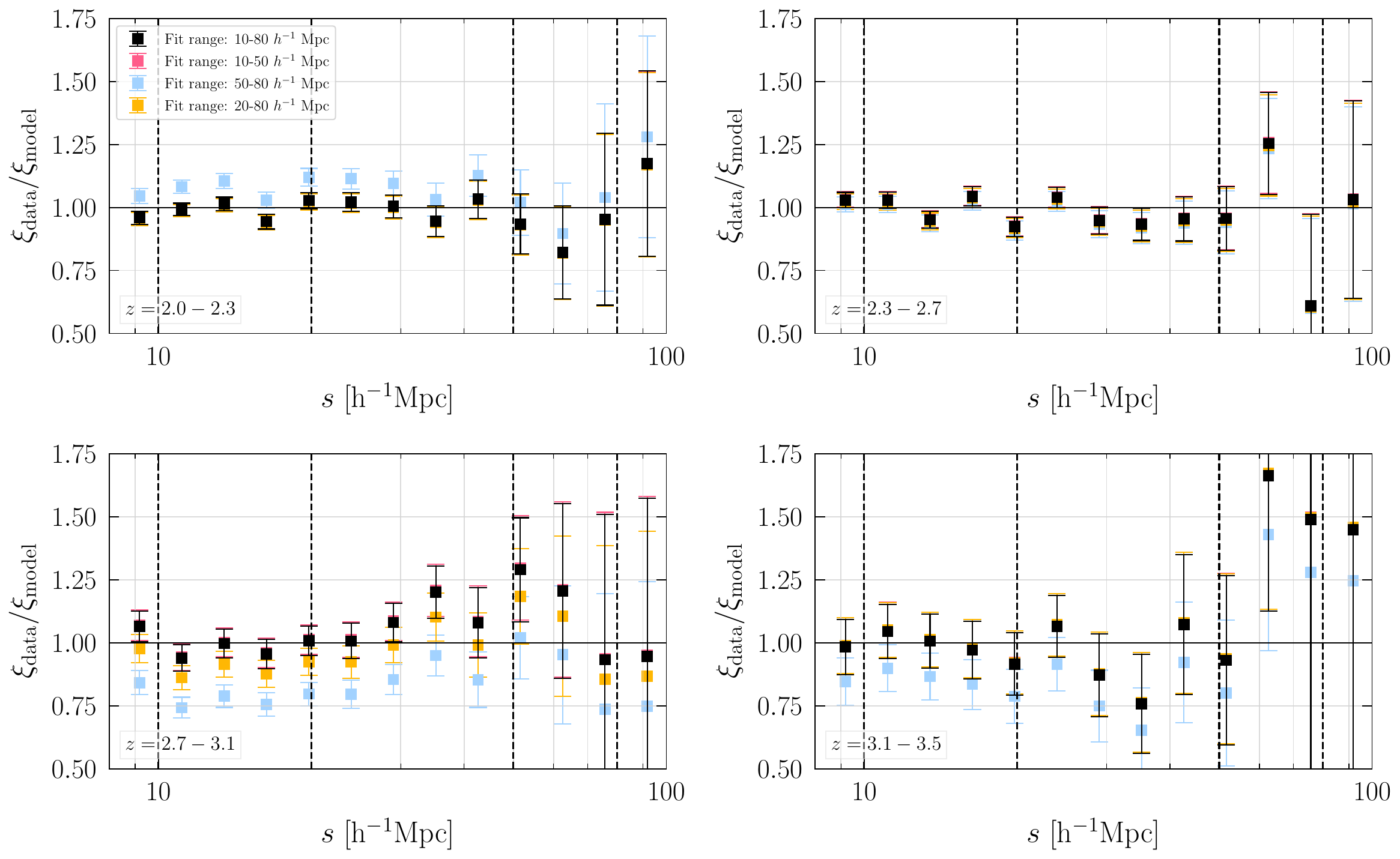}
    \caption{Fractional difference of the quasar auto correlation function compared to the kaiser model fit over three different regions. Each panel corresponds to one of the four redshift bins. The $10-80\ h^{-1}$ Mpc and $10-50\ h^{-1}$ Mpc fits are nearly identical in all redshift bins.}
    \label{fig:acf_residuals_z}
\end{figure}

\begin{figure}[tbp]
    \centering
    \includegraphics[width=0.75\linewidth]{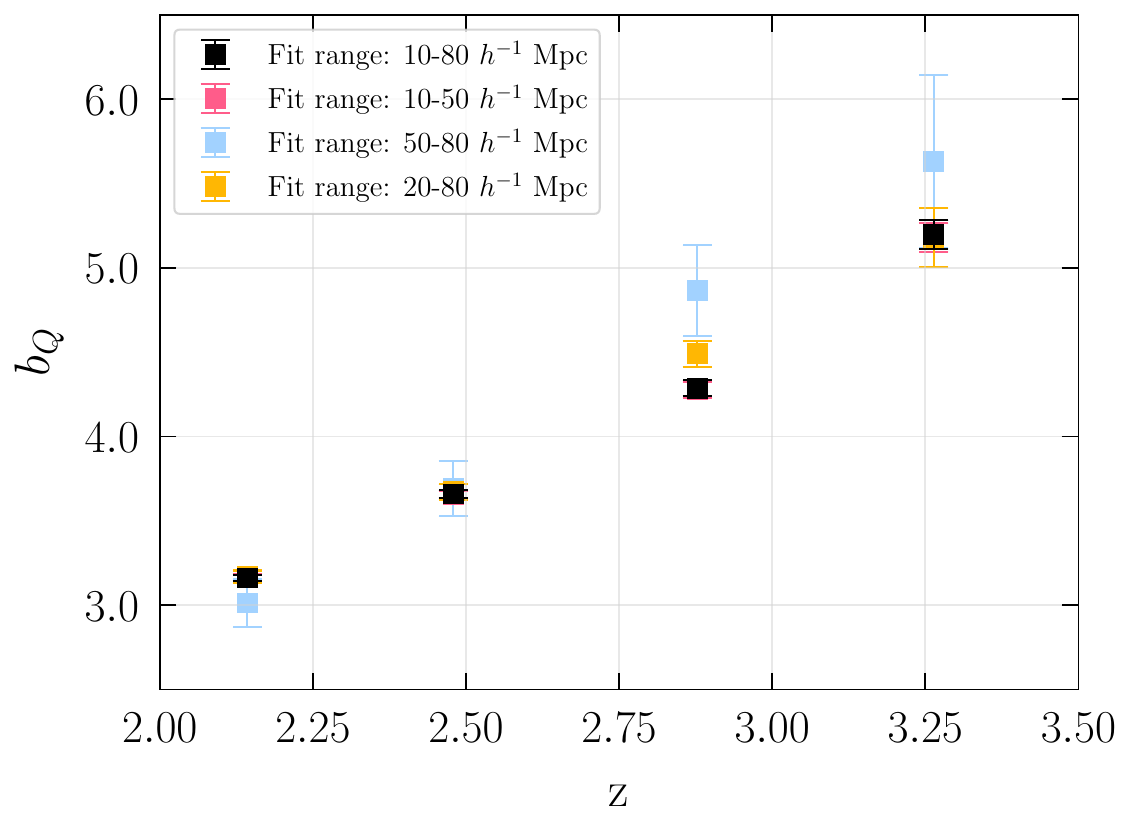}
    \caption{Quasar bias as a function of redshift for the four fit regions. The $10-80\,\hMpc$ fit (black squares) is nearly identical to the $10-50\,\hMpc$ fit (pink squares) and the $20-80\,\hMpc$ fit (yellow squares). The quasar bias determined from the $50-80\ h^{-1}$ Mpc fit (blue squares) deviates slightly more from the results from the other fits, especially at higher redshift.}
    \label{fig:bQ_fit_ranges}
\end{figure}

We compute the quasar bias for the full sample and the individual redshift bins using these alternate fitting ranges. 
Table \ref{tab:bQ_fit_ranges} shows the resulting quasar bias measurements.
Figure \ref{fig:acf_residuals_z} shows the ratio of the measured autocorrelation function to the best-fit Kaiser model for the four fitting ranges in each redshift bin. 
The model is evaluated at the mean redshift of each sample. 
For the full sample, the bias measurements obtained from all four fitting ranges are highly consistent. 
The redshift binned samples are consistent within $1 \sigma$ across all fitting ranges, except for the $z=2.3-2.7$ bin, where the measurements obtained using the $50-80\,\hMpc$ and $20-80\,\hMpc$ ranges are consistent within $2 \sigma$. 
Figure \ref{fig:bQ_fit_ranges} shows the quasar bias measurements as a function of redshift for the alternative fitting ranges.
Overall, we find that restricting the fit to larger scales yields largely consistent bias measurements, demonstrating that our results are robust to the choice of fitting range.

We further tested how these different fitting ranges would impact our ability to measure luminosity dependence. 
For both the $10-80\,\hMpc$ and $10-50\,\hMpc$ fitting ranges, we find consistent evidence for luminosity dependence of quasar bias at $> 4 \sigma$ significance.
For the $20-80\,\hMpc$ fitting range, the `same boundaries' binning scheme yields a significance of $\sim 4 \sigma$, however, the `equal numbers' binning gives a lower-significance detection of luminosity dependence, at only $\sim 1-2 \sigma$. 
The $50-80\,\hMpc$ fitting range yields a much lower significance ($<1.5 \sigma$) for both binning schemes.
The lack of significant evidence for luminosity dependence in this $50-80\,\hMpc$ fitting range is consistent with the reduced statistical constraining power when the analysis is restricted to larger scales, where the clustering signal is weaker and the fit is based on fewer data points. 
This result is consistent with \cite{knox2025}, who found that detecting luminosity dependence requires both high measurement precision and a sample with a sufficiently broad luminosity range. 
Quasar HOD studies and improved characterizations of redshift errors will be particularly useful for extending the analysis to smaller scales, where the luminosity dependence may be more tightly constrained. 

\begin{table}[]
    \centering
    \renewcommand{\arraystretch}{1.25} 
    \begin{tabular}{ccccccc}
        \hline
        $\Delta z$ & \multicolumn{2}{c}{$10-50\,\hMpc$} & \multicolumn{2}{c}{$50-80\,\hMpc$} & \multicolumn{2}{c}{$20-80\,\hMpc$} \\
         & $b_Q$ & $\chi_{\text{red}}^2$ & $b_Q$ & $\chi_{\text{red}}^2$ & $b_Q$ & $\chi_{\text{red}}^2$ \\
        \hline
        \multicolumn{7}{l}{\underline{Full sample}:} \\
        $2.0-3.5$ & $3.60 \pm 0.01$ & $0.86$ & $3.60 \pm 0.10$ & $0.55$ & $3.64 \pm 0.03$ & $0.74$ \\ 
        \multicolumn{7}{l}{\underline{Redshift subsamples}:} \\
        $2.0-2.3$ & $3.16 \pm 0.02$ & $1.12$ & $3.01 \pm 0.14$ & $0.18$ & $3.17 \pm 0.04$ & $0.53$ \\
        $2.3-2.7$ & $3.65 \pm 0.02$ & $1.63$ & $3.69 \pm 0.16$ & $1.81$ & $3.67 \pm 0.05$ & $1.41$ \\
        $2.7-3.1$ & $4.28 \pm 0.05$ & $0.97$ & $4.86 \pm 0.27$ & $0.18$ & $4.49 \pm 0.08$ & $0.62$ \\
        $3.1-3.5$ & $5.18 \pm 0.09$ & $0.50$ & $5.63 \pm 0.51$ & $0.75$ & $5.18 \pm 0.17$ & $0.87$ \\
        \hline
    \end{tabular}
    \caption{Quasar bias values and the corresponding $\chi_{\text{red}}^2$ values for the fit ranges $10-50\,\hMpc$ (7 degrees of freedom), $50-80\,\hMpc$ (2 degrees of freedom), and $20-80\,\hMpc$ (6 degrees of freedom).}
    \label{tab:bQ_fit_ranges}
\end{table}

\section{Systematic error estimation}
\label{app:syserr}

To estimate the systematic errors that come from our methodological assumptions, such as the chosen fitting range and use of the combined NGC + SGC (GCcomb) regions for our measurements, we varied each of these choices and compared the resulting quasar bias measurements. 
We compute $\Delta b_Q$ which is defined as the difference between our fiducial choices ($10-80\,\hMpc$ fitting range and using the combined GCcomb region) and each of our chosen variations.
The variations we include to compare the clustering in different regions are NGC alone and SGC alone, each compared to the bias measured for the combined GCcomb region.
Generally, we find that the clustering is similar in the NGC and SGC, with the exception of the lowest redshift bin where there is a $\sim 1.7 \,\sigma$ difference between the clustering in the SGC and the GCcomb region. 
We also compare the bias measured from our $10-80\,\hMpc$ fit range to the bias from $20-80\,\hMpc$ to include any variations in our fit that come from nonlinear effects around $10\,\hMpc$. 
The bias measurements and the corresponding reduced chi-squared values for the $20-80\,\hMpc$ fitting range is shown in Table \ref{tab:bQ_fit_ranges}, and Table \ref{tab:bQsyserr} shows the results for the NGC and SGC regions.

\begin{table}[]
    \centering
    \renewcommand{\arraystretch}{1.25} 
    \begin{tabular}{ccccc}
        \hline
        $\Delta z$ & \multicolumn{2}{c}{NGC} & \multicolumn{2}{c}{SGC} \\
          & $b_Q$ & $\chi_{\text{red}}^2$ & $b_Q$ & $\chi_{\text{red}}^2$ \\
          \hline
         \multicolumn{5}{l}{\underline{Full sample}:} \\
        $2.0-3.5$ & $3.59 \pm 0.02$ & $0.57$ & $3.62 \pm 0.02$ & $0.30$ \\
        \multicolumn{5}{l}{\underline{Redshift subsamples}:} \\
        $2.0-2.3$ & $3.13 \pm 0.02$ & $0.77$ & $3.22 \pm 0.03$ & $0.65$ \\
        $2.3-2.7$ & $3.67 \pm 0.03$ & $0.94$ & $3.61 \pm 0.04$ & $0.75$ \\
        $2.7-3.1$ & $4.26 \pm 0.05$ & $0.76$ & $4.32 \pm 0.07$ & $1.35$ \\
        $3.1-3.5$ & $5.11 \pm 0.11$ & $1.02$ & $5.26 \pm 0.14$ & $0.89$ \\
        \hline
    \end{tabular}
    \caption{Quasar bias measurements and the corresponding reduced chi squared values for NGC only and SGC only using the fit range $10-80\,\hMpc$. For both the NGC and SGC fits there are 10 degrees of freedom. The results for the $20-80\,\hMpc$ fit range are shown in Table\,\ref{tab:bQ_fit_ranges}.}
    \label{tab:bQsyserr}
\end{table}

To determine the systematic error contribution from the region of sky used, we compute the difference in the quasar bias measured in each of the NGC and SGC regions to that from the combined GCcomb region, 
\begin{align}
    \Delta b_{Q, \text{NGC}} &= b_{Q, \text{NGC}} - b_{Q,\text{GCcomb}} \\
    \Delta b_{Q, \text{SGC}} &= b_{Q, \text{SGC}} - b_{Q, \text{GCcomb}}. 
\end{align}
We then combine these differences to determine $\sigma_{\text{region}}$,
\begin{equation}
    \sigma_{\text{region}} = \sqrt{\frac{\Delta b_{Q, \text{NGC}}^2 + \Delta b_{Q, \text{SGC}}^2}{2}}.
    \label{eq:sigma_region}
\end{equation}

We estimate the systematic uncertainty associated with the choice of fitting range, which includes effects from nonlinearities on small scales, by computing the difference between the measured quasar bias using our fiducial fitting range of $10-80\,\hMpc$ to the variation $20-80\,\hMpc$,
\begin{equation}
    \sigma_{\text{fit}} = \Delta b_{Q, (20-80\,\hMpc)} = b_{Q, (20-80\,\hMpc)} - b_{Q, (10-80\,\hMpc)}.
    \label{eq:sigma_fit}
\end{equation}
The final uncertainty on the quasar bias measurements is then obtained by adding the systematic uncertainties $\sigma_{\text{region}}$ and $\sigma_{\text{fit}}$, in quadrature with the original statistical uncertainties.
When combined with our statistical errors, we find that the systematic error increases the uncertainty on our measurements by a factor of $\sim 1.4 - 5.9 \times$.

\acknowledgments

We thank Chris Hirata for helpful discussions.
MC and PM acknowledge support from the United States Department of Energy, Office of High Energy Physics under Award Number DE-SC0011726.

This material is based upon work supported by the U.S. Department of Energy (DOE), Office of Science, Office of High-Energy Physics, under Contract No. DE–AC02–05CH11231, and by the National Energy Research Scientific Computing Center, a DOE Office of Science User Facility under the same contract. Additional support for DESI was provided by the U.S. National Science Foundation (NSF), Division of Astronomical Sciences under Contract No. AST-0950945 to the NSF’s National Optical-Infrared Astronomy Research Laboratory; the Science and Technology Facilities Council of the United Kingdom; the Gordon and Betty Moore Foundation; the Heising-Simons Foundation; the French Alternative Energies and Atomic Energy Commission (CEA); the National Council of Humanities, Science and Technology of Mexico (CONAHCYT); the Ministry of Science, Innovation and Universities of Spain (MICIU/AEI/10.13039/501100011033), and by the DESI Member Institutions: \url{https://www.desi.lbl.gov/collaborating-institutions}. Any opinions, findings, and conclusions or recommendations expressed in this material are those of the author(s) and do not necessarily reflect the views of the U. S. National Science Foundation, the U. S. Department of Energy, or any of the listed funding agencies.

The authors are honored to be permitted to conduct scientific research on I'oligam Du'ag (Kitt Peak), a mountain with particular significance to the Tohono O’odham Nation.


\bibliographystyle{JHEP}
\bibliography{biblio.bib}

@ARTICLE{rees1984,
       author = {{Rees}, Martin J.},
        title = "{Black Hole Models for Active Galactic Nuclei}",
      journal = {\araa},
         year = 1984,
        month = jan,
       volume = {22},
        pages = {471-506},
          doi = {10.1146/annurev.aa.22.090184.002351},
       adsurl = {https://ui.adsabs.harvard.edu/abs/1984ARA&A..22..471R}
}

@ARTICLE{shanks1987,
       author = {{Shanks}, T. and {Fong}, R. and {Boyle}, B.~J. and {Peterson}, B.~A.},
        title = "{The spatial clustering of QSOs}",
      journal = {\mnras},
         year = 1987,
        month = aug,
       volume = {227},
        pages = {739-748},
          doi = {10.1093/mnras/227.3.739},
       adsurl = {https://ui.adsabs.harvard.edu/abs/1987MNRAS.227..739S}
}

@ARTICLE{richstone1998,
       author = {{Richstone}, D. and {Ajhar}, E.~A. and {Bender}, R. and {Bower}, G. and {Dressler}, A. and {Faber}, S.~M. and {Filippenko}, A.~V. and {Gebhardt}, K. and {Green}, R. and {Ho}, L.~C. and {Kormendy}, J. and {Lauer}, T.~R. and {Magorrian}, J. and {Tremaine}, S.},
        title = "{Supermassive black holes and the evolution of galaxies.}",
      journal = {\nat},
         year = 1998,
        month = oct,
       volume = {385},
       number = {6701},
        pages = {A14},
          doi = {10.48550/arXiv.astro-ph/9810378},
archivePrefix = {arXiv},
       eprint = {astro-ph/9810378},
 primaryClass = {astro-ph},
       adsurl = {https://ui.adsabs.harvard.edu/abs/1998Natur.395A..14R}
}

@ARTICLE{stephens1997,
       author = {{Stephens}, Andrew W. and {Schneider}, Donald P. and {Schmidt}, Maarten and {Gunn}, James E. and {Weinberg}, David H.},
        title = "{A Study of Quasar Clustering at $z>2.7$ From the Palomar Transit GRISM Survey}",
      journal = {\aj},
         year = 1997,
        month = jul,
       volume = {114},
        pages = {41-47},
          doi = {10.1086/118450},
archivePrefix = {arXiv},
       eprint = {astro-ph/9708241},
 primaryClass = {astro-ph},
       adsurl = {https://ui.adsabs.harvard.edu/abs/1997AJ....114...41S}
}

@ARTICLE{jing1998,
       author = {{Jing}, Y.~P. and {Mo}, H.~J. and {B{\"o}rner}, G.},
        title = "{Spatial Correlation Function and Pairwise Velocity Dispersion of Galaxies: Cold Dark Matter Models versus the Las Campanas Survey}",
      journal = {\apj},
         year = 1998,
        month = feb,
       volume = {494},
       number = {1},
        pages = {1-12},
          doi = {10.1086/305209},
archivePrefix = {arXiv},
       eprint = {astro-ph/9707106},
 primaryClass = {astro-ph},
       adsurl = {https://ui.adsabs.harvard.edu/abs/1998ApJ...494....1J}
}

@ARTICLE{peacock2000,
       author = {{Peacock}, J.~A. and {Smith}, R.~E.},
        title = "{Halo occupation numbers and galaxy bias}",
      journal = {\mnras},
         year = 2000,
        month = nov,
       volume = {318},
       number = {4},
        pages = {1144-1156},
          doi = {10.1046/j.1365-8711.2000.03779.x},
archivePrefix = {arXiv},
       eprint = {astro-ph/0005010},
 primaryClass = {astro-ph},
       adsurl = {https://ui.adsabs.harvard.edu/abs/2000MNRAS.318.1144P}
}

@ARTICLE{berlind2002,
       author = {{Berlind}, Andreas A. and {Weinberg}, David H.},
        title = "{The Halo Occupation Distribution: Toward an Empirical Determination of the Relation between Galaxies and Mass}",
      journal = {\apj},
         year = 2002,
        month = aug,
       volume = {575},
       number = {2},
        pages = {587-616},
          doi = {10.1086/341469},
archivePrefix = {arXiv},
       eprint = {astro-ph/0109001},
 primaryClass = {astro-ph},
       adsurl = {https://ui.adsabs.harvard.edu/abs/2002ApJ...575..587B}
}

@ARTICLE{yu2002,
       author = {{Yu}, Qingjuan and {Tremaine}, Scott},
        title = "{Observational constraints on growth of massive black holes}",
      journal = {\mnras},
         year = 2002,
        month = oct,
       volume = {335},
       number = {4},
        pages = {965-976},
          doi = {10.1046/j.1365-8711.2002.05532.x},
archivePrefix = {arXiv},
       eprint = {astro-ph/0203082},
 primaryClass = {astro-ph},
       adsurl = {https://ui.adsabs.harvard.edu/abs/2002MNRAS.335..965Y}
}

@ARTICLE{zheng2005,
       author = {{Zheng}, Zheng and {Berlind}, Andreas A. and {Weinberg}, David H. and {Benson}, Andrew J. and {Baugh}, Carlton M. and {Cole}, Shaun and {Dav{\'e}}, Romeel and {Frenk}, Carlos S. and {Katz}, Neal and {Lacey}, Cedric G.},
        title = "{Theoretical Models of the Halo Occupation Distribution: Separating Central and Satellite Galaxies}",
      journal = {\apj},
         year = 2005,
        month = nov,
       volume = {633},
       number = {2},
        pages = {791-809},
          doi = {10.1086/466510},
archivePrefix = {arXiv},
       eprint = {astro-ph/0408564},
 primaryClass = {astro-ph},
       adsurl = {https://ui.adsabs.harvard.edu/abs/2005ApJ...633..791Z}
}

@ARTICLE{croom2005,
       author = {{Croom}, Scott M. and {Boyle}, B.~J. and {Shanks}, T. and {Smith}, R.~J. and {Miller}, L. and {Outram}, P.~J. and {Loaring}, N.~S. and {Hoyle}, F. and {da {\^A}ngela}, J.},
        title = "{The 2dF QSO Redshift Survey - XIV. Structure and evolution from the two-point correlation function}",
      journal = {\mnras},
         year = 2005,
        month = jan,
       volume = {356},
       number = {2},
        pages = {415-438},
          doi = {10.1111/j.1365-2966.2004.08379.x},
archivePrefix = {arXiv},
       eprint = {astro-ph/0409314},
 primaryClass = {astro-ph},
       adsurl = {https://ui.adsabs.harvard.edu/abs/2005MNRAS.356..415C}
}

@ARTICLE{shen2007,
       author = {{Shen}, Yue and {Strauss}, Michael A. and {Oguri}, Masamune and {Hennawi}, Joseph F. and {Fan}, Xiaohui and {Richards}, Gordon T. and {Hall}, Patrick B. and {Gunn}, James E. and {Schneider}, Donald P. and {Szalay}, Alexander S. and {Thakar}, Anirudda R. and {Vanden Berk}, Daniel E. and {Anderson}, Scott F. and {Bahcall}, Neta A. and {Connolly}, Andrew J. and {Knapp}, Gillian R.},
        title = "{Clustering of High-Redshift (z $\geq$ 2.9) Quasars from the Sloan Digital Sky Survey}",
      journal = {\aj},
         year = 2007,
        month = may,
       volume = {133},
       number = {5},
        pages = {2222-2241},
          doi = {10.1086/513517},
archivePrefix = {arXiv},
       eprint = {astro-ph/0702214},
 primaryClass = {astro-ph},
       adsurl = {https://ui.adsabs.harvard.edu/abs/2007AJ....133.2222S}
}

@ARTICLE{eftekh15,
       author = {{Eftekharzadeh}, Sarah and {Myers}, Adam D. and {White}, Martin and {Weinberg}, David H. and {Schneider}, Donald P. and {Shen}, Yue and {Font-Ribera}, Andreu and {Ross}, Nicholas P. and {Paris}, Isabelle and {Streblyanska}, Alina},
        title = "{Clustering of intermediate redshift quasars using the final SDSS III-BOSS sample}",
      journal = {\mnras},
         year = 2015,
        month = nov,
       volume = {453},
       number = {3},
        pages = {2779-2798},
          doi = {10.1093/mnras/stv1763},
archivePrefix = {arXiv},
       eprint = {1507.08380},
 primaryClass = {astro-ph.CO},
       adsurl = {https://ui.adsabs.harvard.edu/abs/2015MNRAS.453.2779E}
}

@ARTICLE{eftekh19,
       author = {{Eftekharzadeh}, S. and {Myers}, A.~D. and {Kourkchi}, E.},
        title = "{A Halo Occupation Interpretation of Quasars at z {\ensuremath{\sim}} 1.5 Using Very Small-Scale Clustering Information}",
      journal = {\mnras},
         year = 2019,
        month = jun,
       volume = {486},
       number = {1},
        pages = {274-282},
          doi = {10.1093/mnras/stz770},
archivePrefix = {arXiv},
       eprint = {1812.05760},
 primaryClass = {astro-ph.CO},
       adsurl = {https://ui.adsabs.harvard.edu/abs/2019MNRAS.486..274E}
}

@article{laurent2017,
   title={Clustering of quasars in SDSS-IV eBOSS: study of potential systematics and bias determination},
   volume={2017},
   ISSN={1475-7516},
   url={http://dx.doi.org/10.1088/1475-7516/2017/07/017},
   DOI={10.1088/1475-7516/2017/07/017},
   number={07},
   journal={Journal of Cosmology and Astroparticle Physics},
   publisher={IOP Publishing},
   author={Laurent, Pierre and Eftekharzadeh, Sarah and Goff, Jean-Marc Le and Myers, Adam and Burtin, Etienne and White, Martin and Ross, Ashley J. and Tinker, Jeremy and Tojeiro, Rita and Bautista, Julian and Brinkmann, Jonathan and Comparat, Johan and Dawson, Kyle and Bourboux, Hélion du Mas des and Kneib, Jean-Paul and McGreer, Ian D. and Palanque-Delabrouille, Nathalie and Percival, Will J. and Prada, Francisco and Rossi, Graziano and Schneider, Donald P. and Weinberg, David and Yèche, Christophe and Zarrouk, Pauline and Zhao, Gong-Bo},
   year={2017},
   month=jul, pages={017–017} }

@article{myers2007,
   title={Clustering Analyses of 300,000 Photometrically Classified Quasars. I. Luminosity and Redshift Evolution in Quasar Bias},
   volume={658},
   ISSN={1538-4357},
   url={http://dx.doi.org/10.1086/511519},
   DOI={10.1086/511519},
   number={1},
   journal={The Astrophysical Journal},
   publisher={American Astronomical Society},
   author={Myers, Adam D. and Brunner, Robert J. and Nichol, Robert C. and Richards, Gordon T. and Schneider, Donald P. and Bahcall, Neta A.},
   year={2007},
   month=mar, pages={85–98} }

@article{porciani2004,
   title={Cosmic evolution of quasar clustering: implications for the host haloes},
   volume={355},
   ISSN={1365-2966},
   url={http://dx.doi.org/10.1111/j.1365-2966.2004.08408.x},
   DOI={10.1111/j.1365-2966.2004.08408.x},
   number={3},
   journal={Monthly Notices of the Royal Astronomical Society},
   publisher={Oxford University Press (OUP)},
   author={Porciani, C. and Magliocchetti, M. and Norberg, P.},
   year={2004},
   month=dec, pages={1010–1030} }

@ARTICLE{dimatteo2005,
       author = {{Di Matteo}, Tiziana and {Springel}, Volker and {Hernquist}, Lars},
        title = "{Energy input from quasars regulates the growth and activity of black holes and their host galaxies}",
      journal = {\nat},
         year = 2005,
        month = feb,
       volume = {433},
       number = {7026},
        pages = {604-607},
          doi = {10.1038/nature03335},
archivePrefix = {arXiv},
       eprint = {astro-ph/0502199},
 primaryClass = {astro-ph},
       adsurl = {https://ui.adsabs.harvard.edu/abs/2005Natur.433..604D}
}

@article{porciani2006,
   title={Luminosity- and redshift-dependent quasar clustering},
   volume={371},
   ISSN={1365-2966},
   url={http://dx.doi.org/10.1111/j.1365-2966.2006.10813.x},
   DOI={10.1111/j.1365-2966.2006.10813.x},
   number={4},
   journal={Monthly Notices of the Royal Astronomical Society},
   publisher={Oxford University Press (OUP)},
   author={Porciani, C. and Norberg, P.},
   year={2006},
   month=oct, pages={1824–1834} }

@ARTICLE{hennawi2006,
       author = {{Hennawi}, Joseph F. and {Strauss}, Michael A. and {Oguri}, Masamune and {Inada}, Naohisa and {Richards}, Gordon T. and {Pindor}, Bartosz and {Schneider}, Donald P. and {Becker}, Robert H. and {Gregg}, Michael D. and {Hall}, Patrick B. and {Johnston}, David E. and {Fan}, Xiaohui and {Burles}, Scott and {Schlegel}, David J. and {Gunn}, James E. and {Lupton}, Robert H. and {Bahcall}, Neta A. and {Brunner}, Robert J. and {Brinkmann}, Jon},
        title = "{Binary Quasars in the Sloan Digital Sky Survey: Evidence for Excess Clustering on Small Scales}",
      journal = {\aj},
         year = 2006,
        month = jan,
       volume = {131},
       number = {1},
        pages = {1-23},
          doi = {10.1086/498235},
archivePrefix = {arXiv},
       eprint = {astro-ph/0504535},
 primaryClass = {astro-ph},
       adsurl = {https://ui.adsabs.harvard.edu/abs/2006AJ....131....1H}
}

@article{martini2001,
   title={Quasar Clustering and the Lifetime of Quasars},
   volume={547},
   ISSN={1538-4357},
   url={http://dx.doi.org/10.1086/318331},
   DOI={10.1086/318331},
   number={1},
   journal={The Astrophysical Journal},
   publisher={American Astronomical Society},
   author={Martini, Paul and Weinberg, David H.},
   year={2001},
   month=jan, pages={12–26} }

@ARTICLE{white2008,
       author = {{White}, Martin and {Martini}, Paul and {Cohn}, J.~D.},
        title = "{Constraints on the correlation between QSO luminosity and host halo mass from high-redshift quasar clustering}",
      journal = {\mnras},
         year = 2008,
        month = nov,
       volume = {390},
       number = {3},
        pages = {1179-1184},
          doi = {10.1111/j.1365-2966.2008.13817.x},
archivePrefix = {arXiv},
       eprint = {0711.4109},
 primaryClass = {astro-ph},
       adsurl = {https://ui.adsabs.harvard.edu/abs/2008MNRAS.390.1179W}
}

@ARTICLE{shankar2010,
       author = {{Shankar}, Francesco and {Weinberg}, David H. and {Shen}, Yue},
        title = "{Constraints on black hole duty cycles and the black hole-halo relation from SDSS quasar clustering}",
      journal = {\mnras},
         year = 2010,
        month = aug,
       volume = {406},
       number = {3},
        pages = {1959-1966},
          doi = {10.1111/j.1365-2966.2010.16801.x},
archivePrefix = {arXiv},
       eprint = {1004.1173},
 primaryClass = {astro-ph.CO},
       adsurl = {https://ui.adsabs.harvard.edu/abs/2010MNRAS.406.1959S}
}

@article{pizzati2024,
   title={Revisiting the extreme clustering of z $\approx$ 4 quasars with large volume cosmological simulations},
   volume={528},
   ISSN={1365-2966},
   url={http://dx.doi.org/10.1093/mnras/stae329},
   DOI={10.1093/mnras/stae329},
   number={3},
   journal={Monthly Notices of the Royal Astronomical Society},
   publisher={Oxford University Press (OUP)},
   author={Pizzati, Elia and Hennawi, Joseph F and Schaye, Joop and Schaller, Matthieu},
   year={2024},
   month=feb, pages={4466–4489} }

@ARTICLE{lidz2006,
       author = {{Lidz}, Adam and {Hopkins}, Philip F. and {Cox}, Thomas J. and {Hernquist}, Lars and {Robertson}, Brant},
        title = "{The Luminosity Dependence of Quasar Clustering}",
      journal = {\apj},
         year = 2006,
        month = apr,
       volume = {641},
       number = {1},
        pages = {41-49},
          doi = {10.1086/500444},
archivePrefix = {arXiv},
       eprint = {astro-ph/0507361},
 primaryClass = {astro-ph},
       adsurl = {https://ui.adsabs.harvard.edu/abs/2006ApJ...641...41L}
}

@misc{knox2025,
      title={Trinity VII. Predictions for the Observable Correlation Functions of Accreting Black Holes}, 
      author={Oddisey Knox and Haowen Zhang and Peter Behroozi},
      year={2025},
      eprint={2506.05612},
      archivePrefix={arXiv},
      primaryClass={astro-ph.GA},
      url={https://arxiv.org/abs/2506.05612}, 
}

@article{desitargetval,
   title={Target Selection and Validation of DESI Quasars},
   volume={944},
   ISSN={1538-4357},
   url={http://dx.doi.org/10.3847/1538-4357/acb3c2},
   DOI={10.3847/1538-4357/acb3c2},
   number={1},
   journal={The Astrophysical Journal},
   publisher={American Astronomical Society},
   author={Chaussidon, Edmond and Yèche, Christophe and Palanque-Delabrouille, Nathalie and Alexander, David M. and Yang, Jinyi and Ahlen, Steven and Bailey, Stephen and Brooks, David and Cai, Zheng and Chabanier, Solène and Davis, Tamara M. and Dawson, Kyle and de laMacorra, Axel and Dey, Arjun and Dey, Biprateep and Eftekharzadeh, Sarah and Eisenstein, Daniel J. and Fanning, Kevin and Font-Ribera, Andreu and Gaztañaga, Enrique and A Gontcho, Satya Gontcho and Gonzalez-Morales, Alma X. and Guy, Julien and Herrera-Alcantar, Hiram K. and Honscheid, Klaus and Ishak, Mustapha and Jiang, Linhua and Juneau, Stephanie and Kehoe, Robert and Kisner, Theodore and Kovács, Andras and Kremin, Anthony and Lan, Ting-Wen and Landriau, Martin and Le Guillou, Laurent and Levi, Michael E. and Magneville, Christophe and Martini, Paul and Meisner, Aaron M. and Moustakas, John and Muñoz-Gutiérrez, Andrea and Myers, Adam D. and Newman, Jeffrey A. and Nie, Jundan and Percival, Will J. and Poppett, Claire and Prada, Francisco and Raichoor, Anand and Ravoux, Corentin and Ross, Ashley J. and Schlafly, Edward and Schlegel, David and Tan, Ting and Tarlé, Gregory and Zhou, Rongpu and Zhou, Zhimin and Zou, Hu},
   year={2023},
   month=feb, pages={107} }

@ARTICLE{ross2025,
       author = {{Ross}, A.~J. and {Aguilar}, J. and {Ahlen}, S. and {Alam}, S. and {Anand}, A. and {Bailey}, S. and {Bianchi}, D. and {Brieden}, S. and {Brooks}, D. and {Burtin}, E. and {Carnero Rosell}, A. and {Chaussidon}, E. and {Claybaugh}, T. and {Cole}, S. and {Dawson}, K. and {de la Macorra}, A. and {de Mattia}, A. and {Dey}, A. and {Dey}, B. and {Doel}, P. and {Fanning}, K. and {Ferraro}, S. and {Ereza}, J. and {Font-Ribera}, A. and {Forero-Romero}, J.~E. and {Gazta{\~n}aga}, E. and {Gil-Mar{\'\i}n}, H. and {Gontcho A Gontcho}, S. and {Gonzalez-Morales}, A.~X. and {Guy}, J. and {Hahn}, C. and {Heydenreich}, S. and {Honscheid}, K. and {Howlett}, C. and {Ishak}, M. and {Karim}, T. and {Kirkby}, D. and {Kisner}, T. and {Kong}, H. and {Kremin}, A. and {Krolewski}, A. and {Lambert}, A. and {Landriau}, M. and {Lasker}, J. and {Guillou}, L.~L. and {Levi}, M.~E. and {Manera}, M. and {Martini}, P. and {McDonald}, P. and {Meisner}, A. and {Miquel}, R. and {Moon}, J. and {Moustakas}, J. and {Mu{\~n}oz-Guti{\'e}rrez}, A. and {Myers}, A.~D. and {Nadathur}, S. and {Napolitano}, L. and {Newman}, J.~A. and {Nie}, J. and {Niz}, G. and {Palanque-Delabrouille}, N. and {Percival}, W.~J. and {Poppett}, C. and {Prada}, F. and {Raichoor}, A. and {Ravoux}, C. and {Rezaie}, M. and {Rosado-Marin}, A. and {Rossi}, G. and {Samushia}, L. and {Sanchez}, E. and {Schlafly}, E.~F. and {Schlegel}, D. and {Seo}, H. and {Smith}, A. and {Sprayberry}, D. and {Tarl{\'e}}, G. and {Valcin}, D. and {Vargas-Maga{\~n}a}, M. and {Weaver}, B.~A. and {Wilson}, M.~J. and {Yu}, J. and {Zarrouk}, P. and {Zhao}, C. and {Zhou}, R. and {Zou}, H.},
        title = "{The construction of large-scale structure catalogs for the Dark Energy Spectroscopic Instrument}",
      journal = {\jcap},
         year = 2025,
        month = jan,
       volume = {2025},
       number = {1},
          eid = {125},
        pages = {125},
          doi = {10.1088/1475-7516/2025/01/125},
archivePrefix = {arXiv},
       eprint = {2405.16593},
 primaryClass = {astro-ph.CO},
       adsurl = {https://ui.adsabs.harvard.edu/abs/2025JCAP...01..125R}
}

@ARTICLE{bianchi2025,
       author = {{Bianchi}, D. and {Hanif}, M.~M.~S. and {Carnero Rosell}, A. and {Lasker}, J. and {Ross}, A.~J. and {Pinon}, M. and {de Mattia}, A. and {White}, M. and {Ahlen}, S. and {Bailey}, S. and {Brooks}, D. and {Burtin}, E. and {Chaussidon}, E. and {Claybaugh}, T. and {Cole}, S. and {de la Macorra}, A. and {Ferraro}, S. and {Font-Ribera}, A. and {Forero-Romero}, J.~E. and {Gazta{\~n}aga}, E. and {Gontcho}, S. Gontcho A. and {Gutierrez}, G. and {Guy}, J. and {Hahn}, C. and {Honscheid}, K. and {Howlett}, C. and {Juneau}, S. and {Kirkby}, D. and {Kisner}, T. and {Kremin}, A. and {Landriau}, M. and {Le Guillou}, L. and {Levi}, M.~E. and {McDonald}, P. and {Meisner}, A. and {Miquel}, R. and {Moustakas}, J. and {Palanque-Delabrouille}, N. and {Percival}, W.~J. and {Prada}, F. and {P{\'e}rez-R{\`a}fols}, I. and {Raichoor}, A. and {Rossi}, G. and {Sanchez}, E. and {Schlegel}, D. and {Schubnell}, M. and {Sharples}, R. and {Silber}, J. and {Sprayberry}, D. and {Tarl{\'e}}, G. and {Vargas-Maga{\~n}a}, M. and {Weaver}, B.~A. and {Zarrouk}, P. and {Zhou}, R. and {Zou}, H.},
        title = "{Characterization of DESI fiber assignment incompleteness effect on 2-point clustering and mitigation methods for DR1 analysis}",
      journal = {\jcap},
         year = 2025,
        month = apr,
       volume = {2025},
       number = {4},
          eid = {074},
        pages = {074},
          doi = {10.1088/1475-7516/2025/04/074},
archivePrefix = {arXiv},
       eprint = {2411.12025},
 primaryClass = {astro-ph.CO},
       adsurl = {https://ui.adsabs.harvard.edu/abs/2025JCAP...04..074B}
}

@article{tinker2010,
   title={THE LARGE-SCALE BIAS OF DARK MATTER HALOS: NUMERICAL CALIBRATION AND MODEL TESTS},
   volume={724},
   ISSN={1538-4357},
   url={http://dx.doi.org/10.1088/0004-637X/724/2/878},
   DOI={10.1088/0004-637x/724/2/878},
   number={2},
   journal={The Astrophysical Journal},
   publisher={American Astronomical Society},
   author={Tinker, Jeremy L. and Robertson, Brant E. and Kravtsov, Andrey V. and Klypin, Anatoly and Warren, Michael S. and Yepes, Gustavo and Gottlöber, Stefan},
   year={2010},
   month=nov, pages={878–886} }

@article{tinker2008,
   title={Toward a Halo Mass Function for Precision Cosmology: The Limits of Universality},
   volume={688},
   ISSN={1538-4357},
   url={http://dx.doi.org/10.1086/591439},
   DOI={10.1086/591439},
   number={2},
   journal={The Astrophysical Journal},
   publisher={American Astronomical Society},
   author={Tinker, Jeremy and Kravtsov, Andrey V. and Klypin, Anatoly and Abazajian, Kevork and Warren, Michael and Yepes, Gustavo and Gottlöber, Stefan and Holz, Daniel E.},
   year={2008},
   month=dec, pages={709–728} }

@article{richards2006,
   title={The Sloan Digital Sky Survey Quasar Survey: Quasar Luminosity Function from Data Release 3},
   volume={131},
   ISSN={1538-3881},
   url={http://dx.doi.org/10.1086/503559},
   DOI={10.1086/503559},
   number={6},
   journal={The Astronomical Journal},
   publisher={American Astronomical Society},
   author={Richards, Gordon T. and Strauss, Michael A. and Fan, Xiaohui and Hall, Patrick B. and Jester, Sebastian and Schneider, Donald P. and Vanden Berk, Daniel E. and Stoughton, Chris and Anderson, Scott F. and Brunner, Robert J. and Gray, Jim and Gunn, James E. and Ivezić, Željko and Kirkland, Margaret K. and Knapp, G. R. and Loveday, Jon and Meiksin, Avery and Pope, Adrian and Szalay, Alexander S. and Thakar, Anirudda R. and Yanny, Brian and York, Donald G. and Barentine, J. C. and Brewington, Howard J. and Brinkmann, J. and Fukugita, Masataka and Harvanek, Michael and Kent, Stephen M. and Kleinman, S. J. and Krzesiński, Jurek and Long, Daniel C. and Lupton, Robert H. and Nash, Thomas and Neilsen, Jr., Eric H. and Nitta, Atsuko and Schlegel, David J. and Snedden, Stephanie A.},
   year={2006},
   month=jun, pages={2766–2787} }

@ARTICLE{richards2006b,
       author = {{Richards}, Gordon T. and {Lacy}, Mark and {Storrie-Lombardi}, Lisa J. and {Hall}, Patrick B. and {Gallagher}, S.~C. and {Hines}, Dean C. and {Fan}, Xiaohui and {Papovich}, Casey and {Vanden Berk}, Daniel E. and {Trammell}, George B. and {Schneider}, Donald P. and {Vestergaard}, Marianne and {York}, Donald G. and {Jester}, Sebastian and {Anderson}, Scott F. and {Budav{\'a}ri}, Tam{\'a}s and {Szalay}, Alexander S.},
        title = "{Spectral Energy Distributions and Multiwavelength Selection of Type 1 Quasars}",
      journal = {\apjs},
         year = 2006,
        month = oct,
       volume = {166},
       number = {2},
        pages = {470-497},
          doi = {10.1086/506525},
archivePrefix = {arXiv},
       eprint = {astro-ph/0601558},
 primaryClass = {astro-ph},
       adsurl = {https://ui.adsabs.harvard.edu/abs/2006ApJS..166..470R}
}

@ARTICLE{adelberger2005,
       author = {{Adelberger}, Kurt L. and {Steidel}, Charles C.},
        title = "{A Possible Correlation between the Luminosities and Lifetimes of Active Galactic Nuclei}",
      journal = {\apj},
         year = 2005,
        month = sep,
       volume = {630},
       number = {1},
        pages = {50-58},
          doi = {10.1086/431789},
archivePrefix = {arXiv},
       eprint = {astro-ph/0505210},
 primaryClass = {astro-ph},
       adsurl = {https://ui.adsabs.harvard.edu/abs/2005ApJ...630...50A}
}

@ARTICLE{hopkins2005,
       author = {{Hopkins}, Philip F. and {Hernquist}, Lars and {Martini}, Paul and {Cox}, Thomas J. and {Robertson}, Brant and {Di Matteo}, Tiziana and {Springel}, Volker},
        title = "{A Physical Model for the Origin of Quasar Lifetimes}",
      journal = {\apjl},
         year = 2005,
        month = jun,
       volume = {625},
       number = {2},
        pages = {L71-L74},
          doi = {10.1086/431146},
archivePrefix = {arXiv},
       eprint = {astro-ph/0502241},
 primaryClass = {astro-ph},
       adsurl = {https://ui.adsabs.harvard.edu/abs/2005ApJ...625L..71H}
}

@ARTICLE{hopkins2007,
       author = {{Hopkins}, Philip F. and {Lidz}, Adam and {Hernquist}, Lars and {Coil}, Alison L. and {Myers}, Adam D. and {Cox}, Thomas J. and {Spergel}, David N.},
        title = "{The Co-Formation of Spheroids and Quasars Traced in their Clustering}",
      journal = {\apj},
         year = 2007,
        month = jun,
       volume = {662},
       number = {1},
        pages = {110-130},
          doi = {10.1086/517512},
archivePrefix = {arXiv},
       eprint = {astro-ph/0611792},
 primaryClass = {astro-ph},
       adsurl = {https://ui.adsabs.harvard.edu/abs/2007ApJ...662..110H}
}

@article{desiinstrument,
   title={Overview of the Instrumentation for the Dark Energy Spectroscopic Instrument},
   volume={164},
   ISSN={1538-3881},
   url={http://dx.doi.org/10.3847/1538-3881/ac882b},
   DOI={10.3847/1538-3881/ac882b},
   number={5},
   journal={The Astronomical Journal},
   publisher={American Astronomical Society},
   author={DESI Collaboration and Abareshi, B. and Aguilar, J. and Ahlen, S. and Alam, Shadab and Alexander, David M. and Alfarsy, R. and Allen, L. and Allende Prieto, C. and Alves, O. and Ameel, J. and Armengaud, E. and Asorey, J. and Aviles, Alejandro and Bailey, S. and Balaguera-Antolínez, A. and Ballester, O. and Baltay, C. and Bault, A. and Beltran, S. F. and Benavides, B. and BenZvi, S. and Berti, A. and Besuner, R. and Beutler, Florian and Bianchi, D. and Blake, C. and Blanc, P. and Blum, R. and Bolton, A. and Bose, S. and Bramall, D. and Brieden, S. and Brodzeller, A. and Brooks, D. and Brownewell, C. and Buckley-Geer, E. and Cahn, R. N. and Cai, Z. and Canning, R. and Capasso, R. and Carnero Rosell, A. and Carton, P. and Casas, R. and Castander, F. J. and Cervantes-Cota, J. L. and Chabanier, S. and Chaussidon, E. and Chuang, C. and Circosta, C. and Cole, S. and Cooper, A. P. and da Costa, L. and Cousinou, M.-C. and Cuceu, A. and Davis, T. M. and Dawson, K. and de la Cruz-Noriega, R. and de la Macorra, A. and de Mattia, A. and Della Costa, J. and Demmer, P. and Derwent, M. and Dey, A. and Dey, B. and Dhungana, G. and Ding, Z. and Dobson, C. and Doel, P. and Donald-McCann, J. and Donaldson, J. and Douglass, K. and Duan, Y. and Dunlop, P. and Edelstein, J. and Eftekharzadeh, S. and Eisenstein, D. J. and Enriquez-Vargas, M. and Escoffier, S. and Evatt, M. and Fagrelius, P. and Fan, X. and Fanning, K. and Fawcett, V. A. and Ferraro, S. and Ereza, J. and Flaugher, B. and Font-Ribera, A. and Forero-Romero, J. E. and Frenk, C. S. and Fromenteau, S. and Gänsicke, B. T. and Garcia-Quintero, C. and Garrison, L. and Gaztañaga, E. and Gerardi, F. and Gil-Marín, H. and Gontcho A Gontcho, S. and Gonzalez-Morales, Alma X. and Gonzalez-de-Rivera, G. and Gonzalez-Perez, V. and Gordon, C. and Graur, O. and Green, D. and Grove, C. and Gruen, D. and Gutierrez, G. and Guy, J. and Hahn, C. and Harris, S. and Herrera, D. and Herrera-Alcantar, Hiram K. and Honscheid, K. and Howlett, C. and Huterer, D. and Iršič, V. and Ishak, M. and Jelinsky, P. and Jiang, L. and Jimenez, J. and Jing, Y. P. and Joyce, R. and Jullo, E. and Juneau, S. and Karaçaylı, N. G. and Karamanis, M. and Karcher, A. and Karim, T. and Kehoe, R. and Kent, S. and Kirkby, D. and Kisner, T. and Kitaura, F. and Koposov, S. E. and Kovács, A. and Kremin, A. and Krolewski, Alex and L’Huillier, B. and Lahav, O. and Lambert, A. and Lamman, C. and Lan, Ting-Wen and Landriau, M. and Lane, S. and Lang, D. and Lange, J. U. and Lasker, J. and Le Guillou, L. and Leauthaud, A. and Le Van Suu, A. and Levi, Michael E. and Li, T. S. and Magneville, C. and Manera, M. and Manser, Christopher J. and Marshall, B. and Martini, Paul and McCollam, W. and McDonald, P. and Meisner, Aaron M. and Mena-Fernández, J. and Meneses-Rizo, J. and Mezcua, M. and Miller, T. and Miquel, R. and Montero-Camacho, P. and Moon, J. and Moustakas, J. and Mueller, E. and Muñoz-Gutiérrez, Andrea and Myers, Adam D. and Nadathur, S. and Najita, J. and Napolitano, L. and Neilsen, E. and Newman, Jeffrey A. and Nie, J. D. and Ning, Y. and Niz, G. and Norberg, P. and Noriega, Hernán E. and O’Brien, T. and Obuljen, A. and Palanque-Delabrouille, N. and Palmese, A. and Zhiwei, P. and Pappalardo, D. and PENG, X. and Percival, W. J. and Perruchot, S. and Pogge, R. and Poppett, C. and Porredon, A. and Prada, F. and Prochaska, J. and Pucha, R. and Pérez-Fernández, A. and Pérez-Ràfols, I. and Rabinowitz, D. and Raichoor, A. and Ramirez-Solano, S. and Ramírez-Pérez, César and Ravoux, C. and Reil, K. and Rezaie, M. and Rocher, A. and Rockosi, C. and Roe, N. A. and Roodman, A. and Ross, A. J. and Rossi, G. and Ruggeri, R. and Ruhlmann-Kleider, V. and Sabiu, C. G. and Gaines, S. and Said, K. and Saintonge, A. and Salas Catonga, Javier and Samushia, L. and Sanchez, E. and Saulder, C. and Schaan, E. and Schlafly, E. and Schlegel, D. and Schmoll, J. and Scholte, D. and Schubnell, M. and Secroun, A. and Seo, H. and Serrano, S. and Sharples, Ray M. and Sholl, Michael J. and Silber, Joseph Harry and Silva, D. R. and Sirk, M. and Siudek, M. and Smith, A. and Sprayberry, D. and Staten, R. and Stupak, B. and Tan, T. and Tarlé, Gregory and Tie, Suk Sien and Tojeiro, R. and Ureña-López, L. A. and Valdes, F. and Valenzuela, O. and Valluri, M. and Vargas-Magaña, M. and Verde, L. and Walther, M. and Wang, B. and Wang, M. S. and Weaver, B. A. and Weaverdyck, C. and Wechsler, R. and Wilson, Michael J. and Yang, J. and Yu, Y. and Yuan, S. and Yèche, Christophe and Zhang, H. and Zhang, K. and Zhao, Cheng and Zhou, Rongpu and Zhou, Zhimin and Zou, H. and Zou, J. and Zou, S. and Zu, Y.},
   year={2022},
   month=oct, pages={207} }

@article{legacysurvey,
   title={Overview of the DESI Legacy Imaging Surveys},
   volume={157},
   ISSN={1538-3881},
   url={http://dx.doi.org/10.3847/1538-3881/ab089d},
   DOI={10.3847/1538-3881/ab089d},
   number={5},
   journal={The Astronomical Journal},
   publisher={American Astronomical Society},
   author={Dey, Arjun and Schlegel, David J. and Lang, Dustin and Blum, Robert and Burleigh, Kaylan and Fan, Xiaohui and Findlay, Joseph R. and Finkbeiner, Doug and Herrera, David and Juneau, Stéphanie and Landriau, Martin and Levi, Michael and McGreer, Ian and Meisner, Aaron and Myers, Adam D. and Moustakas, John and Nugent, Peter and Patej, Anna and Schlafly, Edward F. and Walker, Alistair R. and Valdes, Francisco and Weaver, Benjamin A. and Yèche, Christophe and Zou, Hu and Zhou, Xu and Abareshi, Behzad and Abbott, T. M. C. and Abolfathi, Bela and Aguilera, C. and Alam, Shadab and Allen, Lori and Alvarez, A. and Annis, James and Ansarinejad, Behzad and Aubert, Marie and Beechert, Jacqueline and Bell, Eric F. and BenZvi, Segev Y. and Beutler, Florian and Bielby, Richard M. and Bolton, Adam S. and Briceño, César and Buckley-Geer, Elizabeth J. and Butler, Karen and Calamida, Annalisa and Carlberg, Raymond G. and Carter, Paul and Casas, Ricard and Castander, Francisco J. and Choi, Yumi and Comparat, Johan and Cukanovaite, Elena and Delubac, Timothée and DeVries, Kaitlin and Dey, Sharmila and Dhungana, Govinda and Dickinson, Mark and Ding, Zhejie and Donaldson, John B. and Duan, Yutong and Duckworth, Christopher J. and Eftekharzadeh, Sarah and Eisenstein, Daniel J. and Etourneau, Thomas and Fagrelius, Parker A. and Farihi, Jay and Fitzpatrick, Mike and Font-Ribera, Andreu and Fulmer, Leah and Gänsicke, Boris T. and Gaztanaga, Enrique and George, Koshy and Gerdes, David W. and A Gontcho, Satya Gontcho and Gorgoni, Claudio and Green, Gregory and Guy, Julien and Harmer, Diane and Hernandez, M. and Honscheid, Klaus and Huang, Lijuan (Wendy) and James, David J. and Jannuzi, Buell T. and Jiang, Linhua and Joyce, Richard and Karcher, Armin and Karkar, Sonia and Kehoe, Robert and Kneib, Jean-Paul and Kueter-Young, Andrea and Lan, Ting-Wen and Lauer, Tod R. and Guillou, Laurent Le and Van Suu, Auguste Le and Lee, Jae Hyeon and Lesser, Michael and Levasseur, Laurence Perreault and Li, Ting S. and Mann, Justin L. and Marshall, Robert and Martínez-Vázquez, C. E. and Martini, Paul and du Mas des Bourboux, Hélion and McManus, Sean and Meier, Tobias Gabriel and Ménard, Brice and Metcalfe, Nigel and Muñoz-Gutiérrez, Andrea and Najita, Joan and Napier, Kevin and Narayan, Gautham and Newman, Jeffrey A. and Nie, Jundan and Nord, Brian and Norman, Dara J. and Olsen, Knut A. G. and Paat, Anthony and Palanque-Delabrouille, Nathalie and Peng, Xiyan and Poppett, Claire L. and Poremba, Megan R. and Prakash, Abhishek and Rabinowitz, David and Raichoor, Anand and Rezaie, Mehdi and Robertson, A. N. and Roe, Natalie A. and Ross, Ashley J. and Ross, Nicholas P. and Rudnick, Gregory and Gaines, Sasha and Saha, Abhijit and Sánchez, F. Javier and Savary, Elodie and Schweiker, Heidi and Scott, Adam and Seo, Hee-Jong and Shan, Huanyuan and Silva, David R. and Slepian, Zachary and Soto, Christian and Sprayberry, David and Staten, Ryan and Stillman, Coley M. and Stupak, Robert J. and Summers, David L. and Tie, Suk Sien and Tirado, H. and Vargas-Magaña, Mariana and Vivas, A. Katherina and Wechsler, Risa H. and Williams, Doug and Yang, Jinyi and Yang, Qian and Yapici, Tolga and Zaritsky, Dennis and Zenteno, A. and Zhang, Kai and Zhang, Tianmeng and Zhou, Rongpu and Zhou, Zhimin},
   year={2019},
   month=apr, pages={168} }

@misc{desisurvey,
      title={The DESI Experiment Part I: Science,Targeting, and Survey Design}, 
      author={DESI Collaboration and Amir Aghamousa and Jessica Aguilar and Steve Ahlen and Shadab Alam and Lori E. Allen and Carlos Allende Prieto and James Annis and Stephen Bailey and Christophe Balland and Otger Ballester and Charles Baltay and Lucas Beaufore and Chris Bebek and Timothy C. Beers and Eric F. Bell and José Luis Bernal and Robert Besuner and Florian Beutler and Chris Blake and Hannes Bleuler and Michael Blomqvist and Robert Blum and Adam S. Bolton and Cesar Briceno and David Brooks and Joel R. Brownstein and Elizabeth Buckley-Geer and Angela Burden and Etienne Burtin and Nicolas G. Busca and Robert N. Cahn and Yan-Chuan Cai and Laia Cardiel-Sas and Raymond G. Carlberg and Pierre-Henri Carton and Ricard Casas and Francisco J. Castander and Jorge L. Cervantes-Cota and Todd M. Claybaugh and Madeline Close and Carl T. Coker and Shaun Cole and Johan Comparat and Andrew P. Cooper and M. -C. Cousinou and Martin Crocce and Jean-Gabriel Cuby and Daniel P. Cunningham and Tamara M. Davis and Kyle S. Dawson and Axel de la Macorra and Juan De Vicente and Timothée Delubac and Mark Derwent and Arjun Dey and Govinda Dhungana and Zhejie Ding and Peter Doel and Yutong T. Duan and Anne Ealet and Jerry Edelstein and Sarah Eftekharzadeh and Daniel J. Eisenstein and Ann Elliott and Stéphanie Escoffier and Matthew Evatt and Parker Fagrelius and Xiaohui Fan and Kevin Fanning and Arya Farahi and Jay Farihi and Ginevra Favole and Yu Feng and Enrique Fernandez and Joseph R. Findlay and Douglas P. Finkbeiner and Michael J. Fitzpatrick and Brenna Flaugher and Samuel Flender and Andreu Font-Ribera and Jaime E. Forero-Romero and Pablo Fosalba and Carlos S. Frenk and Michele Fumagalli and Boris T. Gaensicke and Giuseppe Gallo and Juan Garcia-Bellido and Enrique Gaztanaga and Nicola Pietro Gentile Fusillo and Terry Gerard and Irena Gershkovich and Tommaso Giannantonio and Denis Gillet and Guillermo Gonzalez-de-Rivera and Violeta Gonzalez-Perez and Shelby Gott and Or Graur and Gaston Gutierrez and Julien Guy and Salman Habib and Henry Heetderks and Ian Heetderks and Katrin Heitmann and Wojciech A. Hellwing and David A. Herrera and Shirley Ho and Stephen Holland and Klaus Honscheid and Eric Huff and Timothy A. Hutchinson and Dragan Huterer and Ho Seong Hwang and Joseph Maria Illa Laguna and Yuzo Ishikawa and Dianna Jacobs and Niall Jeffrey and Patrick Jelinsky and Elise Jennings and Linhua Jiang and Jorge Jimenez and Jennifer Johnson and Richard Joyce and Eric Jullo and Stéphanie Juneau and Sami Kama and Armin Karcher and Sonia Karkar and Robert Kehoe and Noble Kennamer and Stephen Kent and Martin Kilbinger and Alex G. Kim and David Kirkby and Theodore Kisner and Ellie Kitanidis and Jean-Paul Kneib and Sergey Koposov and Eve Kovacs and Kazuya Koyama and Anthony Kremin and Richard Kron and Luzius Kronig and Andrea Kueter-Young and Cedric G. Lacey and Robin Lafever and Ofer Lahav and Andrew Lambert and Michael Lampton and Martin Landriau and Dustin Lang and Tod R. Lauer and Jean-Marc Le Goff and Laurent Le Guillou and Auguste Le Van Suu and Jae Hyeon Lee and Su-Jeong Lee and Daniela Leitner and Michael Lesser and Michael E. Levi and Benjamin L'Huillier and Baojiu Li and Ming Liang and Huan Lin and Eric Linder and Sarah R. Loebman and Zarija Lukić and Jun Ma and Niall MacCrann and Christophe Magneville and Laleh Makarem and Marc Manera and Christopher J. Manser and Robert Marshall and Paul Martini and Richard Massey and Thomas Matheson and Jeremy McCauley and Patrick McDonald and Ian D. McGreer and Aaron Meisner and Nigel Metcalfe and Timothy N. Miller and Ramon Miquel and John Moustakas and Adam Myers and Milind Naik and Jeffrey A. Newman and Robert C. Nichol and Andrina Nicola and Luiz Nicolati da Costa and Jundan Nie and Gustavo Niz and Peder Norberg and Brian Nord and Dara Norman and Peter Nugent and Thomas O'Brien and Minji Oh and Knut A. G. Olsen and Cristobal Padilla and Hamsa Padmanabhan and Nikhil Padmanabhan and Nathalie Palanque-Delabrouille and Antonella Palmese and Daniel Pappalardo and Isabelle Pâris and Changbom Park and Anna Patej and John A. Peacock and Hiranya V. Peiris and Xiyan Peng and Will J. Percival and Sandrine Perruchot and Matthew M. Pieri and Richard Pogge and Jennifer E. Pollack and Claire Poppett and Francisco Prada and Abhishek Prakash and Ronald G. Probst and David Rabinowitz and Anand Raichoor and Chang Hee Ree and Alexandre Refregier and Xavier Regal and Beth Reid and Kevin Reil and Mehdi Rezaie and Constance M. Rockosi and Natalie Roe and Samuel Ronayette and Aaron Roodman and Ashley J. Ross and Nicholas P. Ross and Graziano Rossi and Eduardo Rozo and Vanina Ruhlmann-Kleider and Eli S. Rykoff and Cristiano Sabiu and Lado Samushia and Eusebio Sanchez and Javier Sanchez and David J. Schlegel and Michael Schneider and Michael Schubnell and Aurélia Secroun and Uros Seljak and Hee-Jong Seo and Santiago Serrano and Arman Shafieloo and Huanyuan Shan and Ray Sharples and Michael J. Sholl and William V. Shourt and Joseph H. Silber and David R. Silva and Martin M. Sirk and Anze Slosar and Alex Smith and George F. Smoot and Debopam Som and Yong-Seon Song and David Sprayberry and Ryan Staten and Andy Stefanik and Gregory Tarle and Suk Sien Tie and Jeremy L. Tinker and Rita Tojeiro and Francisco Valdes and Octavio Valenzuela and Monica Valluri and Mariana Vargas-Magana and Licia Verde and Alistair R. Walker and Jiali Wang and Yuting Wang and Benjamin A. Weaver and Curtis Weaverdyck and Risa H. Wechsler and David H. Weinberg and Martin White and Qian Yang and Christophe Yeche and Tianmeng Zhang and Gong-Bo Zhao and Yi Zheng and Xu Zhou and Zhimin Zhou and Yaling Zhu and Hu Zou and Ying Zu},
      year={2016},
      eprint={1611.00036},
      archivePrefix={arXiv},
      primaryClass={astro-ph.IM},
      url={https://arxiv.org/abs/1611.00036}, 
}

@book{peebles1980,
  author    = {Peebles, P.~J.~E.},
  title     = {The Large‑Scale Structure of the Universe},
  year      = {1980},
  publisher = {Princeton University Press},
  address   = {Princeton, NJ, USA},
  isbn      = {978-0-691-08240-0}
}

@article{landyszalay,
title = "Bias and variance of angular correlation functions",
author = "Landy, {Stephen D.} and Szalay, {Alexander S.}",
year = "1993",
month = jul,
day = "20",
doi = "10.1086/172900",
language = "English (US)",
volume = "412",
pages = "64--71",
journal = "Astrophysical Journal",
issn = "0004-637X",
publisher = "American Astronomical Society",
number = "1",
}

@article{desi2024IV,
   title={DESI 2024 IV: Baryon Acoustic Oscillations from the Lyman alpha forest},
   volume={2025},
   ISSN={1475-7516},
   url={http://dx.doi.org/10.1088/1475-7516/2025/01/124},
   DOI={10.1088/1475-7516/2025/01/124},
   number={01},
   journal={Journal of Cosmology and Astroparticle Physics},
   publisher={IOP Publishing},
   author={Adame, A.G. and Aguilar, J. and Ahlen, S. and Alam, S. and Alexander, D.M. and Alvarez, M. and Alves, O. and Anand, A. and Andrade, U. and Armengaud, E. and Avila, S. and Aviles, A. and Awan, H. and Bailey, S. and Baltay, C. and Bault, A. and Bautista, J. and Behera, J. and BenZvi, S. and Beutler, F. and Bianchi, D. and Blake, C. and Blum, R. and Brieden, S. and Brodzeller, A. and Brooks, D. and Buckley-Geer, E. and Burtin, E. and Calderon, R. and Canning, R. and Carnero Rosell, A. and Cereskaite, R. and Cervantes-Cota, J.L. and Chabanier, S. and Chaussidon, E. and Chaves-Montero, J. and Chen, S. and Chen, X. and Claybaugh, T. and Cole, S. and Cuceu, A. and Davis, T.M. and Dawson, K. and de la Cruz, R. and de la Macorra, A. and de Mattia, A. and Deiosso, N. and Dey, A. and Dey, B. and Ding, J. and Ding, Z. and Doel, P. and Edelstein, J. and Eftekharzadeh, S. and Eisenstein, D.J. and Elliott, A. and Fagrelius, P. and Fanning, K. and Ferraro, S. and Ereza, J. and Findlay, N. and Flaugher, B. and Font-Ribera, A. and Forero-Sánchez, D. and Forero-Romero, J.E. and Garcia-Quintero, C. and Gaztañaga, E. and Gil-Marín, H. and Gontcho, S.Gontcho A. and Gonzalez-Morales, A.X. and Gonzalez-Perez, V. and Gordon, C. and Green, D. and Gruen, D. and Gsponer, R. and Gutierrez, G. and Guy, J. and Hadzhiyska, B. and Hahn, C. and Hanif, M.M.S. and Herrera-Alcantar, H.K. and Honscheid, K. and Howlett, C. and Huterer, D. and Iršič, V. and Ishak, M. and Juneau, S. and Karaçaylı, N.G. and Kehoe, R. and Kent, S. and Kirkby, D. and Kremin, A. and Krolewski, A. and Lai, Y. and Lan, T.-W. and Landriau, M. and Lang, D. and Lasker, J. and Le Goff, J.M. and Le Guillou, L. and Leauthaud, A. and Levi, M.E. and Li, T.S. and Linder, E. and Lodha, K. and Magneville, C. and Manera, M. and Margala, D. and Martini, P. and Maus, M. and McDonald, P. and Medina-Varela, L. and Meisner, A. and Mena-Fernández, J. and Miquel, R. and Moon, J. and Moore, S. and Moustakas, J. and Mueller, E. and Muñoz-Gutiérrez, A. and Myers, A.D. and Nadathur, S. and Napolitano, L. and Neveux, R. and Newman, J.A. and Nguyen, N.M. and Nie, J. and Niz, G. and Noriega, H.E. and Padmanabhan, N. and Paillas, E. and Palanque-Delabrouille, N. and Pan, J. and Penmetsa, S. and Percival, W.J. and Pieri, M.M. and Pinon, M. and Poppett, C. and Porredon, A. and Prada, F. and Pérez-Fernández, A. and Pérez-Ràfols, I. and Rabinowitz, D. and Raichoor, A. and Ramírez-Pérez, C. and Ramirez-Solano, S. and Rashkovetskyi, M. and Ravoux, C. and Rezaie, M. and Rich, J. and Rocher, A. and Rockosi, C. and Roe, N.A. and Rosado-Marin, A. and Ross, A.J. and Rossi, G. and Ruggeri, R. and Ruhlmann-Kleider, V. and Samushia, L. and Sanchez, E. and Saulder, C. and Schlafly, E.F. and Schlegel, D. and Schubnell, M. and Seo, H. and Sharples, R. and Silber, J. and Sinigaglia, F. and Slosar, A. and Smith, A. and Sprayberry, D. and Tan, T. and Tarlé, G. and Trusov, S. and Vaisakh, R. and Valcin, D. and Valdes, F. and Vargas-Magaña, M. and Verde, L. and Walther, M. and Wang, B. and Wang, M.S. and Weaver, B.A. and Weaverdyck, N. and Wechsler, R.H. and Weinberg, D.H. and White, M. and Yu, J. and Yu, Y. and Yuan, S. and Yèche, C. and Zaborowski, E.A. and Zarrouk, P. and Zhang, H. and Zhao, C. and Zhao, R. and Zhou, R. and Zou, H.},
   year={2025},
   month=jan, pages={124} }

@article{linder2005,
   title={Cosmic growth history and expansion history},
   volume={72},
   ISSN={1550-2368},
   url={http://dx.doi.org/10.1103/PhysRevD.72.043529},
   DOI={10.1103/physrevd.72.043529},
   number={4},
   journal={Physical Review D},
   publisher={American Physical Society (APS)},
   author={Linder, Eric V.},
   year={2005},
   month=aug }

@ARTICLE{corrfunc2020,
    author = {{Sinha}, Manodeep and {Garrison}, Lehman H.},
    title = "{CORRFUNC - a suite of blazing fast correlation functions on the CPU}",
    journal = {\mnras},
    year = "2020",
    month = "Jan",
    volume = {491},
    number = {2},
    pages = {3022-3041},
    doi = {10.1093/mnras/stz3157},
    adsurl =
    {https://ui.adsabs.harvard.edu/abs/2020MNRAS.491.3022S}
}

@inbook{corrfunc2019,
   title={CORRFUNC: Blazing Fast Correlation Functions with AVX512F SIMD Intrinsics},
   ISBN={9789811377297},
   ISSN={1865-0937},
   url={http://dx.doi.org/10.1007/978-981-13-7729-7_1},
   DOI={10.1007/978-981-13-7729-7_1},
   booktitle={Software Challenges to Exascale Computing},
   publisher={Springer Singapore},
   author={Sinha, Manodeep and Garrison, Lehman},
   year={2019},
   pages={3–20} }

@article{planck2018,
   title={Planck2018 results: VI. Cosmological parameters},
   volume={641},
   ISSN={1432-0746},
   url={http://dx.doi.org/10.1051/0004-6361/201833910},
   DOI={10.1051/0004-6361/201833910},
   journal={Astronomy \% Astrophysics},
   publisher={EDP Sciences},
   author={Aghanim, N. and Akrami, Y. and Ashdown, M. and Aumont, J. and Baccigalupi, C. and Ballardini, M. and Banday, A. J. and Barreiro, R. B. and Bartolo, N. and Basak, S. and Battye, R. and Benabed, K. and Bernard, J.-P. and Bersanelli, M. and Bielewicz, P. and Bock, J. J. and Bond, J. R. and Borrill, J. and Bouchet, F. R. and Boulanger, F. and Bucher, M. and Burigana, C. and Butler, R. C. and Calabrese, E. and Cardoso, J.-F. and Carron, J. and Challinor, A. and Chiang, H. C. and Chluba, J. and Colombo, L. P. L. and Combet, C. and Contreras, D. and Crill, B. P. and Cuttaia, F. and de Bernardis, P. and de Zotti, G. and Delabrouille, J. and Delouis, J.-M. and Di Valentino, E. and Diego, J. M. and Doré, O. and Douspis, M. and Ducout, A. and Dupac, X. and Dusini, S. and Efstathiou, G. and Elsner, F. and Enßlin, T. A. and Eriksen, H. K. and Fantaye, Y. and Farhang, M. and Fergusson, J. and Fernandez-Cobos, R. and Finelli, F. and Forastieri, F. and Frailis, M. and Fraisse, A. A. and Franceschi, E. and Frolov, A. and Galeotta, S. and Galli, S. and Ganga, K. and Génova-Santos, R. T. and Gerbino, M. and Ghosh, T. and González-Nuevo, J. and Górski, K. M. and Gratton, S. and Gruppuso, A. and Gudmundsson, J. E. and Hamann, J. and Handley, W. and Hansen, F. K. and Herranz, D. and Hildebrandt, S. R. and Hivon, E. and Huang, Z. and Jaffe, A. H. and Jones, W. C. and Karakci, A. and Keihänen, E. and Keskitalo, R. and Kiiveri, K. and Kim, J. and Kisner, T. S. and Knox, L. and Krachmalnicoff, N. and Kunz, M. and Kurki-Suonio, H. and Lagache, G. and Lamarre, J.-M. and Lasenby, A. and Lattanzi, M. and Lawrence, C. R. and Le Jeune, M. and Lemos, P. and Lesgourgues, J. and Levrier, F. and Lewis, A. and Liguori, M. and Lilje, P. B. and Lilley, M. and Lindholm, V. and López-Caniego, M. and Lubin, P. M. and Ma, Y.-Z. and Macías-Pérez, J. F. and Maggio, G. and Maino, D. and Mandolesi, N. and Mangilli, A. and Marcos-Caballero, A. and Maris, M. and Martin, P. G. and Martinelli, M. and Martínez-González, E. and Matarrese, S. and Mauri, N. and McEwen, J. D. and Meinhold, P. R. and Melchiorri, A. and Mennella, A. and Migliaccio, M. and Millea, M. and Mitra, S. and Miville-Deschênes, M.-A. and Molinari, D. and Montier, L. and Morgante, G. and Moss, A. and Natoli, P. and Nørgaard-Nielsen, H. U. and Pagano, L. and Paoletti, D. and Partridge, B. and Patanchon, G. and Peiris, H. V. and Perrotta, F. and Pettorino, V. and Piacentini, F. and Polastri, L. and Polenta, G. and Puget, J.-L. and Rachen, J. P. and Reinecke, M. and Remazeilles, M. and Renzi, A. and Rocha, G. and Rosset, C. and Roudier, G. and Rubiño-Martín, J. A. and Ruiz-Granados, B. and Salvati, L. and Sandri, M. and Savelainen, M. and Scott, D. and Shellard, E. P. S. and Sirignano, C. and Sirri, G. and Spencer, L. D. and Sunyaev, R. and Suur-Uski, A.-S. and Tauber, J. A. and Tavagnacco, D. and Tenti, M. and Toffolatti, L. and Tomasi, M. and Trombetti, T. and Valenziano, L. and Valiviita, J. and Van Tent, B. and Vibert, L. and Vielva, P. and Villa, F. and Vittorio, N. and Wandelt, B. D. and Wehus, I. K. and White, M. and White, S. D. M. and Zacchei, A. and Zonca, A.},
   year={2020},
   month=sep, pages={A6} }

@article{desidr2lya,
       author = {{DESI Collaboration} and {Abdul-Karim}, M. and {Aguilar}, J. and {Ahlen}, S. and {Allende Prieto}, C. and {Alves}, O. and {Anand}, A. and {Andrade}, U. and {Armengaud}, E. and {Aviles}, A. and {Bailey}, S. and {Bault}, A. and {Behera}, J. and {BenZvi}, S. and {Bianchi}, D. and {Blake}, C. and {Brodzeller}, A. and {Brooks}, D. and {Buckley-Geer}, E. and {Burtin}, E. and {Calderon}, R. and {Canning}, R. and {Carnero Rosell}, A. and {Carrilho}, P. and {Casas}, L. and {Castander}, F.~J. and {Cereskaite}, R. and {Charles}, M. and {Chaussidon}, E. and {Chaves-Montero}, J. and {Chebat}, D. and {Claybaugh}, T. and {Cole}, S. and {Cooper}, A.~P. and {Cuceu}, A. and {Dawson}, K.~S. and {de Belsunce}, R. and {de la Macorra}, A. and {de Mattia}, A. and {Deiosso}, N. and {Della Costa}, J. and {Dey}, A. and {Dey}, B. and {Ding}, Z. and {Doel}, P. and {Edelstein}, J. and {Eisenstein}, D.~J. and {Elbers}, W. and {Fagrelius}, P. and {Fanning}, K. and {Ferraro}, S. and {Font-Ribera}, A. and {Forero-Romero}, J.~E. and {Garcia-Quintero}, C. and {Garrison}, L.~H. and {Gazta{\~n}aga}, E. and {Gil-Mar{\'\i}n}, H. and {Gontcho}, S. Gontcho A and {Gonzalez-Morales}, A.~X. and {Gordon}, C. and {Green}, D. and {Gutierrez}, G. and {Guy}, J. and {Hahn}, C. and {Herbold}, M. and {Herrera-Alcantar}, H.~K. and {Ho}, M. -F. and {Honscheid}, K. and {Howlett}, C. and {Huterer}, D. and {Ishak}, M. and {Juneau}, S. and {Kara{\c{c}}ayl{\i}}, N.~G. and {Kehoe}, R. and {Kent}, S. and {Kirkby}, D. and {Kisner}, T. and {Kitaura}, F. -S. and {Koposov}, S.~E. and {Kremin}, A. and {Lahav}, O. and {Lamman}, C. and {Landriau}, M. and {Lang}, D. and {Lasker}, J. and {Le Goff}, J.~M. and {Le Guillou}, L. and {Leauthaud}, A. and {Levi}, M.~E. and {Li}, Q. and {Li}, T.~S. and {Lodha}, K. and {Lokken}, M. and {Magneville}, C. and {Manera}, M. and {Martini}, P. and {Matthewson}, W.~L. and {McDonald}, P. and {Meisner}, A. and {Mena-Fern{\'a}ndez}, J. and {Miquel}, R. and {Moustakas}, J. and {Mu{\~n}oz Santos}, D. and {Mu{\~n}oz-Guti{\'e}rrez}, A. and {Myers}, A.~D. and {Newman}, J.~A. and {Niz}, G. and {Noriega}, H.~E. and {Paillas}, E. and {Palanque-Delabrouille}, N. and {Pan}, J. and {Percival}, W.~J. and {P{\'e}rez-R{\`a}fols}, I. and {Pieri}, M.~M. and {Poppett}, C. and {Prada}, F. and {Rabinowitz}, D. and {Raichoor}, A. and {Ram{\'\i}rez-P{\'e}rez}, C. and {Rashkovetskyi}, M. and {Ravoux}, C. and {Rich}, J. and {Rockosi}, C. and {Ross}, A.~J. and {Rossi}, G. and {Ruhlmann-Kleider}, V. and {Sanchez}, E. and {Sanders}, N. and {Satyavolu}, S. and {Schlegel}, D. and {Schubnell}, M. and {Seo}, H. and {Shafieloo}, A. and {Sharples}, R. and {Silber}, J. and {Sinigaglia}, F. and {Sprayberry}, D. and {Tan}, T. and {Tarl{\'e}}, G. and {Taylor}, P. and {Turner}, W. and {Valdes}, F. and {Vargas-Maga{\~n}a}, M. and {Walther}, M. and {Weaver}, B.~A. and {Wolfson}, M. and {Y{\`e}che}, C. and {Zarrouk}, P. and {Zhou}, R. and {Zou}, H.},
        title = "{DESI DR2 Results I: Baryon Acoustic Oscillations from the Lyman Alpha Forest}",
      journal = {arXiv e-prints},
         year = 2025,
        month = mar,
          eid = {arXiv:2503.14739},
        pages = {arXiv:2503.14739},
          doi = {10.48550/arXiv.2503.14739},
archivePrefix = {arXiv},
       eprint = {2503.14739},
 primaryClass = {astro-ph.CO},
       adsurl = {https://ui.adsabs.harvard.edu/abs/2025arXiv250314739D}
}

@article{daangela2008,
       author = {{da {\^A}ngela}, J. and {Shanks}, T. and {Croom}, S.~M. and {Weilbacher}, P. and {Brunner}, R.~J. and {Couch}, W.~J. and {Miller}, L. and {Myers}, A.~D. and {Nichol}, R.~C. and {Pimbblet}, K.~A. and {de Propris}, R. and {Richards}, G.~T. and {Ross}, N.~P. and {Schneider}, D.~P. and {Wake}, D.},
        title = "{The 2dF-SDSS LRG and QSO survey: QSO clustering and the L-z degeneracy}",
      journal = {\mnras},
         year = 2008,
        month = jan,
       volume = {383},
       number = {2},
        pages = {565-580},
          doi = {10.1111/j.1365-2966.2007.12552.x},
archivePrefix = {arXiv},
       eprint = {astro-ph/0612401},
 primaryClass = {astro-ph},
       adsurl = {https://ui.adsabs.harvard.edu/abs/2008MNRAS.383..565D}
}

@misc{chaussidon2024,
      title={Constraining primordial non-Gaussianity with DESI 2024 LRG and QSO samples}, 
      author={E. Chaussidon and C. Yèche and A. de Mattia and C. Payerne and P. McDonald and A. J. Ross and S. Ahlen and D. Bianchi and D. Brooks and E. Burtin and T. Claybaugh and A. de la Macorra and P. Doel and S. Ferraro and A. Font-Ribera and J. E. Forero-Romero and E. Gaztañaga and H. Gil-Marín and S. Gontcho A Gontcho and G. Gutierrez and J. Guy and K. Honscheid and C. Howlett and D. Huterer and R. Kehoe and D. Kirkby and T. Kisner and A. Kremin and L. Le Guillou and M. E. Levi and M. Manera and A. Meisner and R. Miquel and J. Moustakas and J. A. Newman and G. Niz and N. Palanque-Delabrouille and W. J. Percival and F. Prada and I. Pérez-Ràfols and C. Ravoux and G. Rossi and E. Sanchez and D. Schlegel and M. Schubnell and H. Seo and D. Sprayberry and G. Tarlé and M. Vargas-Magaña and B. A. Weaver and C. Zhao and H. Zou},
      year={2024},
      eprint={2411.17623},
      archivePrefix={arXiv},
      primaryClass={astro-ph.CO},
      url={https://arxiv.org/abs/2411.17623}, 
}

@article{yeche2020,
   title={Preliminary Target Selection for the DESI Quasar (QSO) Sample},
   volume={4},
   ISSN={2515-5172},
   url={http://dx.doi.org/10.3847/2515-5172/abc01a},
   DOI={10.3847/2515-5172/abc01a},
   number={10},
   journal={Research Notes of the AAS},
   publisher={American Astronomical Society},
   author={Yèche, Christophe and Palanque-Delabrouille, Nathalie and Claveau, Charles-Antoine and Brooks, David D. and Chaussidon, Edmond and Davis, Tamara M. and Dawson, Kyle S. and Dey, Arjun and Duan, Yutong and Eftekharzadeh, Sarah and Eisenstein, Daniel J. and Gaztañaga, Enrique and Kehoe, Robert and Landriau, Martin and Lang, Dustin and Levi, Michael E. and Meisner, Aaron M. and Myers, Adam D. and Newman, Jeffrey A. and Poppett, Claire and Prada, Francisco and Raichoor, Anand and Schlegel, David J. and Schubnell, Michael and Staten, Ryan and Tarlé, Gregory and Zhou, Rongpu},
   year={2020},
   month=oct, pages={179} }

@article{mohammad2022,
   title={Creating jackknife and bootstrap estimates of the covariance matrix for the two-point correlation function},
   volume={514},
   ISSN={1365-2966},
   url={http://dx.doi.org/10.1093/mnras/stac1458},
   DOI={10.1093/mnras/stac1458},
   number={1},
   journal={Monthly Notices of the Royal Astronomical Society},
   publisher={Oxford University Press (OUP)},
   author={Mohammad, Faizan G and Percival, Will J},
   year={2022},
   month=may, pages={1289–1301} }

@ARTICLE{white2012,
       author = {{White}, Martin and {Myers}, Adam D. and {Ross}, Nicholas P. and {Schlegel}, David J. and {Hennawi}, Joseph F. and {Shen}, Yue and {McGreer}, Ian and {Strauss}, Michael A. and {Bolton}, Adam S. and {Bovy}, Jo and {Fan}, X. and {Miralda-Escude}, Jordi and {Palanque-Delabrouille}, N. and {Paris}, I. and {Petitjean}, P. and {Schneider}, D.~P. and {Viel}, M. and {Weinberg}, David H. and {Yeche}, Ch. and {Zehavi}, I. and {Pan}, K. and {Snedden}, S. and {Bizyaev}, D. and {Brewington}, H. and {Brinkmann}, J. and {Malanushenko}, V. and {Malanushenko}, E. and {Oravetz}, D. and {Simmons}, A. and {Sheldon}, A. and {Weaver}, Benjamin A.},
        title = "{The clustering of intermediate-redshift quasars as measured by the Baryon Oscillation Spectroscopic Survey}",
      journal = {\mnras},
         year = 2012,
        month = aug,
       volume = {424},
       number = {2},
        pages = {933-950},
          doi = {10.1111/j.1365-2966.2012.21251.x},
archivePrefix = {arXiv},
       eprint = {1203.5306},
 primaryClass = {astro-ph.CO},
       adsurl = {https://ui.adsabs.harvard.edu/abs/2012MNRAS.424..933W}
}

@misc{redrock,
    author = {{Bailey et al. in preparation, (2026)}},
    note = {in preparation}
}

@MISC{fastspecfit,
  author = {{Moustakas}, John and {Scholte}, Dirk and {Dey}, Biprateep and {Khederlarian}, Ashod},
  title = "{FastSpecFit: Fast spectral synthesis and emission-line fitting of DESI spectra}",
   howpublished = {Astrophysics Source Code Library, record ascl:2308.005},
  year = 2023,
  month = aug,
  eid = {ascl:2308.005},
  pages = {ascl:2308.005},
  archivePrefix = {ascl},
  eprint = {2308.005},
  adsurl = {https://ui.adsabs.harvard.edu/abs/2023ascl.soft08005M}
}

@software{camb,
       author = {{Lewis}, Antony and {Challinor}, Anthony},
        title = "{CAMB: Code for Anisotropies in the Microwave Background}",
 howpublished = {Astrophysics Source Code Library, record ascl:1102.026},
         year = 2011,
        month = feb,
          eid = {ascl:1102.026},
       adsurl = {https://ui.adsabs.harvard.edu/abs/2011ascl.soft02026L}
}

@ARTICLE{haiman2001,
       author = {{Haiman}, Zolt{\'a}n and {Hui}, Lam},
        title = "{Constraining the Lifetime of Quasars from Their Spatial Clustering}",
      journal = {\apj},
         year = 2001,
        month = jan,
       volume = {547},
       number = {1},
        pages = {27-38},
          doi = {10.1086/318330},
archivePrefix = {arXiv},
       eprint = {astro-ph/0002190},
 primaryClass = {astro-ph},
       adsurl = {https://ui.adsabs.harvard.edu/abs/2001ApJ...547...27H}
}

@ARTICLE{desi2024III,
    author = {{DESI Collaboration} and {Adame}, A.~G. and {Aguilar}, J. and {Ahlen}, S. and {Alam}, S. and {Alexander}, D.~M. and {Alvarez}, M. and {Alves}, O. and {Anand}, A. and {Andrade}, U. and {Armengaud}, E. and {Avila}, S. and {Aviles}, A. and {Awan}, H. and {Bailey}, S. and {Baltay}, C. and {Bault}, A. and {Behera}, J. and {BenZvi}, S. and {Beutler}, F. and {Bianchi}, D. and {Blake}, C. and {Blum}, R. and {Brieden}, S. and {Brodzeller}, A. and {Brooks}, D. and {Buckley-Geer}, E. and {Burtin}, E. and {Calderon}, R. and {Canning}, R. and {Carnero Rosell}, A. and {Cereskaite}, R. and {Cervantes-Cota}, J.~L. and {Chabanier}, S. and {Chaussidon}, E. and {Chaves-Montero}, J. and {Chen}, S. and {Chen}, X. and {Claybaugh}, T. and {Cole}, S. and {Cuceu}, A. and {Davis}, T.~M. and {Dawson}, K. and {de la Macorra}, A. and {de Mattia}, A. and {Deiosso}, N. and {Dey}, A. and {Dey}, B. and {Ding}, Z. and {Doel}, P. and {Edelstein}, J. and {Eftekharzadeh}, S. and {Eisenstein}, D.~J. and {Elliott}, A. and {Fagrelius}, P. and {Fanning}, K. and {Ferraro}, S. and {Ereza}, J. and {Findlay}, N. and {Flaugher}, B. and {Font-Ribera}, A. and {Forero-S{\'a}nchez}, D. and {Forero-Romero}, J.~E. and {Garcia-Quintero}, C. and {Gazta{\~n}aga}, E. and {Gil-Mar{\'\i}n}, H. and {Gontcho}, S. Gontcho A and {Gonzalez-Morales}, A.~X. and {Gonzalez-Perez}, V. and {Gordon}, C. and {Green}, D. and {Gruen}, D. and {Gsponer}, R. and {Gutierrez}, G. and {Guy}, J. and {Hadzhiyska}, B. and {Hahn}, C. and {Hanif}, M.~M. S and {Herrera-Alcantar}, H.~K. and {Honscheid}, K. and {Howlett}, C. and {Huterer}, D. and {Ir{\v{s}}i{\v{c}}}, V. and {Ishak}, M. and {Juneau}, S. and {Kara{\c{c}}ayl{\i}}, N.~G. and {Kehoe}, R. and {Kent}, S. and {Kirkby}, D. and {Kremin}, A. and {Krolewski}, A. and {Lai}, Y. and {Lan}, T. -W. and {Landriau}, M. and {Lang}, D. and {Lasker}, J. and {Le Goff}, J.~M. and {Le Guillou}, L. and {Leauthaud}, A. and {Levi}, M.~E. and {Li}, T.~S. and {Linder}, E. and {Lodha}, K. and {Magneville}, C. and {Manera}, M. and {Margala}, D. and {Martini}, P. and {Maus}, M. and {McDonald}, P. and {Medina-Varela}, L. and {Meisner}, A. and {Mena-Fern{\'a}ndez}, J. and {Miquel}, R. and {Moon}, J. and {Moore}, S. and {Moustakas}, J. and {Mudur}, N. and {Mueller}, E. and {Mu{\~n}oz-Guti{\'e}rrez}, A. and {Myers}, A.~D. and {Nadathur}, S. and {Napolitano}, L. and {Neveux}, R. and {Newman}, J.~A. and {Nguyen}, N.~M. and {Nie}, J. and {Niz}, G. and {Noriega}, H.~E. and {Padmanabhan}, N. and {Paillas}, E. and {Palanque-Delabrouille}, N. and {Pan}, J. and {Penmetsa}, S. and {Percival}, W.~J. and {Pieri}, M. and {Pinon}, M. and {Poppett}, C. and {Porredon}, A. and {Prada}, F. and {P{\'e}rez-Fern{\'a}ndez}, A. and {P{\'e}rez-R{\`a}fols}, I. and {Rabinowitz}, D. and {Raichoor}, A. and {Ram{\'\i}rez-P{\'e}rez}, C. and {Ramirez-Solano}, S. and {Rashkovetskyi}, M. and {Rezaie}, M. and {Rich}, J. and {Rocher}, A. and {Rockosi}, C. and {Roe}, N.~A. and {Rosado-Marin}, A. and {Ross}, A.~J. and {Rossi}, G. and {Ruggeri}, R. and {Ruhlmann-Kleider}, V. and {Samushia}, L. and {Sanchez}, E. and {Saulder}, C. and {Schlafly}, E.~F. and {Schlegel}, D. and {Schubnell}, M. and {Seo}, H. and {Sharples}, R. and {Silber}, J. and {Slosar}, A. and {Smith}, A. and {Sprayberry}, D. and {Swanson}, J. and {Tan}, T. and {Tarl{\'e}}, G. and {Trusov}, S. and {Vaisakh}, R. and {Valcin}, D. and {Valdes}, F. and {Vargas-Maga{\~n}a}, M. and {Verde}, L. and {Walther}, M. and {Wang}, B. and {Wang}, M.~S. and {Weaver}, B.~A. and {Weaverdyck}, N. and {Wechsler}, R.~H. and {Weinberg}, D.~H. and {White}, M. and {Yu}, J. and {Yu}, Y. and {Yuan}, S. and {Y{\`e}che}, C. and {Zaborowski}, E.~A. and {Zarrouk}, P. and {Zhang}, H. and {Zhao}, C. and {Zhao}, R. and {Zhou}, R. and {Zou}, H.},
    title = "{DESI 2024 III: Baryon Acoustic Oscillations from Galaxies and Quasars}",
    journal = {arXiv e-prints},
    year = 2024,
    month = apr,
    eid = {arXiv:2404.03000},
    pages = {arXiv:2404.03000},
    doi = {10.48550/arXiv.2404.03000},
    archivePrefix = {arXiv},
    eprint = {2404.03000},
    primaryClass = {astro-ph.CO},
    adsurl = {https://ui.adsabs.harvard.edu/abs/2024arXiv240403000D}
}

@ARTICLE{desi2024VI,
    author = {{DESI Collaboration} and {Adame}, A.~G. and {Aguilar}, J. and {Ahlen}, S. and {Alam}, S. and {Alexander}, D.~M. and {Alvarez}, M. and {Alves}, O. and {Anand}, A. and {Andrade}, U. and {Armengaud}, E. and {Avila}, S. and {Aviles}, A. and {Awan}, H. and {Bahr-Kalus}, B. and {Bailey}, S. and {Baltay}, C. and {Bault}, A. and {Behera}, J. and {BenZvi}, S. and {Bera}, A. and {Beutler}, F. and {Bianchi}, D. and {Blake}, C. and {Blum}, R. and {Brieden}, S. and {Brodzeller}, A. and {Brooks}, D. and {Buckley-Geer}, E. and {Burtin}, E. and {Calderon}, R. and {Canning}, R. and {Carnero Rosell}, A. and {Cereskaite}, R. and {Cervantes-Cota}, J.~L. and {Chabanier}, S. and {Chaussidon}, E. and {Chaves-Montero}, J. and {Chen}, S. and {Chen}, X. and {Claybaugh}, T. and {Cole}, S. and {Cuceu}, A. and {Davis}, T.~M. and {Dawson}, K. and {de la Macorra}, A. and {de Mattia}, A. and {Deiosso}, N. and {Dey}, A. and {Dey}, B. and {Ding}, Z. and {Doel}, P. and {Edelstein}, J. and {Eftekharzadeh}, S. and {Eisenstein}, D.~J. and {Elliott}, A. and {Fagrelius}, P. and {Fanning}, K. and {Ferraro}, S. and {Ereza}, J. and {Findlay}, N. and {Flaugher}, B. and {Font-Ribera}, A. and {Forero-S{\'a}nchez}, D. and {Forero-Romero}, J.~E. and {Frenk}, C.~S. and {Garcia-Quintero}, C. and {Gazta{\~n}aga}, E. and {Gil-Mar{\'\i}n}, H. and {Gontcho}, S. Gontcho A and {Gonzalez-Morales}, A.~X. and {Gonzalez-Perez}, V. and {Gordon}, C. and {Green}, D. and {Gruen}, D. and {Gsponer}, R. and {Gutierrez}, G. and {Guy}, J. and {Hadzhiyska}, B. and {Hahn}, C. and {Hanif}, M.~M. S and {Herrera-Alcantar}, H.~K. and {Honscheid}, K. and {Howlett}, C. and {Huterer}, D. and {Ir{\v{s}}i{\v{c}}}, V. and {Ishak}, M. and {Juneau}, S. and {Kara{\c{c}}ayl{\i}}, N.~G. and {Kehoe}, R. and {Kent}, S. and {Kirkby}, D. and {Kremin}, A. and {Krolewski}, A. and {Lai}, Y. and {Lan}, T. -W. and {Landriau}, M. and {Lang}, D. and {Lasker}, J. and {Le Goff}, J.~M. and {Le Guillou}, L. and {Leauthaud}, A. and {Levi}, M.~E. and {Li}, T.~S. and {Linder}, E. and {Lodha}, K. and {Magneville}, C. and {Manera}, M. and {Margala}, D. and {Martini}, P. and {Maus}, M. and {McDonald}, P. and {Medina-Varela}, L. and {Meisner}, A. and {Mena-Fern{\'a}ndez}, J. and {Miquel}, R. and {Moon}, J. and {Moore}, S. and {Moustakas}, J. and {Mudur}, N. and {Mueller}, E. and {Mu{\~n}oz-Guti{\'e}rrez}, A. and {Myers}, A.~D. and {Nadathur}, S. and {Napolitano}, L. and {Neveux}, R. and {Newman}, J.~A. and {Nguyen}, N.~M. and {Nie}, J. and {Niz}, G. and {Noriega}, H.~E. and {Padmanabhan}, N. and {Paillas}, E. and {Palanque-Delabrouille}, N. and {Pan}, J. and {Penmetsa}, S. and {Percival}, W.~J. and {Pieri}, M.~M. and {Pinon}, M. and {Poppett}, C. and {Porredon}, A. and {Prada}, F. and {P{\'e}rez-Fern{\'a}ndez}, A. and {P{\'e}rez-R{\`a}fols}, I. and {Rabinowitz}, D. and {Raichoor}, A. and {Ram{\'\i}rez-P{\'e}rez}, C. and {Ramirez-Solano}, S. and {Ravoux}, C. and {Rashkovetskyi}, M. and {Rezaie}, M. and {Rich}, J. and {Rocher}, A. and {Rockosi}, C. and {Roe}, N.~A. and {Rosado-Marin}, A. and {Ross}, A.~J. and {Rossi}, G. and {Ruggeri}, R. and {Ruhlmann-Kleider}, V. and {Samushia}, L. and {Sanchez}, E. and {Saulder}, C. and {Schlafly}, E.~F. and {Schlegel}, D. and {Schubnell}, M. and {Seo}, H. and {Shafieloo}, A. and {Sharples}, R. and {Silber}, J. and {Slosar}, A. and {Smith}, A. and {Sprayberry}, D. and {Tan}, T. and {Tarl{\'e}}, G. and {Taylor}, P. and {Trusov}, S. and {Ure{\~n}a-L{\'o}pez}, L.~A. and {Vaisakh}, R. and {Valcin}, D. and {Valdes}, F. and {Vargas-Maga{\~n}a}, M. and {Verde}, L. and {Walther}, M. and {Wang}, B. and {Wang}, M.~S. and {Weaver}, B.~A. and {Weaverdyck}, N. and {Wechsler}, R.~H. and {Weinberg}, D.~H. and {White}, M. and {Yu}, J. and {Yu}, Y. and {Yuan}, S. and {Y{\`e}che}, C. and {Zaborowski}, E.~A. and {Zarrouk}, P. and {Zhang}, H. and {Zhao}, C. and {Zhao}, R. and {Zhou}, R. and {Zhuang}, T. and {Zou}, H.},
    title = "{DESI 2024 VI: Cosmological Constraints from the Measurements of Baryon Acoustic Oscillations}",
    journal = {arXiv e-prints},
    year = 2024,
    month = apr,
    eid = {arXiv:2404.03002},
    pages = {arXiv:2404.03002},
    doi = {10.48550/arXiv.2404.03002},
    archivePrefix = {arXiv},
    eprint = {2404.03002},
    primaryClass = {astro-ph.CO},
    adsurl = {https://ui.adsabs.harvard.edu/abs/2024arXiv240403002D}
}

@ARTICLE{desiinstrumentdesign,
       author = {{DESI Collaboration} and {Aghamousa}, Amir and {Aguilar}, Jessica and {Ahlen}, Steve and {Alam}, Shadab and {Allen}, Lori E. and {Allende Prieto}, Carlos and {Annis}, James and {Bailey}, Stephen and {Balland}, Christophe and {Ballester}, Otger and {Baltay}, Charles and {Beaufore}, Lucas and {Bebek}, Chris and {Beers}, Timothy C. and {Bell}, Eric F. and {Bernal}, Jos{\'e} Luis and {Besuner}, Robert and {Beutler}, Florian and {Blake}, Chris and {Bleuler}, Hannes and {Blomqvist}, Michael and {Blum}, Robert and {Bolton}, Adam S. and {Briceno}, Cesar and {Brooks}, David and {Brownstein}, Joel R. and {Buckley-Geer}, Elizabeth and {Burden}, Angela and {Burtin}, Etienne and {Busca}, Nicolas G. and {Cahn}, Robert N. and {Cai}, Yan-Chuan and {Cardiel-Sas}, Laia and {Carlberg}, Raymond G. and {Carton}, Pierre-Henri and {Casas}, Ricard and {Castander}, Francisco J. and {Cervantes-Cota}, Jorge L. and {Claybaugh}, Todd M. and {Close}, Madeline and {Coker}, Carl T. and {Cole}, Shaun and {Comparat}, Johan and {Cooper}, Andrew P. and {Cousinou}, M. -C. and {Crocce}, Martin and {Cuby}, Jean-Gabriel and {Cunningham}, Daniel P. and {Davis}, Tamara M. and {Dawson}, Kyle S. and {de la Macorra}, Axel and {De Vicente}, Juan and {Delubac}, Timoth{\'e}e and {Derwent}, Mark and {Dey}, Arjun and {Dhungana}, Govinda and {Ding}, Zhejie and {Doel}, Peter and {Duan}, Yutong T. and {Ealet}, Anne and {Edelstein}, Jerry and {Eftekharzadeh}, Sarah and {Eisenstein}, Daniel J. and {Elliott}, Ann and {Escoffier}, St{\'e}phanie and {Evatt}, Matthew and {Fagrelius}, Parker and {Fan}, Xiaohui and {Fanning}, Kevin and {Farahi}, Arya and {Farihi}, Jay and {Favole}, Ginevra and {Feng}, Yu and {Fernandez}, Enrique and {Findlay}, Joseph R. and {Finkbeiner}, Douglas P. and {Fitzpatrick}, Michael J. and {Flaugher}, Brenna and {Flender}, Samuel and {Font-Ribera}, Andreu and {Forero-Romero}, Jaime E. and {Fosalba}, Pablo and {Frenk}, Carlos S. and {Fumagalli}, Michele and {Gaensicke}, Boris T. and {Gallo}, Giuseppe and {Garcia-Bellido}, Juan and {Gaztanaga}, Enrique and {Pietro Gentile Fusillo}, Nicola and {Gerard}, Terry and {Gershkovich}, Irena and {Giannantonio}, Tommaso and {Gillet}, Denis and {Gonzalez-de-Rivera}, Guillermo and {Gonzalez-Perez}, Violeta and {Gott}, Shelby and {Graur}, Or and {Gutierrez}, Gaston and {Guy}, Julien and {Habib}, Salman and {Heetderks}, Henry and {Heetderks}, Ian and {Heitmann}, Katrin and {Hellwing}, Wojciech A. and {Herrera}, David A. and {Ho}, Shirley and {Holland}, Stephen and {Honscheid}, Klaus and {Huff}, Eric and {Hutchinson}, Timothy A. and {Huterer}, Dragan and {Hwang}, Ho Seong and {Illa Laguna}, Joseph Maria and {Ishikawa}, Yuzo and {Jacobs}, Dianna and {Jeffrey}, Niall and {Jelinsky}, Patrick and {Jennings}, Elise and {Jiang}, Linhua and {Jimenez}, Jorge and {Johnson}, Jennifer and {Joyce}, Richard and {Jullo}, Eric and {Juneau}, St{\'e}phanie and {Kama}, Sami and {Karcher}, Armin and {Karkar}, Sonia and {Kehoe}, Robert and {Kennamer}, Noble and {Kent}, Stephen and {Kilbinger}, Martin and {Kim}, Alex G. and {Kirkby}, David and {Kisner}, Theodore and {Kitanidis}, Ellie and {Kneib}, Jean-Paul and {Koposov}, Sergey and {Kovacs}, Eve and {Koyama}, Kazuya and {Kremin}, Anthony and {Kron}, Richard and {Kronig}, Luzius and {Kueter-Young}, Andrea and {Lacey}, Cedric G. and {Lafever}, Robin and {Lahav}, Ofer and {Lambert}, Andrew and {Lampton}, Michael and {Landriau}, Martin and {Lang}, Dustin and {Lauer}, Tod R. and {Le Goff}, Jean-Marc and {Le Guillou}, Laurent and {Le Van Suu}, Auguste and {Lee}, Jae Hyeon and {Lee}, Su-Jeong and {Leitner}, Daniela and {Lesser}, Michael and {Levi}, Michael E. and {L'Huillier}, Benjamin and {Li}, Baojiu and {Liang}, Ming and {Lin}, Huan and {Linder}, Eric and {Loebman}, Sarah R. and {Luki{\'c}}, Zarija and {Ma}, Jun and {MacCrann}, Niall and {Magneville}, Christophe and {Makarem}, Laleh and {Manera}, Marc and {Manser}, Christopher J. and {Marshall}, Robert and {Martini}, Paul and {Massey}, Richard and {Matheson}, Thomas and {McCauley}, Jeremy and {McDonald}, Patrick and {McGreer}, Ian D. and {Meisner}, Aaron and {Metcalfe}, Nigel and {Miller}, Timothy N. and {Miquel}, Ramon and {Moustakas}, John and {Myers}, Adam and {Naik}, Milind and {Newman}, Jeffrey A. and {Nichol}, Robert C. and {Nicola}, Andrina and {Nicolati da Costa}, Luiz and {Nie}, Jundan and {Niz}, Gustavo and {Norberg}, Peder and {Nord}, Brian and {Norman}, Dara and {Nugent}, Peter and {O'Brien}, Thomas and {Oh}, Minji and {Olsen}, Knut A.~G.},
        title = "{The DESI Experiment Part II: Instrument Design}",
      journal = {arXiv e-prints},
         year = 2016,
        month = oct,
          eid = {arXiv:1611.00037},
        pages = {arXiv:1611.00037},
          doi = {10.48550/arXiv.1611.00037},
archivePrefix = {arXiv},
       eprint = {1611.00037},
 primaryClass = {astro-ph.IM},
       adsurl = {https://ui.adsabs.harvard.edu/abs/2016arXiv161100037D}
}

@ARTICLE{desicorrector,
       author = {{Miller}, Timothy N. and {Doel}, Peter and {Gutierrez}, Gaston and {Besuner}, Robert and {Brooks}, David and {Gallo}, Giuseppe and {Heetderks}, Henry and {Jelinsky}, Patrick and {Kent}, Stephen M. and {Lampton}, Michael and {Levi}, Michael E. and {Liang}, Ming and {Meisner}, Aaron and {Sholl}, Michael J. and {Silber}, Joseph Harry and {Sprayberry}, David and {Aguilar}, Jessica Nicole and {de la Macorra}, Axel and {Eisenstein}, Daniel and {Fanning}, Kevin and {Font-Ribera}, Andreu and {Gazta{\~n}aga}, Enrique and {Gontcho A Gontcho}, Satya and {Honscheid}, Klaus and {Jimenez}, Jorge and {Joyce}, Dick and {Kehoe}, Robert and {Kisner}, Theodore and {Kremin}, Anthony and {Landriau}, Martin and {Le Guillou}, Laurent and {Magneville}, Christophe and {Martini}, Paul and {Miquel}, Ramon and {Moustakas}, John and {Nie}, Jundan and {Percival}, Will and {Poppett}, Claire and {Prada}, Francisco and {Rossi}, Graziano and {Schlegel}, David and {Schubnell}, Michael and {Seo}, Hee-Jong and {Sharples}, Ray and {Tarl{\'e}}, Gregory and {Vargas-Maga{\~n}a}, Mariana and {Zhou}, Zhimin and {the DESI Collaboration}},
        title = "{The Optical Corrector for the Dark Energy Spectroscopic Instrument}",
      journal = {\aj},
         year = 2024,
        month = aug,
       volume = {168},
       number = {2},
          eid = {95},
        pages = {95},
          doi = {10.3847/1538-3881/ad45fe},
archivePrefix = {arXiv},
       eprint = {2306.06310},
 primaryClass = {astro-ph.IM},
       adsurl = {https://ui.adsabs.harvard.edu/abs/2024AJ....168...95M}
}

@ARTICLE{desifibers,
       author = {{Poppett}, Claire and {Tyas}, Luke and {Aguilar}, J. and {Bebek}, Christopher and {Bramall}, D. and {Claybaugh}, T. and {Edelstein}, J. and {Fagrelius}, P. and {Heetderks}, H. and {Jelinsky}, P. and {Jelinsky}, S. and {Lafever}, Robin and {Lambert}, A. and {Lampton}, M. and {Levi}, Michael E. and {Martini}, P. and {Rockosi}, C. and {Schmoll}, J. and {Sharples}, Ray M. and {Sirk}, Martin and {Wishnow}, Edward and {Yu}, Jiaxi and {Ahlen}, S. and {Bault}, A. and {BenZvi}, S. and {Brooks}, D. and {Cole}, S. and {de la Macorra}, A. and {Dey}, Arjun and {Doel}, P. and {Fanning}, K. and {Font-Ribera}, A. and {Forero-Romero}, J.~E. and {Gazta{\~n}aga}, E. and {Gontcho A Gontcho}, S. and {Gonzalez-Morales}, A.~X. and {Hahn}, C. and {Honscheid}, K. and {Jimenez}, J. and {Juneau}, S. and {Kirkby}, D. and {Kremin}, A. and {Landriau}, M. and {Le Guillou}, L. and {Manera}, M. and {Meisner}, A. and {Miquel}, R. and {Moustakas}, J. and {Mueller}, E. and {Mu{\~n}oz-Guti{\'e}rrez}, A. and {Myers}, A.~D. and {Nie}, J. and {Niz}, G. and {Palanque-Delabrouille}, N. and {Percival}, W.~J. and {Prada}, F. and {Rabinowitz}, D. and {Rezaie}, M. and {Rossi}, G. and {Sanchez}, E. and {Schlafly}, Edward F. and {Schlegel}, D. and {Schubnell}, M. and {Seo}, H. and {Sprayberry}, D. and {Tarl{\'e}}, G. and {Vargas-Maga{\~n}a}, M. and {Weaver}, B.~A. and {Zhou}, R.},
        title = "{Overview of the Fiber System for the Dark Energy Spectroscopic Instrument}",
      journal = {\aj},
         year = 2024,
        month = dec,
       volume = {168},
       number = {6},
          eid = {245},
        pages = {245},
          doi = {10.3847/1538-3881/ad76a4},
       adsurl = {https://ui.adsabs.harvard.edu/abs/2024AJ....168..245P}
}

@ARTICLE{desioperations,
       author = {{Schlafly}, Edward F. and {Kirkby}, David and {Schlegel}, David J. and {Myers}, Adam D. and {Raichoor}, Anand and {Dawson}, Kyle and {Aguilar}, Jessica and {Allende Prieto}, Carlos and {Bailey}, Stephen and {BenZvi}, Segev and {Bermejo-Climent}, Jose and {Brooks}, David and {de la Macorra}, Axel and {Dey}, Arjun and {Doel}, Peter and {Fanning}, Kevin and {Font-Ribera}, Andreu and {Forero-Romero}, Jaime E. and {Garc{\'\i}a-Bellido}, Juan and {Gontcho A Gontcho}, Satya and {Guy}, Julien and {Hahn}, ChangHoon and {Honscheid}, Klaus and {Ishak}, Mustapha and {Juneau}, St{\'e}phanie and {Kehoe}, Robert and {Kisner}, Theodore and {Kremin}, Anthony and {Landriau}, Martin and {Lang}, Dustin A. and {Lasker}, James and {Levi}, Michael E. and {Magneville}, Christophe and {Manser}, Christopher J. and {Martini}, Paul and {Meisner}, Aaron M. and {Miquel}, Ramon and {Moustakas}, John and {Newman}, Jeffrey A. and {Nie}, Jundan and {Palanque-Delabrouille}, Nathalie. and {Percival}, Will J. and {Poppett}, Claire and {Rockosi}, Constance and {Ross}, Ashley J. and {Rossi}, Graziano and {Tarl{\'e}}, Gregory and {Weaver}, Benjamin A. and {Y{\`e}che}, Christophe and {Zhou}, Rongpu and {DESI Collaboration}},
        title = "{Survey Operations for the Dark Energy Spectroscopic Instrument}",
      journal = {\aj},
         year = 2023,
        month = dec,
       volume = {166},
       number = {6},
          eid = {259},
        pages = {259},
          doi = {10.3847/1538-3881/ad0832},
archivePrefix = {arXiv},
       eprint = {2306.06309},
 primaryClass = {astro-ph.CO},
       adsurl = {https://ui.adsabs.harvard.edu/abs/2023AJ....166..259S}
}

@ARTICLE{guy2023,
       author = {{Guy}, J. and {Bailey}, S. and {Kremin}, A. and {Alam}, Shadab and {Alexander}, D.~M. and {Allende Prieto}, C. and {BenZvi}, S. and {Bolton}, A.~S. and {Brooks}, D. and {Chaussidon}, E. and {Cooper}, A.~P. and {Dawson}, K. and {de la Macorra}, A. and {Dey}, A. and {Dey}, Biprateep and {Dhungana}, G. and {Eisenstein}, D.~J. and {Font-Ribera}, A. and {Forero-Romero}, J.~E. and {Gazta{\~n}aga}, E. and {Gontcho A Gontcho}, S. and {Green}, D. and {Honscheid}, K. and {Ishak}, M. and {Kehoe}, R. and {Kirkby}, D. and {Kisner}, T. and {Koposov}, Sergey E. and {Lan}, Ting-Wen and {Landriau}, M. and {Le Guillou}, L. and {Levi}, Michael E. and {Magneville}, C. and {Manser}, Christopher J. and {Martini}, P. and {Meisner}, Aaron M. and {Miquel}, R. and {Moustakas}, J. and {Myers}, Adam D. and {Newman}, Jeffrey A. and {Nie}, Jundan and {Palanque-Delabrouille}, N. and {Percival}, W.~J. and {Poppett}, C. and {Prada}, F. and {Raichoor}, A. and {Ravoux}, C. and {Ross}, A.~J. and {Schlafly}, E.~F. and {Schlegel}, D. and {Schubnell}, M. and {Sharples}, Ray M. and {Tarl{\'e}}, Gregory and {Weaver}, B.~A. and {Y{\'e}che}, Christophe and {Zhou}, Rongpu and {Zhou}, Zhimin and {Zou}, H.},
        title = "{The Spectroscopic Data Processing Pipeline for the Dark Energy Spectroscopic Instrument}",
      journal = {\aj},
         year = 2023,
        month = apr,
       volume = {165},
       number = {4},
          eid = {144},
        pages = {144},
          doi = {10.3847/1538-3881/acb212},
archivePrefix = {arXiv},
       eprint = {2209.14482},
 primaryClass = {astro-ph.IM},
       adsurl = {https://ui.adsabs.harvard.edu/abs/2023AJ....165..144G}
}

@ARTICLE{desiDR1,
       author = {{DESI Collaboration} and {Abdul-Karim}, M. and {Adame}, A.~G. and {Aguado}, D. and {Aguilar}, J. and {Ahlen}, S. and {Alam}, S. and {Aldering}, G. and {Alexander}, D.~M. and {Alfarsy}, R. and {Allen}, L. and {Allende Prieto}, C. and {Alves}, O. and {Anand}, A. and {Andrade}, U. and {Armengaud}, E. and {Avila}, S. and {Aviles}, A. and {Awan}, H. and {Bailey}, S. and {Baleato Lizancos}, A. and {Ballester}, O. and {Bault}, A. and {Bautista}, J. and {BenZvi}, S. and {Beraldo e Silva}, L. and {Bermejo-Climent}, J.~R. and {Beutler}, F. and {Bianchi}, D. and {Blake}, C. and {Blum}, R. and {Bolton}, A.~S. and {Bonici}, M. and {Brieden}, S. and {Brodzeller}, A. and {Brooks}, D. and {Buckley-Geer}, E. and {Burtin}, E. and {Canning}, R. and {Carnero Rosell}, A. and {Carr}, A. and {Carrilho}, P. and {Casas}, L. and {Castander}, F.~J. and {Cereskaite}, R. and {Cervantes-Cota}, J.~L. and {Chaussidon}, E. and {Chaves-Montero}, J. and {Chen}, S. and {Chen}, X. and {Claybaugh}, T. and {Cole}, S. and {Cooper}, A.~P. and {Cousinou}, M. -C. and {Cuceu}, A. and {Davis}, T.~M. and {Dawson}, K.~S. and {de Belsunce}, R. and {de la Cruz}, R. and {de la Macorra}, A. and {de Mattia}, A. and {Deiosso}, N. and {Della Costa}, J. and {Demina}, R. and {Demirbozan}, U. and {DeRose}, J. and {Dey}, A. and {Dey}, B. and {Ding}, J. and {Ding}, Z. and {Doel}, P. and {Douglass}, K. and {Dowicz}, M. and {Ebina}, H. and {Edelstein}, J. and {Eisenstein}, D.~J. and {Elbers}, W. and {Emas}, N. and {Escoffier}, S. and {Fagrelius}, P. and {Fan}, X. and {Fanning}, K. and {Fawcett}, V.~A. and {Fern\textbackslash'andez-Garc\textbackslash'ia}, E. and {Ferraro}, S. and {Findlay}, N. and {Font-Ribera}, A. and {Forero-Romero}, J.~E. and {Forero-S\textbackslash'anchez}, D. and {Frenk}, C.~S. and {G\textbackslash''ansicke}, B.~T. and {Galbany}, L. and {Garc\textbackslash'ia-Bellido}, J. and {Garcia-Quintero}, C. and {Garrison}, L.~H. and {Gazta\textbackslash\raisebox{-0.5ex}\textasciitildenaga}, E. and {Gil-Mar\textbackslash'in}, H. and {Gnedin}, O.~Y. and {Gontcho}, S. Gontcho A and {Gonzalez-Morales}, A.~X. and {Gonzalez-Perez}, V. and {Gordon}, C. and {Graur}, O. and {Green}, D. and {Gruen}, D. and {Gsponer}, R. and {Guandalin}, C. and {Gutierrez}, G. and {Guy}, J. and {Hahn}, C. and {Han}, J.~J. and {Han}, J. and {He}, S. and {Herrera-Alcantar}, H.~K. and {Honscheid}, K. and {Hou}, J. and {Howlett}, C. and {Huterer}, D. and {Ir\textbackslashv\{s\}i\textbackslashv\{c\}}, V. and {Ishak}, M. and {Jacques}, A. and {Jimenez}, J. and {Jing}, Y.~P. and {Joachimi}, B. and {Joudaki}, S. and {Joyce}, R. and {Jullo}, E. and {Juneau}, S. and {Kara\textbackslashc\{c\}ayl\{\textbackslashi\}}, N.~G. and {Karim}, T. and {Kehoe}, R. and {Kent}, S. and {Khederlarian}, A. and {Kirkby}, D. and {Kisner}, T. and {Kitaura}, F. -S. and {Kizhuprakkat}, N. and {Kong}, H. and {Koposov}, S.~E. and {Kremin}, A. and {Krolewski}, A. and {Lahav}, O. and {Lai}, Y. and {Lamman}, C. and {Lan}, T. -W. and {Landriau}, M. and {Lang}, D. and {Lange}, J.~U. and {Lasker}, J. and {Le Goff}, J.~M. and {Le Guillou}, L. and {Leauthaud}, A. and {Levi}, M.~E. and {Li}, S. and {Li}, T.~S. and {Lodha}, K. and {Lokken}, M. and {Luo}, Y. and {Magneville}, C. and {Manera}, M. and {Manser}, C.~J. and {Margala}, D. and {Martini}, P. and {Maus}, M. and {McCullough}, J. and {McDonald}, P. and {Medina}, G.~E. and {Medina-Varela}, L. and {Meisner}, A. and {Mena-Fern\textbackslash'andez}, J. and {Menegas}, A. and {Mezcua}, M. and {Miquel}, R. and {Montero-Camacho}, P. and {Moon}, J. and {Moustakas}, J. and {Mu\textbackslash\raisebox{-0.5ex}\textasciitildenoz-Guti\textbackslash'errez}, A. and {Mu\textbackslash\raisebox{-0.5ex}\textasciitildenoz-Santos}, D. and {Myers}, A.~D. and {Myles}, J. and {Nadathur}, S. and {Najita}, J. and {Napolitano}, L. and {Newman}, J.~A. and {Nikakhtar}, F. and {Nikutta}, R. and {Niz}, G. and {Noriega}, H.~E. and {Padmanabhan}, N. and {Paillas}, E. and {Palanque-Delabrouille}, N. and {Palmese}, A. and {Pan}, J. and {Pan}, Z. and {Parkinson}, D. and {Peacock}, J. and {Percival}, W.~J. and {P\textbackslash'erez-Fern\textbackslash'andez}, A. and {P\textbackslash'erez-R\textbackslash`afols}, I. and {Peterson}, P.},
        title = "{Data Release 1 of the Dark Energy Spectroscopic Instrument}",
      journal = {arXiv e-prints},
         year = 2025,
        month = mar,
          eid = {arXiv:2503.14745},
        pages = {arXiv:2503.14745},
          doi = {10.48550/arXiv.2503.14745},
archivePrefix = {arXiv},
       eprint = {2503.14745},
 primaryClass = {astro-ph.CO},
       adsurl = {https://ui.adsabs.harvard.edu/abs/2025arXiv250314745D}
}

@ARTICLE{desi2025VII,
       author = {{Adame}, A.~G. and {Aguilar}, J. and {Ahlen}, S. and {Alam}, S. and {Alexander}, D.~M. and {Allende Prieto}, C. and {Alvarez}, M. and {Alves}, O. and {Anand}, A. and {Andrade}, U. and {Armengaud}, E. and {Avila}, S. and {Aviles}, A. and {Awan}, H. and {Bahr-Kalus}, B. and {Bailey}, S. and {Baltay}, C. and {Bault}, A. and {Behera}, J. and {BenZvi}, S. and {Beutler}, F. and {Bianchi}, D. and {Blake}, C. and {Blum}, R. and {Bonici}, M. and {Brieden}, S. and {Brodzeller}, A. and {Brooks}, D. and {Buckley-Geer}, E. and {Burtin}, E. and {Calderon}, R. and {Canning}, R. and {Carnero Rosell}, A. and {Cereskaite}, R. and {Cervantes-Cota}, J.~L. and {Chabanier}, S. and {Chaussidon}, E. and {Chaves-Montero}, J. and {Chebat}, D. and {Chen}, S. and {Chen}, X. and {Claybaugh}, T. and {Cole}, S. and {Cuceu}, A. and {Davis}, T.~M. and {Dawson}, K. and {de la Macorra}, A. and {de Mattia}, A. and {Deiosso}, N. and {Dey}, A. and {Dey}, B. and {Ding}, Z. and {Doel}, P. and {Edelstein}, J. and {Eftekharzadeh}, S. and {Eisenstein}, D.~J. and {Elbers}, W. and {Elliott}, A. and {Fagrelius}, P. and {Fanning}, K. and {Ferraro}, S. and {Ereza}, J. and {Findlay}, N. and {Flaugher}, B. and {Font-Ribera}, A. and {Forero-S{\'a}nchez}, D. and {Forero-Romero}, J.~E. and {Frenk}, C.~S. and {Garcia-Quintero}, C. and {Garrison}, L.~H. and {Gazta{\~n}aga}, E. and {Gil-Mar{\'\i}n}, H. and {Gontcho}, S. Gontcho A. and {Gonzalez-Morales}, A.~X. and {Gonzalez-Perez}, V. and {Gordon}, C. and {Green}, D. and {Gruen}, D. and {Gsponer}, R. and {Gutierrez}, G. and {Guy}, J. and {Hadzhiyska}, B. and {Hahn}, C. and {Hanif}, M.~M.~S. and {Herrera-Alcantar}, H.~K. and {Honscheid}, K. and {Howlett}, C. and {Huterer}, D. and {Ir{\v{s}}i{\v{c}}}, V. and {Ishak}, M. and {Joyce}, R. and {Juneau}, S. and {Kara{\c{c}}ayl{\i}}, N.~G. and {Kehoe}, R. and {Kent}, S. and {Kirkby}, D. and {Kong}, H. and {Koposov}, S.~E. and {Kremin}, A. and {Krolewski}, A. and {Lahav}, O. and {Lai}, Y. and {Lan}, T. -W. and {Landriau}, M. and {Lang}, D. and {Lasker}, J. and {Le Goff}, J.~M. and {Le Guillou}, L. and {Leauthaud}, A. and {Levi}, M.~E. and {Li}, T.~S. and {Lodha}, K. and {Magneville}, C. and {Manera}, M. and {Margala}, D. and {Martini}, P. and {Matthewson}, W. and {Maus}, M. and {McDonald}, P. and {Medina-Varela}, L. and {Meisner}, A. and {Mena-Fern{\'a}ndez}, J. and {Miquel}, R. and {Moon}, J. and {Moore}, S. and {Moustakas}, J. and {Mudur}, N. and {Mueller}, E. and {Mu{\~n}oz-Guti{\'e}rrez}, A. and {Myers}, A.~D. and {Nadathur}, S. and {Napolitano}, L. and {Neveux}, R. and {Newman}, J.~A. and {Nguyen}, N.~M. and {Nie}, J. and {Niz}, G. and {Noriega}, H.~E. and {Padmanabhan}, N. and {Paillas}, E. and {Palanque-Delabrouille}, N. and {Pan}, J. and {Penmetsa}, S. and {Percival}, W.~J. and {Pieri}, M.~M. and {Pinon}, M. and {Poppett}, C. and {Porredon}, A. and {Prada}, F. and {P{\'e}rez-Fern{\'a}ndez}, A. and {P{\'e}rez-R{\`a}fols}, I. and {Rabinowitz}, D. and {Raichoor}, A. and {Ram{\'\i}rez-P{\'e}rez}, C. and {Ramirez-Solano}, S. and {Rashkovetskyi}, M. and {Ravoux}, C. and {Rezaie}, M. and {Rich}, J. and {Rocher}, A. and {Rockosi}, C. and {Roe}, N.~A. and {Rosado-Marin}, A. and {Ross}, A.~J. and {Rossi}, G. and {Ruggeri}, R. and {Ruhlmann-Kleider}, V. and {Samushia}, L. and {Sanchez}, E. and {Saulder}, C. and {Schlafly}, E.~F. and {Schlegel}, D. and {Schubnell}, M. and {Seo}, H. and {Shafieloo}, A. and {Sharples}, R. and {Silber}, J. and {Slosar}, A. and {Smith}, A. and {Sprayberry}, D. and {Tan}, T. and {Tarl{\'e}}, G. and {Taylor}, P. and {Trusov}, S. and {Vaisakh}, R. and {Valcin}, D. and {Valdes}, F. and {Valogiannis}, G. and {Vargas-Maga{\~n}a}, M. and {Verde}, L. and {Walther}, M. and {Wang}, B. and {Wang}, M.~S. and {Weaver}, B.~A. and {Weaverdyck}, N. and {Wechsler}, R.~H. and {Weinberg}, D.~H. and {White}, M. and {Wilson}, M.~J. and {Yi}, L.},
        title = "{DESI 2024 VII: cosmological constraints from the full-shape modeling of clustering measurements}",
      journal = {\jcap},
         year = 2025,
        month = jul,
       volume = {2025},
       number = {7},
          eid = {028},
        pages = {028},
          doi = {10.1088/1475-7516/2025/07/028},
archivePrefix = {arXiv},
       eprint = {2411.12022},
 primaryClass = {astro-ph.CO},
       adsurl = {https://ui.adsabs.harvard.edu/abs/2025JCAP...07..028A}
}

@ARTICLE{desisurveyval,
       author = {{Alexander}, David M. and {Davis}, Tamara M. and {Chaussidon}, E. and {Fawcett}, V.~A. and {X. Gonzalez-Morales}, Alma and {Lan}, Ting-Wen and {Y{\`e}che}, Christophe and {Ahlen}, S. and {Aguilar}, J.~N. and {Armengaud}, E. and {Bailey}, S. and {Brooks}, D. and {Cai}, Z. and {Canning}, R. and {Carr}, A. and {Chabanier}, S. and {Cousinou}, Marie-Claude and {Dawson}, K. and {de la Macorra}, A. and {Dey}, A. and {Dey}, Biprateep and {Dhungana}, G. and {Edge}, A.~C. and {Eftekharzadeh}, S. and {Fanning}, K. and {Farr}, James and {Font-Ribera}, A. and {Garcia-Bellido}, J. and {Garrison}, Lehman and {Gazta{\~n}aga}, E. and {A Gontcho}, Satya Gontcho and {Gordon}, C. and {Medellin Gonzalez}, Stefany Guadalupe and {Guy}, J. and {Herrera-Alcantar}, Hiram K. and {Jiang}, L. and {Juneau}, S. and {Kara{\c{c}}ayl{\i}}, N.~G. and {Kehoe}, R. and {Kisner}, T. and {Kov{\'a}cs}, A. and {Landriau}, M. and {Levi}, Michael E. and {Magneville}, C. and {Martini}, P. and {Meisner}, Aaron M. and {Mezcua}, M. and {Miquel}, R. and {Camacho}, P. Montero and {Moustakas}, J. and {Mu{\~n}oz-Guti{\'e}rrez}, Andrea and {Myers}, Adam D. and {Nadathur}, S. and {Napolitano}, L. and {Nie}, J.~D. and {Palanque-Delabrouille}, N. and {Pan}, Z. and {Percival}, W.~J. and {P{\'e}rez-R{\`a}fols}, I. and {Poppett}, C. and {Prada}, F. and {Ram{\'\i}rez-P{\'e}rez}, C{\'e}sar and {Ravoux}, C. and {Rosario}, D.~J. and {Schubnell}, M. and {Tarl{\'e}}, Gregory and {Walther}, M. and {Weiner}, B. and {Youles}, S. and {Zhou}, Zhimin and {Zou}, H. and {Zou}, Siwei},
        title = "{The DESI Survey Validation: Results from Visual Inspection of the Quasar Survey Spectra}",
      journal = {\aj},
         year = 2023,
        month = mar,
       volume = {165},
       number = {3},
          eid = {124},
        pages = {124},
          doi = {10.3847/1538-3881/acacfc},
archivePrefix = {arXiv},
       eprint = {2208.08517},
 primaryClass = {astro-ph.GA},
       adsurl = {https://ui.adsabs.harvard.edu/abs/2023AJ....165..124A}
}

@ARTICLE{colossus2018,
       author = {{Diemer}, Benedikt},
        title = "{COLOSSUS: A Python Toolkit for Cosmology, Large-scale Structure, and Dark Matter Halos}",
      journal = {\apjs},
         year = 2018,
        month = dec,
       volume = {239},
       number = {2},
          eid = {35},
        pages = {35},
          doi = {10.3847/1538-4365/aaee8c},
archivePrefix = {arXiv},
       eprint = {1712.04512},
 primaryClass = {astro-ph.CO},
       adsurl = {https://ui.adsabs.harvard.edu/abs/2018ApJS..239...35D}
}

@article{alam2020,
   title={Multitracer extension of the halo model: probing quenching and conformity in eBOSS},
   volume={497},
   ISSN={1365-2966},
   url={http://dx.doi.org/10.1093/mnras/staa1956},
   DOI={10.1093/mnras/staa1956},
   number={1},
   journal={Monthly Notices of the Royal Astronomical Society},
   publisher={Oxford University Press (OUP)},
   author={Alam, Shadab and Peacock, John A and Kraljic, Katarina and Ross, Ashley J and Comparat, Johan},
   year={2020},
   month=jul, pages={581–595} }

@article{richardson2012,
   title={THE HALO OCCUPATION DISTRIBUTION OF SDSS QUASARS},
   volume={755},
   ISSN={1538-4357},
   url={http://dx.doi.org/10.1088/0004-637X/755/1/30},
   DOI={10.1088/0004-637x/755/1/30},
   number={1},
   journal={The Astrophysical Journal},
   publisher={American Astronomical Society},
   author={Richardson, Jonathan and Zheng, Zheng and Chatterjee, Suchetana and Nagai, Daisuke and Shen, Yue},
   year={2012},
   month=jul, pages={30} }

@article{shen2010,
   title={BINARY QUASARS AT HIGH REDSHIFT. II. SUB-Mpc CLUSTERING AT $z \sim 3-4$},
   volume={719},
   ISSN={1538-4357},
   url={http://dx.doi.org/10.1088/0004-637X/719/2/1693},
   DOI={10.1088/0004-637x/719/2/1693},
   number={2},
   journal={The Astrophysical Journal},
   publisher={American Astronomical Society},
   author={Shen, Yue and Hennawi, Joseph F. and Shankar, Francesco and Myers, Adam D. and Strauss, Michael A. and Djorgovski, S. G. and Fan, Xiaohui and Giocoli, Carlo and Mahabal, Ashish and Schneider, Donald P. and Weinberg, David H.},
   year={2010},
   month=aug, pages={1693–1698} }

@ARTICLE{yuan2024,
       author = {{Yuan}, Sihan and {Zhang}, Hanyu and {Ross}, Ashley J. and {Donald-McCann}, Jamie and {Hadzhiyska}, Boryana and {Wechsler}, Risa H. and {Zheng}, Zheng and {Alam}, Shadab and {Gonzalez-Perez}, Violeta and {Aguilar}, Jessica Nicole and {Ahlen}, Steven and {Bianchi}, Davide and {Brooks}, David and {de la Macorra}, Axel and {Fanning}, Kevin and {Forero-Romero}, Jaime E. and {Honscheid}, Klaus and {Ishak}, Mustapha and {Kehoe}, Robert and {Lasker}, James and {Landriau}, Martin and {Manera}, Marc and {Martini}, Paul and {Meisner}, Aaron and {Miquel}, Ramon and {Moustakas}, John and {Nadathur}, Seshadri and {Newman}, Jeffrey A. and {Nie}, Jundan and {Percival}, Will and {Poppett}, Claire and {Rocher}, Antoine and {Rossi}, Graziano and {Sanchez}, Eusebio and {Samushia}, Lado and {Schubnell}, Michael and {Seo}, Hee-Jong and {Tarl{\'e}}, Gregory and {Weaver}, Benjamin Alan and {Yu}, Jiaxi and {Zhou}, Zhimin and {Zou}, Hu},
        title = "{The DESI one-per cent survey: exploring the halo occupation distribution of luminous red galaxies and quasi-stellar objects with ABACUSSUMMIT}",
      journal = {\mnras},
         year = 2024,
        month = may,
       volume = {530},
       number = {1},
        pages = {947-965},
          doi = {10.1093/mnras/stae359},
archivePrefix = {arXiv},
       eprint = {2306.06314},
 primaryClass = {astro-ph.CO},
       adsurl = {https://ui.adsabs.harvard.edu/abs/2024MNRAS.530..947Y}
}

@ARTICLE{bault2025,
       author = {{Bault}, Abby and {Kirkby}, David and {Guy}, Julien and {Brodzeller}, Allyson and {Aguilar}, J. and {Ahlen}, S. and {Bailey}, S. and {Brooks}, D. and {Cabayol-Garcia}, L. and {Chaves-Montero}, J. and {Claybaugh}, T. and {Cuceu}, A. and {Dawson}, K. and {de la Cruz}, R. and {de la Macorra}, A. and {Dey}, A. and {Doel}, P. and {Filbert}, S. and {Font-Ribera}, A. and {Forero-Romero}, J.~E. and {Gazta{\~n}aga}, E. and {Gontcho A Gontcho}, S. and {Gordon}, C. and {Herrera-Alcantar}, H.~K. and {Honscheid}, K. and {Ir{\v{s}}i{\v{c}}}, V. and {Kara{\c{c}}ayl{\i}}, N.~G. and {Kehoe}, R. and {Kisner}, T. and {Kremin}, A. and {Lambert}, A. and {Landriau}, M. and {Le Guillou}, L. and {Levi}, M.~E. and {Manera}, M. and {Martini}, P. and {Meisner}, A. and {Miquel}, R. and {Montero-Camacho}, P. and {Moustakas}, J. and {Mu{\~n}oz-Guti{\'e}rrez}, A. and {Nie}, J. and {Niz}, G. and {Palanque-Delabrouille}, N. and {Percival}, W.~J. and {P{\'e}rez-R{\`a}fols}, I. and {Poppett}, C. and {Prada}, F. and {Ram{\'\i}rez-P{\'e}rez}, C. and {Ravoux}, C. and {Rezaie}, M. and {Rossi}, G. and {Sanchez}, E. and {Schlafly}, E.~F. and {Schlegel}, D. and {Schubnell}, M. and {Silber}, J. and {Tan}, T. and {Tarl{\'e}}, G. and {Walther}, M. and {Weaver}, B.~A. and {Zhou}, Z.},
        title = "{Impact of systematic redshift errors on the cross-correlation of the Lyman-{\ensuremath{\alpha}} forest with quasars at small scales using DESI Early Data}",
      journal = {\jcap},
         year = 2025,
        month = jan,
       volume = {2025},
       number = {1},
          eid = {130},
        pages = {130},
          doi = {10.1088/1475-7516/2025/01/130},
archivePrefix = {arXiv},
       eprint = {2402.18009},
 primaryClass = {astro-ph.CO},
       adsurl = {https://ui.adsabs.harvard.edu/abs/2025JCAP...01..130B}
}

@ARTICLE{wyithe2003,
       author = {{Wyithe}, J. Stuart B. and {Loeb}, Abraham},
        title = "{Self-regulated Growth of Supermassive Black Holes in Galaxies as the Origin of the Optical and X-Ray Luminosity Functions of Quasars}",
      journal = {\apj},
         year = 2003,
        month = oct,
       volume = {595},
       number = {2},
        pages = {614-623},
          doi = {10.1086/377475},
archivePrefix = {arXiv},
       eprint = {astro-ph/0304156},
 primaryClass = {astro-ph},
       adsurl = {https://ui.adsabs.harvard.edu/abs/2003ApJ...595..614W}
}

@ARTICLE{osmer1981,
       author = {{Osmer}, P.~S.},
        title = "{The three-dimensional distribution of quasars in the CTIO surveys}",
      journal = {\apj},
         year = 1981,
        month = aug,
       volume = {247},
        pages = {762-773},
          doi = {10.1086/159087},
       adsurl = {https://ui.adsabs.harvard.edu/abs/1981ApJ...247..762O}
}

@misc{busca2018,
      title={QuasarNET: Human-level spectral classification and redshifting with Deep Neural Networks}, 
      author={Nicolas Busca and Christophe Balland},
      year={2018},
      eprint={1808.09955},
      archivePrefix={arXiv},
      primaryClass={astro-ph.IM},
      url={https://arxiv.org/abs/1808.09955}, 
}

@article{farr2020,
   title={Optimal strategies for identifying quasars in DESI},
   volume={2020},
   ISSN={1475-7516},
   url={http://dx.doi.org/10.1088/1475-7516/2020/11/015},
   DOI={10.1088/1475-7516/2020/11/015},
   number={11},
   journal={Journal of Cosmology and Astroparticle Physics},
   publisher={IOP Publishing},
   author={Farr, James and Font-Ribera, Andreu and Pontzen, Andrew},
   year={2020},
   month=nov, pages={015–015} }

@ARTICLE{bianchi2017,
       author = {{Bianchi}, Davide and {Percival}, Will J.},
        title = "{Unbiased clustering estimation in the presence of missing observations}",
      journal = {\mnras},
         year = 2017,
        month = nov,
       volume = {472},
       number = {1},
        pages = {1106-1118},
          doi = {10.1093/mnras/stx2053},
archivePrefix = {arXiv},
       eprint = {1703.02070},
 primaryClass = {astro-ph.CO},
       adsurl = {https://ui.adsabs.harvard.edu/abs/2017MNRAS.472.1106B}
}

@ARTICLE{haring2004,
       author = {{H{\"a}ring}, Nadine and {Rix}, Hans-Walter},
        title = "{On the Black Hole Mass-Bulge Mass Relation}",
      journal = {\apjl},
         year = 2004,
        month = apr,
       volume = {604},
       number = {2},
        pages = {L89-L92},
          doi = {10.1086/383567},
archivePrefix = {arXiv},
       eprint = {astro-ph/0402376},
 primaryClass = {astro-ph},
       adsurl = {https://ui.adsabs.harvard.edu/abs/2004ApJ...604L..89H}
}

@ARTICLE{kauffmann2000,
       author = {{Kauffmann}, Guinevere and {Haehnelt}, Martin},
        title = "{A unified model for the evolution of galaxies and quasars}",
      journal = {\mnras},
         year = 2000,
        month = jan,
       volume = {311},
       number = {3},
        pages = {576-588},
          doi = {10.1046/j.1365-8711.2000.03077.x},
archivePrefix = {arXiv},
       eprint = {astro-ph/9906493},
 primaryClass = {astro-ph},
       adsurl = {https://ui.adsabs.harvard.edu/abs/2000MNRAS.311..576K}
}

@ARTICLE{mo1996,
       author = {{Mo}, H.~J. and {White}, S.~D.~M.},
        title = "{An analytic model for the spatial clustering of dark matter haloes}",
      journal = {\mnras},
         year = 1996,
        month = sep,
       volume = {282},
       number = {2},
        pages = {347-361},
          doi = {10.1093/mnras/282.2.347},
archivePrefix = {arXiv},
       eprint = {astro-ph/9512127},
 primaryClass = {astro-ph},
       adsurl = {https://ui.adsabs.harvard.edu/abs/1996MNRAS.282..347M}
}

@ARTICLE{kaiser1984,
       author = {{Kaiser}, N.},
        title = "{On the spatial correlations of Abell clusters.}",
      journal = {\apjl},
         year = 1984,
        month = sep,
       volume = {284},
        pages = {L9-L12},
          doi = {10.1086/184341},
       adsurl = {https://ui.adsabs.harvard.edu/abs/1984ApJ...284L...9K}
}

@ARTICLE{tinker2017,
       author = {{Tinker}, Jeremy L. and {Brownstein}, Joel R. and {Guo}, Hong and {Leauthaud}, Alexie and {Maraston}, Claudia and {Masters}, Karen and {Montero-Dorta}, Antonio D. and {Thomas}, Daniel and {Tojeiro}, Rita and {Weiner}, Benjamin and {Zehavi}, Idit and {Olmstead}, Matthew D.},
        title = "{The Correlation between Halo Mass and Stellar Mass for the Most Massive Galaxies in the Universe}",
      journal = {\apj},
         year = 2017,
        month = apr,
       volume = {839},
       number = {2},
          eid = {121},
        pages = {121},
          doi = {10.3847/1538-4357/aa6845},
archivePrefix = {arXiv},
       eprint = {1607.04678},
 primaryClass = {astro-ph.GA},
       adsurl = {https://ui.adsabs.harvard.edu/abs/2017ApJ...839..121T}
}

@ARTICLE{behroozi2010,
       author = {{Behroozi}, Peter S. and {Conroy}, Charlie and {Wechsler}, Risa H.},
        title = "{A Comprehensive Analysis of Uncertainties Affecting the Stellar Mass-Halo Mass Relation for 0 < z < 4}",
      journal = {\apj},
         year = 2010,
        month = jul,
       volume = {717},
       number = {1},
        pages = {379-403},
          doi = {10.1088/0004-637X/717/1/379},
archivePrefix = {arXiv},
       eprint = {1001.0015},
 primaryClass = {astro-ph.CO},
       adsurl = {https://ui.adsabs.harvard.edu/abs/2010ApJ...717..379B}
}

@ARTICLE{kollmeier2006,
       author = {{Kollmeier}, Juna A. and {Onken}, Christopher A. and {Kochanek}, Christopher S. and {Gould}, Andrew and {Weinberg}, David H. and {Dietrich}, Matthias and {Cool}, Richard and {Dey}, Arjun and {Eisenstein}, Daniel J. and {Jannuzi}, Buell T. and {Le Floc'h}, Emeric and {Stern}, Daniel},
        title = "{Black Hole Masses and Eddington Ratios at 0.3 < z < 4}",
      journal = {\apj},
         year = 2006,
        month = sep,
       volume = {648},
       number = {1},
        pages = {128-139},
          doi = {10.1086/505646},
archivePrefix = {arXiv},
       eprint = {astro-ph/0508657},
 primaryClass = {astro-ph},
       adsurl = {https://ui.adsabs.harvard.edu/abs/2006ApJ...648..128K}
}

@ARTICLE{arita+2023,
       author = {{Arita}, Junya and {Kashikawa}, Nobunari and {Matsuoka}, Yoshiki and {He}, Wanqiu and {Ito}, Kei and {Liang}, Yongming and {Ishimoto}, Rikako and {Yoshioka}, Takehiro and {Takeda}, Yoshihiro and {Iwasawa}, Kazushi and {Onoue}, Masafusa and {Toba}, Yoshiki and {Imanishi}, Masatoshi},
        title = "{Subaru High-z Exploration of Low-luminosity Quasars (SHELLQs). XVIII. The Dark Matter Halo Mass of Quasars at z   6}",
      journal = {\apj},
         year = 2023,
        month = sep,
       volume = {954},
       number = {2},
          eid = {210},
        pages = {210},
          doi = {10.3847/1538-4357/ace43a},
archivePrefix = {arXiv},
       eprint = {2307.02531},
 primaryClass = {astro-ph.GA},
       adsurl = {https://ui.adsabs.harvard.edu/abs/2023ApJ...954..210A}
}

@ARTICLE{lasker+2025,
       author = {{Lasker}, J. and {Carnero Rosell}, A. and {Myers}, A.~D. and {Ross}, A.~J. and {Bianchi}, D. and {Hanif}, M.~M.~S. and {Kehoe}, R. and {de Mattia}, A. and {Napolitano}, L. and {Percival}, W.~J. and {Staten}, R. and {Aguilar}, J. and {Ahlen}, S. and {Bigwood}, L. and {Brooks}, D. and {Claybaugh}, T. and {Cole}, S. and {de la Macorra}, A. and {Ding}, Z. and {Doel}, P. and {Fanning}, K. and {Forero-Romero}, J.~E. and {Gazta{\~n}aga}, E. and {Gontcho A Gontcho}, S. and {Gutierrez}, G. and {Honscheid}, K. and {Howlett}, C. and {Juneau}, S. and {Kremin}, A. and {Landriau}, M. and {Le Guillou}, L. and {Levi}, M.~E. and {Manera}, M. and {Meisner}, A. and {Miquel}, R. and {Moustakas}, J. and {Mueller}, E. and {Nie}, J. and {Niz}, G. and {Oh}, M. and {Palanque-Delabrouille}, N. and {Poppett}, C. and {Prada}, F. and {Rezaie}, M. and {Rossi}, G. and {Sanchez}, E. and {Schlegel}, D. and {Schubnell}, M. and {Seo}, H. and {Sprayberry}, D. and {Tarl{\'e}}, G. and {Vargas-Maga{\~n}a}, M. and {Weaver}, B.~A. and {Wilson}, Michael J. and {Zheng}, Y. and {DESI Collaboration}},
        title = "{Production of alternate realizations of DESI fiber assignment for unbiased clustering measurement in data and simulations}",
      journal = {\jcap},
         year = 2025,
        month = jan,
       volume = {2025},
       number = {1},
          eid = {127},
        pages = {127},
          doi = {10.1088/1475-7516/2025/01/127},
archivePrefix = {arXiv},
       eprint = {2404.03006},
 primaryClass = {astro-ph.CO},
       adsurl = {https://ui.adsabs.harvard.edu/abs/2025JCAP...01..127L}
}

@ARTICLE{cole1989,
       author = {{Cole}, Shaun and {Kaiser}, Nick},
        title = "{Biased clustering in the cold dark matter cosmogony.}",
      journal = {\mnras},
         year = 1989,
        month = apr,
       volume = {237},
        pages = {1127-1146},
          doi = {10.1093/mnras/237.4.1127},
       adsurl = {https://ui.adsabs.harvard.edu/abs/1989MNRAS.237.1127C}
}

@ARTICLE{bardeen1986,
       author = {{Bardeen}, J.~M. and {Bond}, J.~R. and {Kaiser}, N. and {Szalay}, A.~S.},
        title = "{The Statistics of Peaks of Gaussian Random Fields}",
      journal = {\apj},
         year = 1986,
        month = may,
       volume = {304},
        pages = {15},
          doi = {10.1086/164143},
       adsurl = {https://ui.adsabs.harvard.edu/abs/1986ApJ...304...15B}
}

@ARTICLE{DESI-II,
       author = {{Schlegel}, David J. and {Ferraro}, Simone and {Aldering}, Greg and {Baltay}, Charles and {BenZvi}, Segev and {Besuner}, Robert and {Blanc}, Guillermo A. and {Bolton}, Adam S. and {Bonaca}, Ana and {Brooks}, David and {Buckley-Geer}, Elizabeth and {Cai}, Zheng and {DeRose}, Joseph and {Dey}, Arjun and {Doel}, Peter and {Drlica-Wagner}, Alex and {Fan}, Xiaohui and {Gutierrez}, Gaston and {Green}, Daniel and {Guy}, Julien and {Huterer}, Dragan and {Infante}, Leopoldo and {Jelinsky}, Patrick and {Karagiannis}, Dionysios and {Kent}, Stephen M. and {Kim}, Alex G. and {Kneib}, Jean-Paul and {Kollmeier}, Juna A. and {Kremin}, Anthony and {Lahav}, Ofer and {Landriau}, Martin and {Lang}, Dustin and {Leauthaud}, Alexie and {Levi}, Michael E. and {Linder}, Eric V. and {Magneville}, Christophe and {Martini}, Paul and {McDonald}, Patrick and {Miller}, Christopher J. and {Myers}, Adam D. and {Newman}, Jeffrey A. and {Nugent}, Peter E. and {Palanque-Delabrouille}, Nathalie and {Padmanabhan}, Nikhil and {Palmese}, Antonella and {Poppett}, Claire and {Prochaska}, Jason X. and {Raichoor}, Anand and {Ramirez}, Solange and {Sailer}, Noah and {Schaan}, Emmanuel and {Schubnell}, Michael and {Seljak}, Uros and {Seo}, Hee-Jong and {Silber}, Joseph and {Simon}, Joshua D. and {Slepian}, Zachary and {Soares-Santos}, Marcelle and {Tarle}, Greg and {Valluri}, Monica and {Weaverdyck}, Noah J. and {Wechsler}, Risa H. and {White}, Martin and {Yeche}, Christophe and {Zhou}, Rongpu},
        title = "{A Spectroscopic Road Map for Cosmic Frontier: DESI, DESI-II, Stage-5}",
      journal = {arXiv e-prints},
         year = 2022,
        month = sep,
          eid = {arXiv:2209.03585},
        pages = {arXiv:2209.03585},
          doi = {10.48550/arXiv.2209.03585},
archivePrefix = {arXiv},
       eprint = {2209.03585},
 primaryClass = {astro-ph.CO},
       adsurl = {https://ui.adsabs.harvard.edu/abs/2022arXiv220903585S}
}

@ARTICLE{myers2023,
       author = {{Myers}, Adam D. and {Moustakas}, John and {Bailey}, Stephen and {Weaver}, Benjamin A. and {Cooper}, Andrew P. and {Forero-Romero}, Jaime E. and {Abolfathi}, Bela and {Alexander}, David M. and {Brooks}, David and {Chaussidon}, Edmond and {Chuang}, Chia-Hsun and {Dawson}, Kyle and {Dey}, Arjun and {Dey}, Biprateep and {Dhungana}, Govinda and {Doel}, Peter and {Fanning}, Kevin and {Gazta{\~n}aga}, Enrique and {Gontcho A Gontcho}, Satya and {Gonzalez-Morales}, Alma X. and {Hahn}, ChangHoon and {Herrera-Alcantar}, Hiram K. and {Honscheid}, Klaus and {Ishak}, Mustapha and {Karim}, Tanveer and {Kirkby}, David and {Kisner}, Theodore and {Koposov}, Sergey E. and {Kremin}, Anthony and {Lan}, Ting-Wen and {Landriau}, Martin and {Lang}, Dustin and {Levi}, Michael E. and {Magneville}, Christophe and {Napolitano}, Lucas and {Martini}, Paul and {Meisner}, Aaron and {Newman}, Jeffrey A. and {Palanque-Delabrouille}, Nathalie and {Percival}, Will and {Poppett}, Claire and {Prada}, Francisco and {Raichoor}, Anand and {Ross}, Ashley J. and {Schlafly}, Edward F. and {Schlegel}, David and {Schubnell}, Michael and {Tan}, Ting and {Tarle}, Gregory and {Wilson}, Michael J. and {Y{\`e}che}, Christophe and {Zhou}, Rongpu and {Zhou}, Zhimin and {Zou}, Hu},
        title = "{The Target-selection Pipeline for the Dark Energy Spectroscopic Instrument}",
      journal = {\aj},
         year = 2023,
        month = feb,
       volume = {165},
       number = {2},
          eid = {50},
        pages = {50},
          doi = {10.3847/1538-3881/aca5f9},
archivePrefix = {arXiv},
       eprint = {2208.08518},
 primaryClass = {astro-ph.IM},
       adsurl = {https://ui.adsabs.harvard.edu/abs/2023AJ....165...50M}
}

@ARTICLE{ross2009,
       author = {{Ross}, Nicholas P. and {Shen}, Yue and {Strauss}, Michael A. and {Vanden Berk}, Daniel E. and {Connolly}, Andrew J. and {Richards}, Gordon T. and {Schneider}, Donald P. and {Weinberg}, David H. and {Hall}, Patrick B. and {Bahcall}, Neta A. and {Brunner}, Robert J.},
        title = "{Clustering of Low-redshift (z <= 2.2) Quasars from the Sloan Digital Sky Survey}",
      journal = {\apj},
         year = 2009,
        month = jun,
       volume = {697},
       number = {2},
        pages = {1634-1655},
          doi = {10.1088/0004-637X/697/2/1634},
archivePrefix = {arXiv},
       eprint = {0903.3230},
 primaryClass = {astro-ph.CO},
       adsurl = {https://ui.adsabs.harvard.edu/abs/2009ApJ...697.1634R}
}

@ARTICLE{conroy2013,
       author = {{Conroy}, Charlie and {White}, Martin},
        title = "{A Simple Model for Quasar Demographics}",
      journal = {\apj},
         year = 2013,
        month = jan,
       volume = {762},
       number = {2},
          eid = {70},
        pages = {70},
          doi = {10.1088/0004-637X/762/2/70},
archivePrefix = {arXiv},
       eprint = {1208.3198},
 primaryClass = {astro-ph.CO},
       adsurl = {https://ui.adsabs.harvard.edu/abs/2013ApJ...762...70C}
}





\end{document}